%% file: B2G-16-024_temp.tex
\pdfoutput=1

\documentclass[11pt,twoside,a4paper,cmspaper,final,collab]{cms-tdr}
\def\svnVersion{435508P}\def\cmsCernNoTag{CERN-EP-2017-107}\def\cmsCernDate{\today}\def\cmsMessage{Published in the Journal of High Energy Physics as \href{http://dx.doi.org/10.1007/JHEP11(2017)085}{\doi{10.1007/JHEP11(2017)085.}}}

\begin{document}\cmsNoteHeader{B2G-16-024}

\hyphenation{had-ron-i-za-tion}
\hyphenation{cal-or-i-me-ter}
\hyphenation{de-vices}
\RCS$Revision: 435317 $
\RCS$HeadURL: svn+ssh://svn.cern.ch/reps/tdr2/papers/B2G-16-024/trunk/B2G-16-024.tex $
\RCS$Id: B2G-16-024.tex 435317 2017-11-20 16:17:15Z jmhogan $

\providecommand{\PQT}{\ensuremath{\cmsSymbolFace{T}}\xspace}
\providecommand{\PQB}{\ensuremath{\cmsSymbolFace{B}}\xspace}
\providecommand{\PAQT}{\ensuremath{\overline{\cmsSymbolFace{T}}}\xspace}
\providecommand{\PAQB}{\ensuremath{\overline{\cmsSymbolFace{B}}}\xspace}
\newcommand{\ST}{\ensuremath{S_\mathrm{T}}\xspace}
\newcommand{\TTbar}{\ensuremath{\PQT\PAQT}\xspace}
\newcommand{\BBbar}{\ensuremath{\PQB\PAQB}\xspace}
\newcommand{\qsec}[1]{Section~\ref{#1}}
\newcommand{\qfig}[1]{Fig.~\ref{#1}}
\newcommand{\qtab}[1]{Table~\ref{#1}}
\newcommand{\boostedW}{boosted W\xspace}
\newcommand{\boostedHiggs}{boosted H\xspace}

\cmsNoteHeader{B2G-16-024}
\title{Search for pair production of vector-like T and B quarks in single-lepton final states using boosted jet substructure in proton-proton collisions at $\sqrt{s} = 13$\TeV}

\date{\today}

\abstract{A search for pair production of massive vector-like T and B quarks in proton-proton collisions at $\sqrt{s} = 13$\TeV is presented. The data set was collected in 2015 by the CMS experiment at the LHC and corresponds to an integrated luminosity of up to 2.6\fbinv. The T and B quarks are assumed to decay through three possible channels into a heavy boson (either a W, Z or Higgs boson) and a third generation quark. This search is performed in final states with one charged lepton and several jets, exploiting techniques to identify W or Higgs bosons decaying hadronically with large transverse momenta. No excess over the predicted standard model background is observed. Upper limits at 95\% confidence level on the T quark pair production cross section are set that exclude T quark masses below 860\GeV in the singlet, and below 830\GeV in the doublet branching fraction scenario. For other branching fraction combinations with $\mathcal{B}(\PQT \to \PQt\PH) + \mathcal{B}(\PQT \to \PQb\PW) \geq 0.4$, lower limits on the T quark range from 790 to 940\GeV. Limits are also set on pair production of singlet vector-like B quarks, which can be excluded up to a mass of 730\GeV.
The techniques showcased here for understanding highly-boosted final states are important as the sensitivity to new particles is extended to higher masses.}

\hypersetup{%
pdfauthor={CMS Collaboration},
pdftitle={Search for pair production of vector-like T and B quarks in single-lepton final states using boosted jet substructure in proton-proton collisions at sqrt(s) = 13 TeV},%
pdfsubject={CMS},%
pdfkeywords={CMS, physics}}

\maketitle \section{Introduction}

The discovery of a light mass Higgs boson (H)~\cite{Aad20121,Chatrchyan201230,Chatrchyan:2013lba} motivates searches for new interactions and particles at the LHC~\cite{LHC}. Cancellation of the loop corrections to the Higgs boson mass without precise fine tuning of parameters requires new particles at the TeV scale. Such new particles are the bosonic partners of the top quark, in supersymmetric models, or the fermionic top quark partners predicted by many other theories,
such as little Higgs~\cite{PhysRevD.69.075002,Matsedonskyi2013} and composite Higgs~\cite{PhysRevD.75.055014, compHiggs, KAPLAN1991259, Dugan:1984hq} models. These heavy quark partners predominantly mix with the third-generation quarks of the standard model (SM)~\cite{vecQuarkMix,PhysRevLett.82.1628} and have vector-like transformation properties under the SM gauge group $\mathrm{SU(2)_L \times U(1)_Y \times SU(3)_C}$, hence the term ``vector-like quarks'' (VLQ). While a chiral extension of the SM quark family has been strongly disfavored by precision electroweak studies at electron-positron colliders~\cite{LEP-2,PhysRevLett.109.241802} and by observed production cross sections and branching fractions of the Higgs boson~\cite{Djouadi2012310}, models with VLQs are not excluded by present data.

\begin{figure}[bp]
\centering
\includegraphics[width=0.32\textwidth]{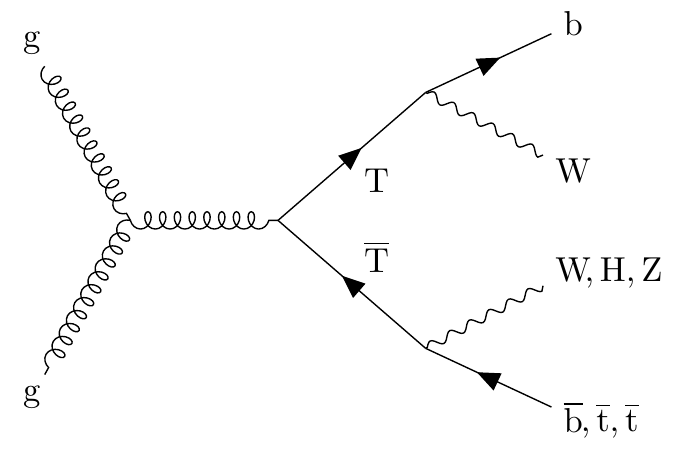}
\includegraphics[width=0.32\textwidth]{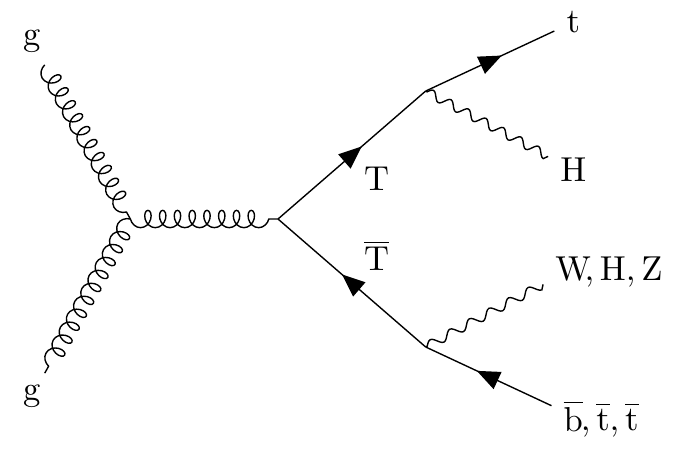}
\includegraphics[width=0.32\textwidth]{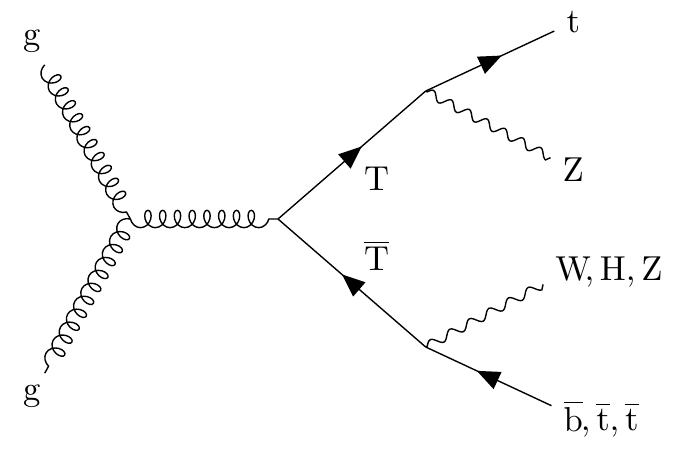}
\caption{Examples of leading-order Feynman diagrams showing production of a \TTbar pair with the T quark decaying to bW (left), tH (middle), and tZ (right).}
\label{fig:diagrams}

\end{figure}

We search for a vector-like T quark with charge 2/3 (in units of the electron charge) that is produced via the strong interaction in proton-proton collisions along with its antiquark, \PAQT. Many models in which VLQs appear assume that T quarks decay to three final states: bW, tZ, or tH~\cite{PhysRevD.88.094010}. Leading-order Feynman diagrams of these three processes are shown in \qfig{fig:diagrams}, created with the tools of Ref.~\cite{Ellis:2016jkw}. The partial decay widths depend on the particular model \cite{DeSimone2013}, so that the branching fractions of these decay modes can take on various possible values, with the sum of all three branching fractions equal to unity. An electroweak isospin singlet T quark is expected to have a branching fraction of approximately 50\% for $\PQT\to\PQb\PW$, and 25\% for each of $\PQT\to\PQt\Z$ and tH, and is used as a benchmark for figures and tables. A T quark in a weak isospin doublet has no decays to bW and equal branching fractions for tZ and tH decays~\cite{delAguila:1989rq, DeSimone2013,signalxsec}. As these are, however, not the only possible representations of T quarks, the final results are interpreted for many allowed branching fraction combinations.

Though this search is optimized for \TTbar production, decays of vector-like bottom quark partners (B quarks) can produce similar topologies and \BBbar production is also considered. The B quark with charge $-1/3$ is expected to decay to tW, bH, or bZ and can also transform either as a singlet or doublet under the electroweak symmetry group. The respective branching fractions are equal to those of the corresponding T quark decays to the same SM bosons. For this search we assume that only one new particle is present, either the T or B quark.

Most recently, searches for pair-produced T and B quarks were performed by both the ATLAS and CMS collaborations at $\sqrt{s} = 8$\TeV~\cite{CMScombo2014,Run1anal,PhysRevD.92.112007,Aad2015, Khachatryan:2015gza, Aad:2015mba}. Depending on the assumed combination of branching fractions to the three decay modes, the CMS collaboration observed lower limits on the T quark mass with values ranging from 720 to 920\GeV and on the B quark mass with values ranging from 740 to 900\GeV at 95\% confidence level (CL)~\cite{CMScombo2014,Khachatryan:2015gza}. The ATLAS collaboration found similar lower mass limits, so that vector-like T and B quarks with masses below 720\GeV are already excluded for all possible branching fraction combinations. We therefore only consider VLQ masses above 700\GeV in this search. The ATLAS collaboration has also searched for pair production of T and B quarks at $\sqrt{s} = 13$\TeV~\cite{Aaboud:2017qpr,Aaboud:2017zfn}.

We require one electron or one muon in the final state, along with several jets. All decay modes of the T and B quarks produce t quarks and/or W bosons, which are the dominant sources of leptons. In the high mass region that we consider, the decay products can have a large Lorentz boost and result in highly collinear final state particles.
This search makes use of techniques to identify b quark jets and reconstruct hadronic decays of massive particles that are highly Lorentz-boosted in the reference frame of the \TTbar system. The data are analyzed in two channels that are optimized for sensitivity to either boosted W or Higgs bosons, referred to as the ``\boostedW'' and ``\boostedHiggs'' channels. The \boostedW channel is most sensitive to scenarios where the T quark has a large branching fraction for bW decays (such as the electroweak singlet benchmark) while the \boostedHiggs channel has the highest sensitivity to scenarios with a large branching fraction to tH (such as the electroweak doublet benchmark). The $\PQT\to\PQt\Z$ decay mode is not a particular target of this search, but
Lorentz-boosted Z bosons decaying hadronically can be selected in either channel since the signatures are similar to those of boosted hadronic W or Higgs boson decays, thus providing some sensitivity to the tZ decay mode.

\section{The CMS detector and event reconstruction}
\label{sec:cms}

The central feature of the CMS apparatus is a superconducting solenoid of 6\unit{m} internal diameter, providing a magnetic field of 3.8\unit{T}. Within the solenoid volume are a silicon pixel and strip tracker, a lead tungstate crystal electromagnetic calorimeter (ECAL), and a brass and scintillator hadron calorimeter (HCAL), each composed of a barrel and two endcap sections. Forward calorimeters extend the pseudorapidity ($\eta$)~\cite{Chatrchyan:2008zzk} coverage provided by the barrel and endcap detectors. Muons are measured in gas-ionization detectors embedded in the steel flux-return yoke outside the solenoid.

A particle-flow (PF) algorithm~\cite{CMS-PRF-14-001} is used to reconstruct and identify each individual particle in an event with an optimized combination of information from the various elements of the CMS detector. The energy of photons is directly obtained from the ECAL measurement, corrected for zero-suppression effects.
The energy of electrons is determined from a combination of the electron momentum at the primary interaction vertex as determined by the tracker, the energy of the corresponding ECAL cluster, and the energy sum of all bremsstrahlung photons spatially compatible with originating from the electron track. The momentum resolution for electrons with transverse momentum $\pt \approx 45$\GeV from $\Z \to \Pep \Pem$ decays ranges from 1.7\% for low-bremsstrahlung electrons in the barrel region to 4.5\% for showering electrons in the endcaps~\cite{Khachatryan:2015hwa}. The energy of muons is obtained from the curvature of the corresponding track. Matching muons to tracks measured in the silicon tracker results in a relative transverse momentum resolution for muons with $20 <\pt < 100\GeV$ of 1.3--2.0\% in the barrel and better than 6\% in the endcaps. The \pt resolution in the barrel is better than 10\% for muons with \pt up to 1\TeV~\cite{Chatrchyan:2012xi}. The energy of charged hadrons is determined from a combination of their momenta measured in the tracker and the matching ECAL and HCAL energy deposits, corrected for zero-suppression effects and for the response function of the calorimeters to hadronic showers. Finally, the energy of neutral hadrons is obtained from the corresponding corrected ECAL and HCAL energy.

Jets are reconstructed from the individual particles produced by the PF event algorithm, clustered using the anti-\kt algorithm~\cite{Cacciari:2008gp, Cacciari:2011ma} with distance parameters of 0.4 (``AK4 jets'') or 0.8 (``AK8 jets''). Jet momentum is defined as the vectorial sum of all particle momenta in the jet, and is found from simulation to be within 5 to 10\% of the true momentum over the whole \pt spectrum and detector acceptance. All jets are required to have $\abs{\eta} < 2.5$ and AK4 (AK8) jets must have $\pt > 30\,(200)\GeV$. An offset correction is applied to jet energies to take into account the contribution from additional proton-proton interactions within the same or nearby bunch crossings (pileup)~\cite{Cacciari:2008gn}. Jet energy corrections are derived from simulation, and are confirmed with in situ measurements of the energy balance in dijet and photon/$\Z(\to \Pe\Pe/\mu\mu)$ + jet events~\cite{Khachatryan:2016kdb}. A smearing of the jet energy is applied to simulated events to mimic the energy resolution observed in data, typically 15\% at 10\GeV, 8\% at 100\GeV, and 4\% at 1\TeV. Additional selection criteria are applied to each event to remove spurious jet-like features originating from isolated noise patterns in the HCAL~\cite{Chatrchyan:2011ds}, anomalously high energy deposits in certain regions of the ECAL, and cosmic ray and beam halo particles that are detected in the muon chambers.

The missing transverse momentum vector is defined as the projection on the plane perpendicular to the beams of the negative vector sum of the momenta of all reconstructed particles in an event. Its magnitude is referred to as \ETmiss. The energy scale corrections applied to jets are propagated to \ETmiss.

A more detailed description of the CMS detector, together with a definition of the coordinate system used and the relevant kinematic variables, can be found in Ref.~\cite{Chatrchyan:2008zzk}.

\section{Data and simulated samples}
\label{sec:samples}

The data used in this analysis were collected during 2015 when the LHC collided protons at $\sqrt{s} = 13$\TeV with a bunch spacing of 25\unit{ns}.
The data set for the \boostedW channel corresponds to an integrated luminosity of 2.3\fbinv. The data set for the \boostedHiggs channel in the electron (muon) channel corresponds to 2.5 (2.6)\fbinv and includes additional data collected with poor forward calorimeter performance where the \MET has been re-computed excluding the affected region of the detector.

To compare the SM expectation with the experimental data, samples of events for all relevant SM background processes and the \TTbar signal are produced using Monte Carlo (MC) simulation. Background processes are simulated using several matrix element generators. The \POWHEG~v2 generator~\cite{Nason:2004rx,Frixione:2007vw,Alioli:2010xd,Frixione:2007nw} is used to simulate \ttbar events, as well as single top quark events in the tW channel at next-to-leading order (NLO). The \MGvATNLO 2.2.2 generator~\cite{MADGRAPH} is used for generation at NLO of Drell--Yan\,+\,jets and \ttbar+\,W events, as well as \ttbar+\,Z events, and $s$- and $t$-channel production of single top quarks. The FxFx scheme~\cite{FXFX} for merging matrix element generation to the parton shower is used. The \MADGRAPH v5.2.2.2 generator is used with the MLM scheme~\cite{MLMmatching} to generate W\,+\,jets, Drell--Yan\,+\,jets, and multijet events at leading order.
\PYTHIA~8.212~\cite{Sjostrand:2006za,Sjostrand:2014zea} is used for the simulation of multijet and diboson events.

The \boostedW channel uses the NLO Drell--Yan + jets simulation and the \MADGRAPH multijet simulation. The \boostedHiggs channel uses the \MADGRAPH Drell--Yan + jets simulation, and the \PYTHIA multijet simulation which is filtered for processes likely to pass the lepton selection in this channel. Background samples are grouped into three categories for presentation: ``TOP'', dominated by \ttbar and including single top quark and \ttbar+\,W/Z samples; ``EW'', dominated by W\,+\,jets and including Drell--Yan\,+\,jets and diboson samples; and ``QCD'', including multijet samples.

Signal samples for both \TTbar and \BBbar production are simulated using \MADGRAPH for mass points between 700 and 1800\GeV  in steps of 100\GeV . A narrow width of 10\GeV is assumed for the vector-like quarks. Predicted cross sections, which depend only on the vector-like quark mass, are computed at next-to-next-to-leading order (NNLO) with the \textsc{Top++}2.0 program~\cite{TPRIMEXSEC,MITOV1,MITOV2,MITOV3,BARNREUTHER,NNLL} and are listed in \qtab{tab:signal}.

Parton showering and the underlying event for all simulated samples are obtained with \PYTHIA using the CUETP8M1 tune~\cite{Skands:2014pea,Khachatryan:2015pea}. To simulate the momentum spectrum of partons inside the colliding protons, the NNPDF3.0~\cite{NNPDF30} parton distribution functions (PDFs) are used. Detector simulation for all MC samples is performed with \GEANTfour~\cite{GEANT4} and includes the effect of pileup.

\begin{table}[tb]
  \centering
    \topcaption{Predicted cross sections for pair production of T or B quarks for various masses. Uncertainties include contributions from energy scale variations and from the PDFs.}
    \renewcommand{\arraystretch}{1.2}
    \begin{tabular}{c|ccr@{\,$\pm$\,}>{$}l<{$}}
      T or B quark mass [\GeVns{}]  &   \multicolumn{4}{c}{Cross section [fb]} \\ \hline
      700 &&&  455 & 19  \\
      800 &&&  196 & 8 \\
      900 &&&  90 & 4 \\
      1000 &&&  44 & 2 \\
      1100 &&&  22 & 1 \\
      1200 &&&  11.8 & 0.6 \\
      1300 &&&  6.4 & ^{0.4}_{0.3} \\
      1400 &&&  3.5 & 0.2 \\
      1500 &&&  2.0 & 0.1 \\
      1600 &&&  1.15 & ^{0.09}_{0.07} \\
      1700 &&&  0.67 & ^{0.06}_{0.04} \\
      1800 &&&  0.39 & ^{0.04}_{0.03} \\
          \end{tabular}
    \label{tab:signal}

\end{table}

\section{Reconstruction methods}\label{sec:objects}

We perform a search for T quarks that decay to final states with an electron or a muon, and jets. Selected events must have one or more pp interaction vertices within the luminous region (longitudinal position $|z| < 24$ cm and radial position $\rho < 2$ cm), reconstructed using a deterministic annealing filter algorithm~\cite{Chatrchyan:2014fea}. The primary interaction vertex is the vertex with the largest $\sum \pt^2$ from its associated jets, leptons, and \ETmiss. The number of pileup interactions differs between data and simulation, so simulated events are weighted to reflect the pileup distribution expected in data given a total inelastic cross section of 69\unit{mb}~\cite{ATLASpileup}.

Two observables that are useful in discriminating signal from background events, exploiting the fact that the decays of T quarks to single-lepton final states produce a large number of hadronic objects, are the following: the quantity \HT, defined as the scalar \pt sum of all reconstructed AK4 jets with $\pt >30$\GeV and $\abs{\eta} < 2.4$, and the quantity \ST, defined as the scalar sum of \ETmiss, the \pt of the lepton, and \HT.

\subsection{Lepton reconstruction and selection}
This search requires one charged lepton, either an electron or a muon, to be reconstructed within the acceptance region of $\abs{\eta} < 2.4$. The event must satisfy a single-electron or single-muon trigger. The choice of triggers is adapted to the particular final state targeted in each channel. In $\PQT\to\PQb\PW$ decays, the W boson is generally well separated from the associated bottom quark since the T quark has low \pt compared to its mass, leading to a low level of hadronic activity in close proximity to the lepton. In contrast, a lepton originating from a top quark decay (\eg, from a $\PQT \to \PQt\PH$ decay) becomes increasingly collinear with the associated bottom quark as the T quark mass increases and the Lorentz boost of the top quark rises.

As a consequence of the above, the \boostedW channel uses triggers selecting leptons that are isolated with respect to nearby PF candidates, either electron candidates with $\pt > 27$\GeV  and $\abs{\eta} < 2.1$, or muon candidates with $\pt > 20\GeV$. The triggers used in the \boostedHiggs channel do not require that the leptons are isolated. In the electron channel, events with at least one electron candidate with $\pt > 45\GeV$, one AK4 jet with $\pt > 200\GeV$, and another AK4 jet with $\pt > 50\GeV$ are selected by the trigger. The muon channel trigger selects events with a muon candidate with $\pt > 45\GeV$ and $\abs{\eta} <$ 2.1. Methods to evaluate lepton isolation efficiency after trigger selection are described below.

Additional lepton identification quality criteria are required to reduce the contribution from background events containing other particles misidentified as leptons. For electrons these quality requirements~\cite{Khachatryan:2015hwa} combine variables measuring track quality, the association between the track and electromagnetic shower, shower shape, and the likelihood of the electron to originate from a photon. Electrons are identified in the \boostedHiggs channel using a set of selection criteria with an efficiency of ${\approx}88\%$ and misidentification rate of ${\approx}7\%$. In the \boostedW channel, two working points are defined based on a multivariate identification algorithm: a tight level with ${\approx}88\%$ efficiency (${\approx}4\%$ misidentification rate) and a loose level with ${\approx}95$\% efficiency (${\approx}5\%$ misidentification rate).

Muons are reconstructed by fitting hits in the silicon tracker together with hits in the muon detectors~\cite{Chatrchyan:2012xi}. Identification algorithms consider the quality of this fit, the number or fraction of valid hits in the trackers and muon detectors, track kinks, and the minimum distance between the extrapolated track from the silicon tracker and the primary interaction vertex. Several working points are defined: the \boostedW channel uses so-called ``tight'' (``loose'') muons with ${\approx}97$\% (100\%) efficiency in the barrel region, and the \boostedHiggs channel uses ``medium'' muons with ${\approx}99$\% efficiency in the barrel region. All muon identification working points have hadron misidentification rates of $<$1\%.

Leptons that pass the requirements in the two channels are removed from jets that have an angular separation of $\Delta{R} < 0.4$ from the lepton. This is done by matching PF candidates identified as leptons to the ones identified as jets and subtracting the four-momentum of a matched lepton candidate from the jet four-momentum.

In order to reduce the rate of background events that contain a soft lepton (\eg, from semileptonic bottom quark decays in multijet events), several metrics can be used to evaluate the isolation of a lepton from surrounding particles.
In the \boostedHiggs channel, either an angular separation of $\Delta{R}(\ell,j) > 0.4$, or $\pt^\text{rel}(\ell,j) > 40$\GeV is required. Here, $\ell$ denotes the highest \pt lepton, $j$ is the jet closest to that lepton in angular separation, and $\pt^\text{rel}(\ell,j)$ is the projection of the lepton momentum on the direction perpendicular to the jet momentum in the $\ell$-$j$ plane. These criteria, also referred to as ``2D isolation'', ensure a high signal efficiency for decays such as  $\PQT\to\PQt\PH$, with leptons produced close to jets, while rejecting a large fraction of the multijet background.

In the \boostedW channel, where fewer leptons with nearby b quarks are expected, isolation is evaluated using mini-isolation ($I_\text{mini}$), defined as the sum of the transverse momenta of PF candidates within a \pt-dependent cone around the lepton, corrected for the effects of pileup and divided by the lepton \pt. The radius of the isolation cone, $R_I$, is defined as:
\begin{equation}
R_I = \frac{10\GeV}{\min(\max(\pt,50\GeV),200\GeV)}.
\end{equation}

Using a \pt-dependent cone size allows for greater efficiency at high energies where jets and leptons are more likely to overlap. ``Tight'' electrons (muons) must have $I_\text{mini} < 0.1\,(0.2)$ while ``loose'' electrons and muons satisfy $I_\text{mini} < 0.4$. In addition, the 2D isolation requirement is applied to remove any residual overlap between mini-isolated leptons and jets.

Scale factors that account for selection efficiency differences between data and simulation are calculated as a function of lepton \pt and $\eta$ using a ``tag-and-probe'' method~\cite{Khachatryan:2015hwa, Chatrchyan:2012xi,Khachatryan:2010xn}.
These were calculated in separate measurements for the single-lepton trigger, lepton identification, and $I_\text{mini}$ requirements.

These scale factors are applied to simulated events for both lepton flavors. For the 2D isolation requirement, no significant difference is found between the selection efficiencies in data and simulation and hence no scale factor is applied.

\subsection{Hadronic W and H tagging}
\begin{figure}[tbp]
\centering
\includegraphics[width=0.48\textwidth]{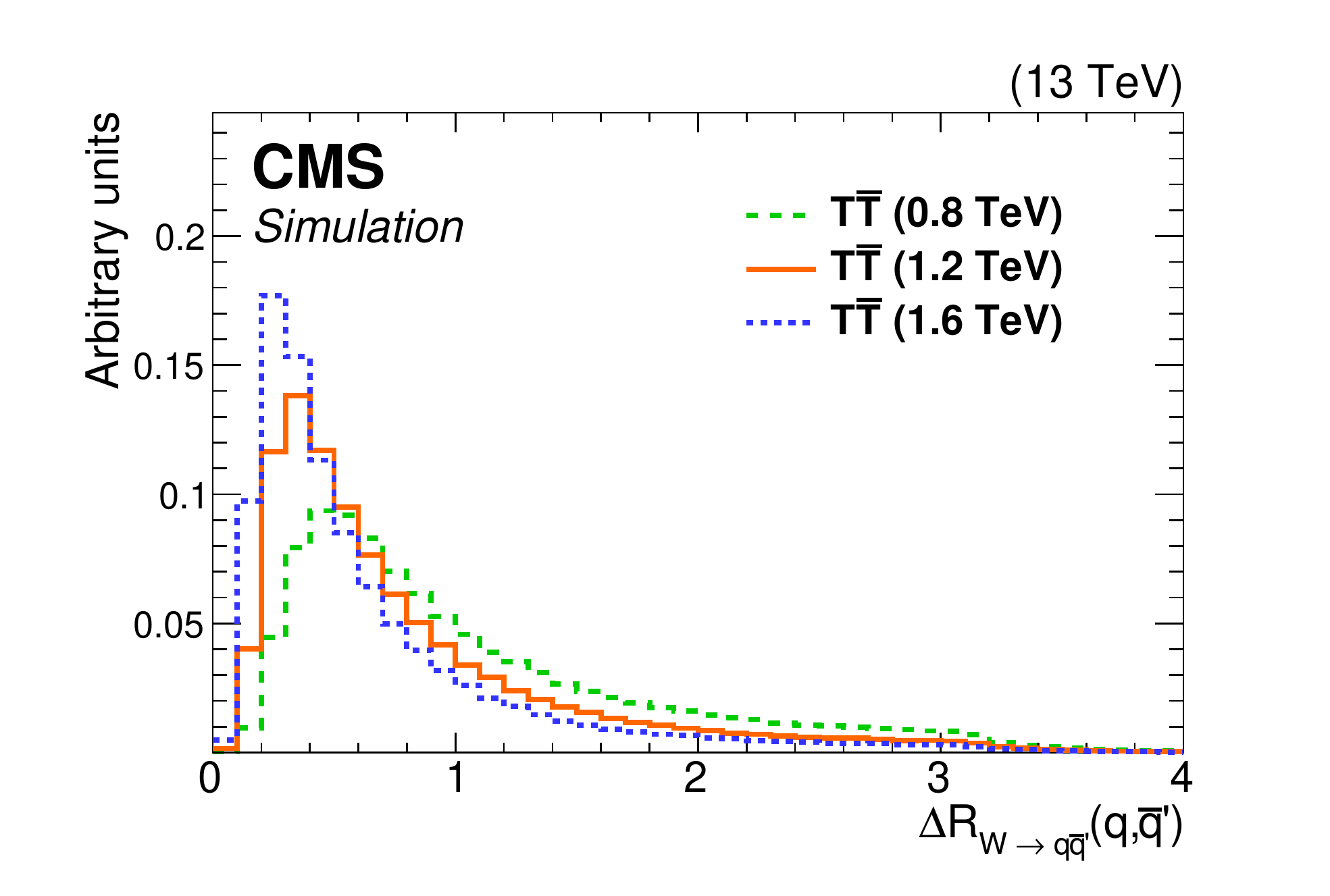}
\includegraphics[width=0.48\textwidth]{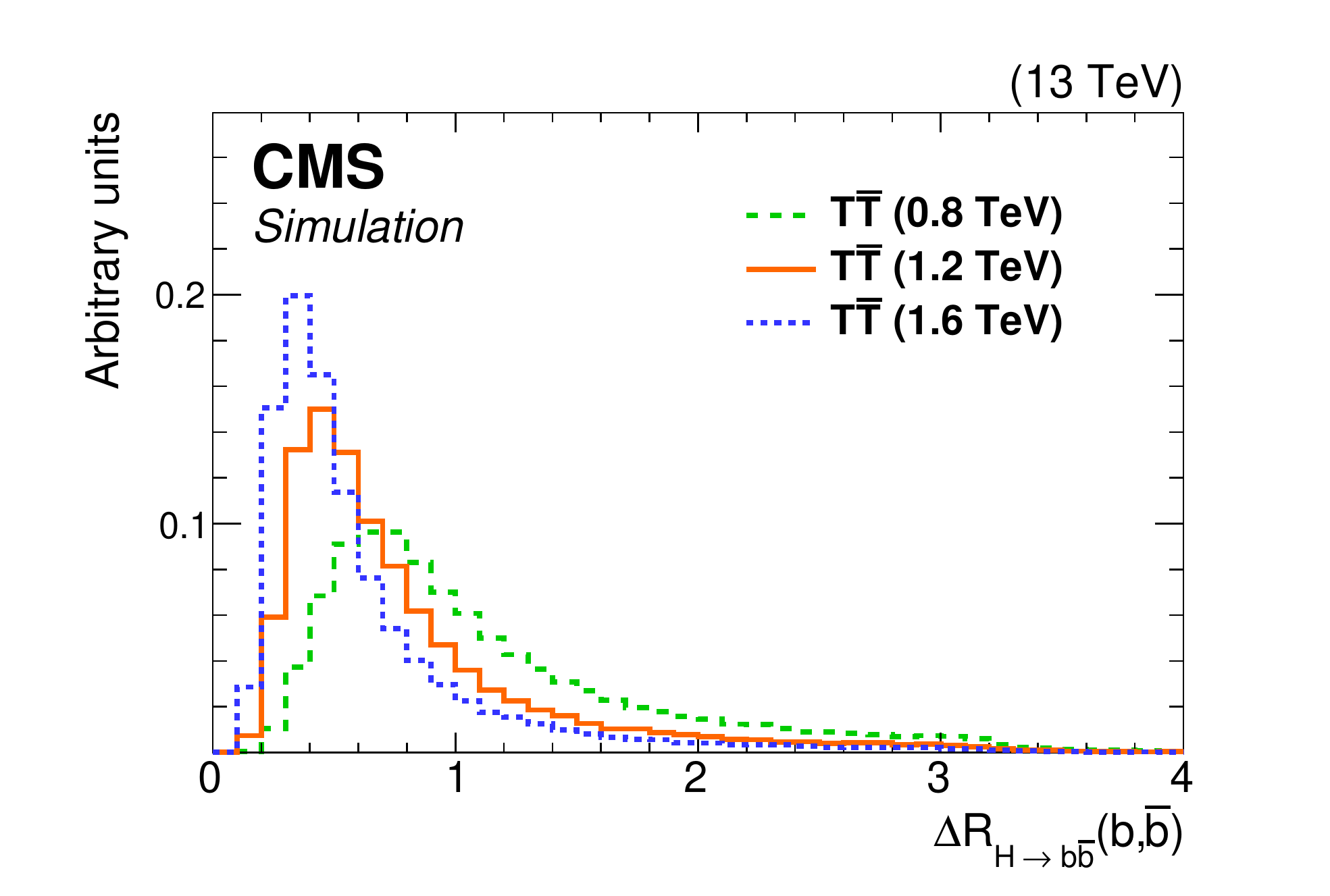}
\caption{Angular separations $\Delta{R}$ between the products of simulated  $\PW\to\PQq\PAQq'$ (left) and  $\PH\to \bbbar$ (right) decay processes for three different mass points of the T quark. Even for the lowest mass point shown, the final state particles are typically emitted with a separation of $\Delta{R} < 0.8$ and are merged into an AK8 jet.}
\label{fig:gen_dr_plots}

\end{figure}

In the decay of a heavy T quark, particles are produced with high momentum and large Lorentz boost. The decay products of top quarks and W, Z, or Higgs bosons are therefore often collimated. This can be seen in \qfig{fig:gen_dr_plots} in which the angular separation $\Delta{R}$ between the products of simulated $\PW \to \PQq\PAQq'$ and $\PH\to \bbbar$ decays are shown for several T quark masses. Even for the lightest considered mass point this separation often has values of $\Delta{R} < 0.8$, where the decay products of heavy bosons can merge into a single AK8 jet.

A jet shape variable called ``$N$-subjettiness''~\cite{NSUBJETS}, denoted as $\tau_N$, is defined as the sum of the transverse momenta of $k$ constituent particles weighted by their minimum angular separation from one of $N$ subjet candidates ($\Delta R_{N,k}$), which are in a jet of characteristic radius $R_0$:
\begin{equation}
\tau_N = \frac{1}{R_0\sum_k p_{\mathrm{T},k}}\sum_k p_{\mathrm{T},k} \min(\Delta R_{1,k},\Delta R_{2,k},\ldots,\Delta R_{N,k}).
\end{equation}
This variable quantifies the consistency of a jet with originating from an $N$-prong particle decay.
The ratio $\tau_2/\tau_1$ provides high sensitivity to two-prong decays such as $\PW\to\PQq\PQq'$.
Jet grooming techniques (``pruning'' and ``soft drop'') are used to remove soft and wide-angle radiation so that the mass of the hard constituents can be measured more precisely~\cite{PRUNING, SOFTDROP}. The pruning procedure reclusters the jet, removing soft or large-angle particles, while the soft drop algorithm recursively declusters the jet, removing sub-clusters until two subjets are identified within the AK8 jet. AK8 jets are reconstructed independently of AK4 jets, so they will frequently overlap. Unless otherwise stated, such overlapping jets are not removed when applying selections based on jet multiplicity.

The AK4 jets and subjets of AK8 jets can be tagged as originating from b quarks based on information about secondary vertices and displaced tracks within the jet. The efficiency for tagging b hadron jets in simulation is approximately 65\%, averaged over jet \pt (slightly lower for subjets of AK8 jets), and the probability of mistagging a charm (light) quark jet is 13\% (1\%)~\cite{BTAG}. Scale factors, which are functions of jet \pt and flavor, are applied to account for efficiency differences between data and simulation.

An AK8 jet is labeled as ``W tagged'' if it has $\pt > 200$\GeV, $\abs{\eta}<2.4$, pruned jet mass between 65 and 105\GeV, and the ratio $\tau_2/\tau_1 < 0.6$. Differences in the pruned jet mass distribution and $\tau_2/\tau_1$ selection efficiency between data and simulation have been evaluated in Ref.~\cite{WTAGSFS}. To account for these differences, pruned jet mass scale factors and mass resolution smearing factors are applied in simulation to all AK8 jets. A $\tau_2/\tau_1$ selection scale factor is applied in simulation to jets that are spatially matched to true boosted products of a hadronic W boson decay.

Higgs boson candidate jets are reconstructed by exploiting the significant branching fraction of the Higgs boson to \bbbar pairs. AK8 jets are marked as ``H tagged'' if they have $\pt > 300\GeV$, soft drop jet mass in the range 60--160\GeV, and if at least one of the two subjets from the soft drop algorithm is tagged as a bottom subjet.

\section{Boosted H channel}\label{sec:boostedH}

\subsection{Event selection and categorization}
\label{sec:ev_selection}

In this channel, one electron with $\pt > 50\GeV$ and $\abs{\eta} < 2.4$, or one muon with $\pt > 47\GeV$ and $\abs{\eta} < 2.1$ is required. In events with an electron,
at least one AK4 jet with $\pt > 250\GeV$ and a second AK4 jet with $\pt > 70\GeV$ are required to select events with a nearly constant trigger efficiency. Furthermore, selected events must have $\ST > 800$\GeV, at least three AK4 jets, and at least two AK8 jets, since we expect a hadronic decay of a boosted Higgs boson in each event along with at least one other hadronic t quark, W, Z, or further Higgs boson decay.
For the rejection of non top quark backgrounds, at least one b-tagged AK4 jet is required.

\begin{figure}[tbp]
\centering
\includegraphics[width=0.48\textwidth]{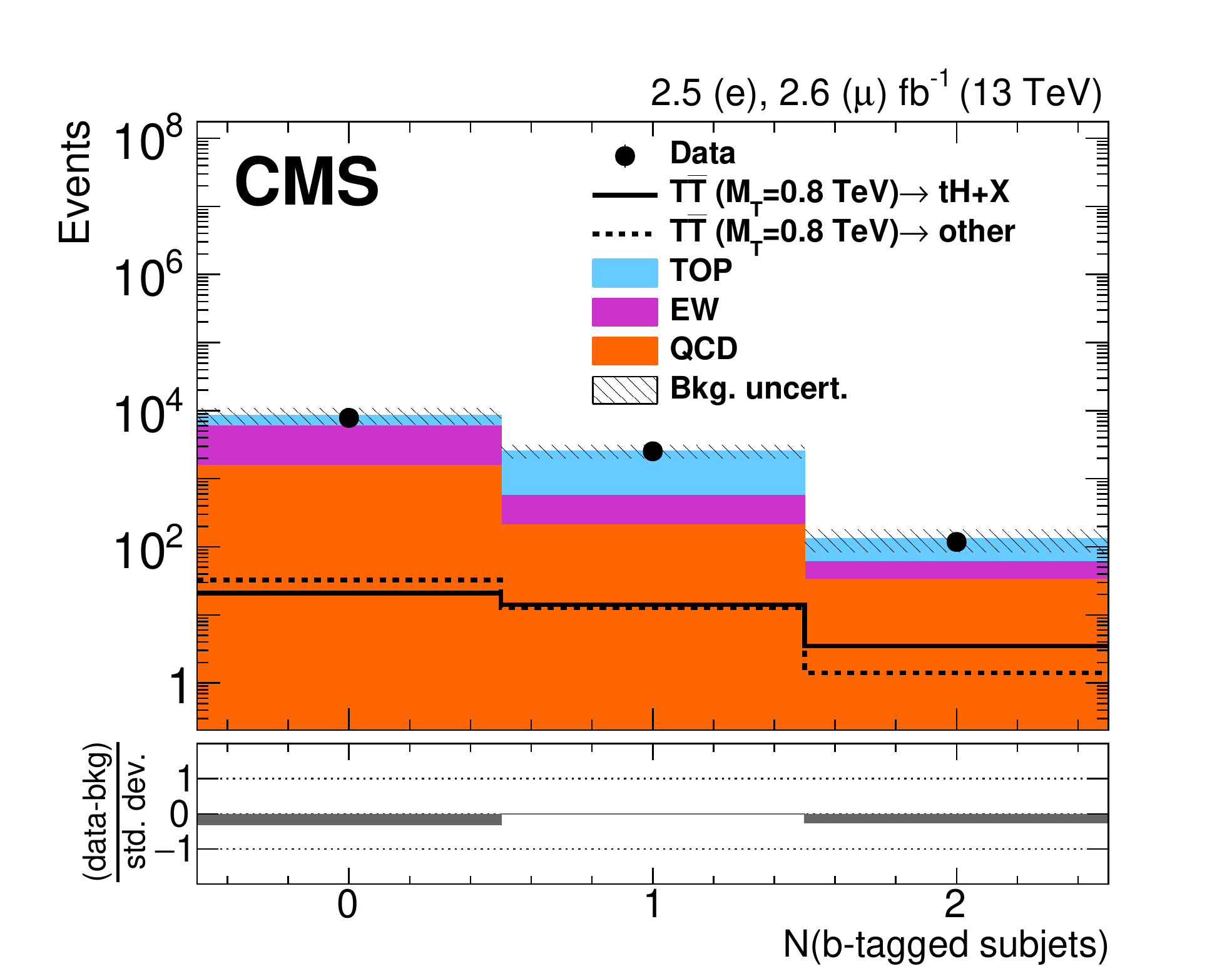}
\includegraphics[width=0.48\textwidth]{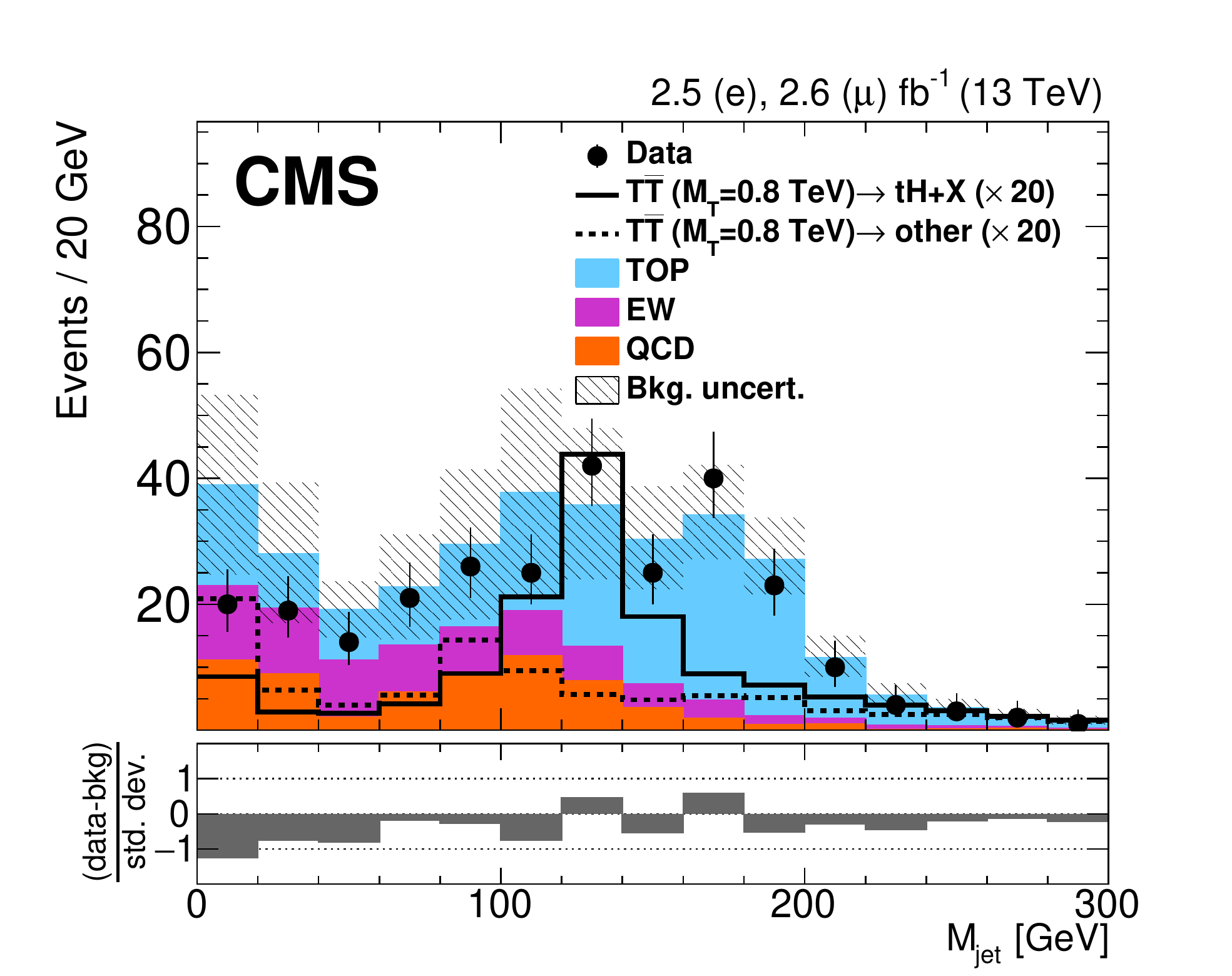}
\caption{
Distributions of the number of b-tagged subjets of the highest \pt H-tagged jet candidate with $\pt > 300\GeV$ and $M_\text{jet}$ in the range [60, 160]\GeV (left), and $M_\text{jet}$ of the highest \pt H-tagged jet candidate with $\pt > 300\GeV$ and two subjet b tags (right). A T quark signal with M(T) = 0.8\TeV is shown (right), normalized to the predicted cross section and scaled by a factor of 20, with the singlet benchmark branching fractions assumed. The solid (dashed) curve shows \TTbar events with at least one (zero) Higgs boson decay, where contributions from each decay mode are weighted to reflect the singlet branching fraction scenario. The uncertainty in the background includes the statistical and systematic uncertainties described in \qsec{sec:systs}.}
\label{fig:higgsvar_plots}
\end{figure}

Distributions of the variables used in the H-tagging algorithm, as described in \qsec{sec:objects}, are shown in \qfig{fig:higgsvar_plots}. These distributions are from events that pass all selection criteria outlined above except for the b-tagging requirement, and that have the corrections described in \qsec{sec:bkg_model} applied.
The distribution of the number of b-tagged subjets for the highest \pt AK8 jet with soft drop jet mass within 60--160\GeV is shown along with the mass of the highest \pt AK8 jet with two b-tagged subjets, before the mass requirement. To illustrate the sensitivity of the H-tagging algorithm to the presence of boosted Higgs bosons, the \TTbar signal with a mass of 1200\GeV is split into two curves: the solid curve shows \TTbar events where at least one Higgs boson is present in the decay chain and the dashed curve shows \TTbar events with only $\PQT\to\PQt\Z$ or $\PQT\to\PQb\PW$ decays. It can be seen that signal events with at least one $\PQT\to\PQt\PH$ decay produce a clear peak at 125\GeV in the mass distribution of the H-tagged jet. Signal events without a Higgs boson in the decay chain have a less pronounced increase at 90\GeV because of hadronic Z boson decays.

After passing the selection defined above, events are split into two exclusive categories, which depend on the number of b-tagged subjets of H-tagged jets, and are defined as follows:
\begin{itemize}
    \item{H2b:} events with at least one H-tagged jet with exactly \textit{two} b-tagged subjets.
    \item {H1b:} events with at least one H-tagged jet with exactly \textit{one} b-tagged subjet. \end{itemize}

To avoid an overlap between the two categories, any event is first checked whether it falls into the H2b category and only if it does not, it can enter into the H1b category.

\subsection{Background modeling}
\label{sec:bkg_model}

\begin{figure}[tbp]
\centering
\includegraphics[width=0.48\textwidth]{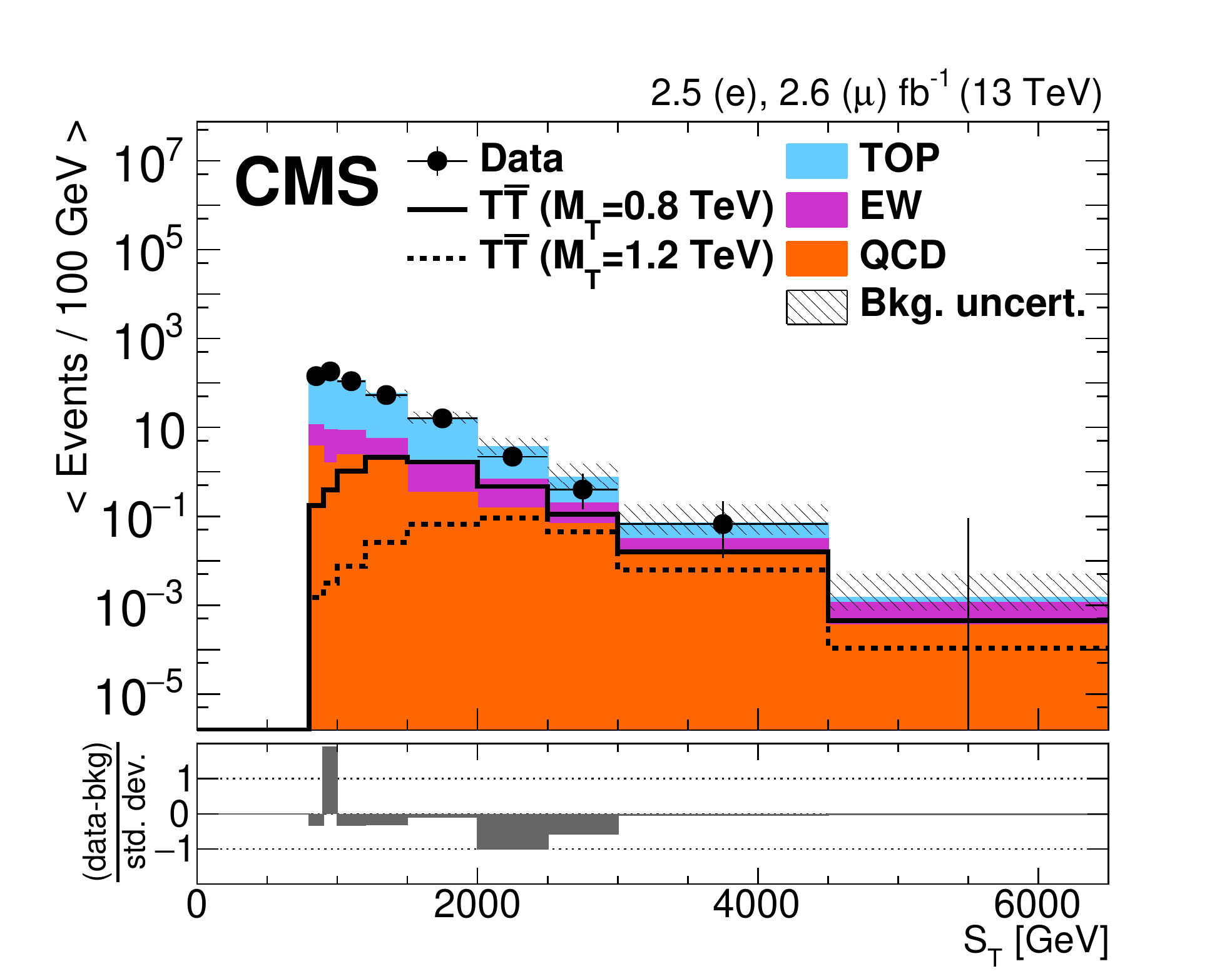}
\includegraphics[width=0.48\textwidth]{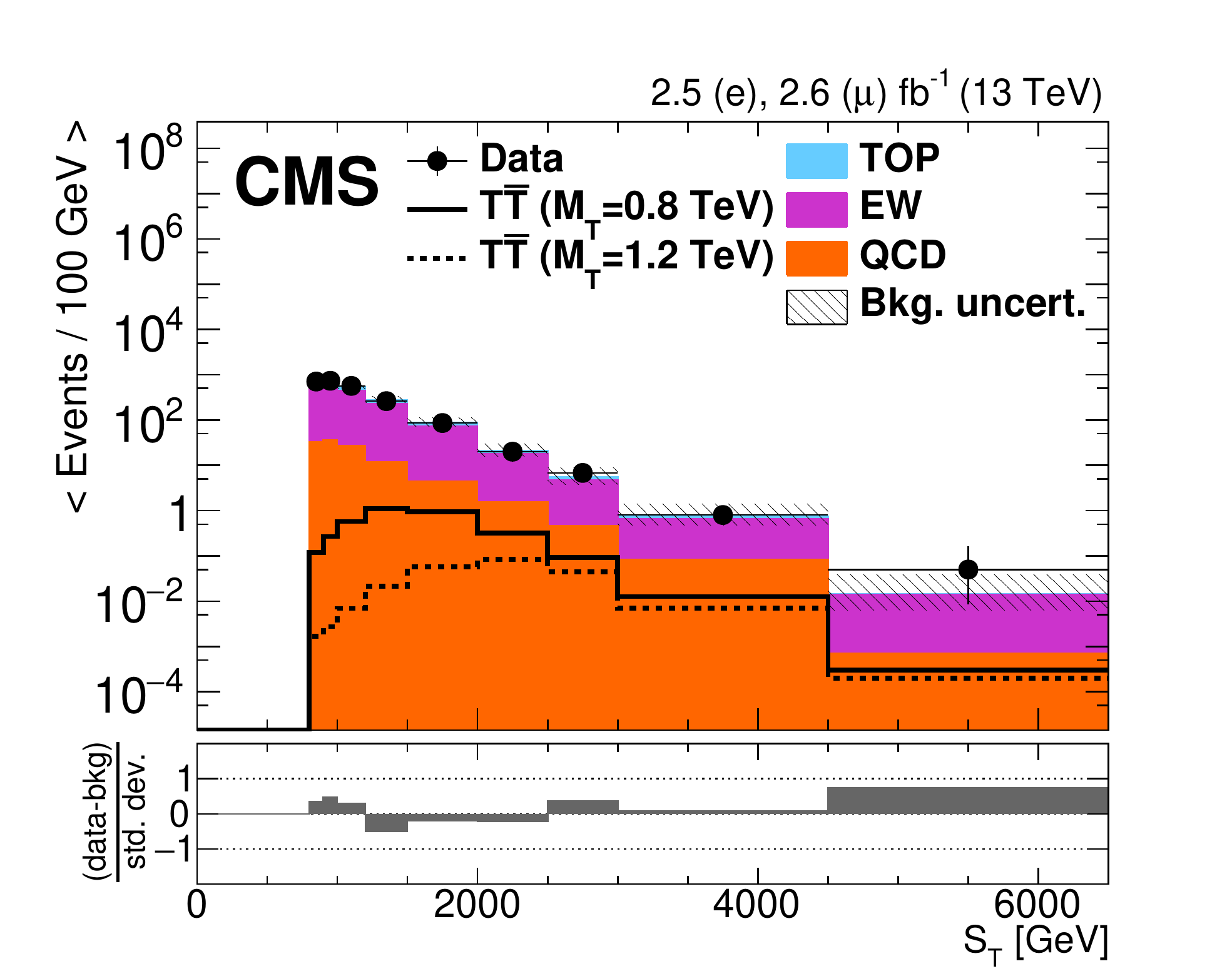}
\caption{
Distributions of \ST in the \ttbar (left) and W\,+\,jets (right) control regions of the boosted H channel after applying all corrections to their shape and normalization. The \TTbar signal, shown for T quark masses of 0.8 and 1.2\TeV, is normalized to the theoretical cross section and the singlet benchmark branching fractions are assumed. The uncertainty in the background includes statistical and systematic uncertainties described in \qsec{sec:systs}.
}
\label{fig_cr}
\end{figure}

To evaluate the modeling of \ttbar and W\,+\,jets production, the dominant background processes, two control regions that are enriched in events from these processes are defined by modifying the event selection defined in \qsec{sec:ev_selection}. In the \ttbar control region, at least two b-tagged jets are required instead of at least one. In the W\,+\,jets control region, the requirement of at least one b-tagged jet is inverted and events with any b-tagged jets are rejected. Events with an H-tagged jet are rejected in both control regions to reduce the signal contribution in these regions, and $\MET > 100$\GeV is required to reject events from multijet production. The signal to background ratio is about six times smaller than the one in the H2b category in the \ttbar control region and about 30 times smaller in the W\,+\,jets control region.
Events are corrected for all known sources of discrepancies between the data and simulation such as differing reconstruction or tagging efficiencies. It is observed that jets have a harder \pt spectrum in simulation, leading to significant discrepancies from observed distributions of quantities such as \HT. The discrepancies in both control regions are well described by 2-parameter linear fits with negative slopes to the ratio between data and simulation in the \HT distributions~\cite{B2G16005,B2G16006}. Modeling of the \ttbar and W\,+\,jets background samples is corrected using the results of these fits.
The \ST distributions for both control regions are shown in \qfig{fig_cr} with all corrections applied.

To evaluate the uncertainty in the normalization of the \ttbar and W\,+\,jets background processes, a binned maximum likelihood fit~\cite{PDGSTATS} of the background-only hypothesis is performed in the two control regions using the \textsc{Theta} framework~\cite{THETA}. All systematic uncertainties (discussed in more detail in \qsec{sec:systs}) are accounted for, except for uncertainties in the rate of \ttbar and W\,+\,jets backgrounds that are constrained using this fit.
The resulting uncertainties in the normalizations of the two backgrounds are 8.7\% for \ttbar and 6\% for W\,+\,jets. These uncertainties are included in the final statistical interpretation of the results (discussed in \qsec{sec:results}) as rate uncertainties.
In both control regions, data and simulation agree within the systematic uncertainties described in \qsec{sec:systs}.

\section{Boosted W channel}\label{sec:boostedW}
\subsection{Event selection}

The selection in this channel is optimized for the identification of boosted W boson decays. Selected events are required to have no H-tagged jets ensuring that the event sample in this channel is complementary to that for the boosted Higgs channel, allowing a straightforward combination of the two channels.
Events are selected that have one electron or muon, usually from the decay of a W boson in the $\PQT \to \PQb\PW$ decay mode or from a leptonic top quark decay in the $\PQT \to \PQt\Z$ or tH decay modes. Electrons (muons) must have $\pt > 40\GeV$, $\abs{\eta} < 2.1\,(2.4)$ and pass the tight identification and isolation requirements described in \qsec{sec:objects}. Events having additional loose electrons or muons with $\pt > 10\GeV$ are rejected.

\begin{figure}[tbp]
\centering
\includegraphics[width=0.49\textwidth]{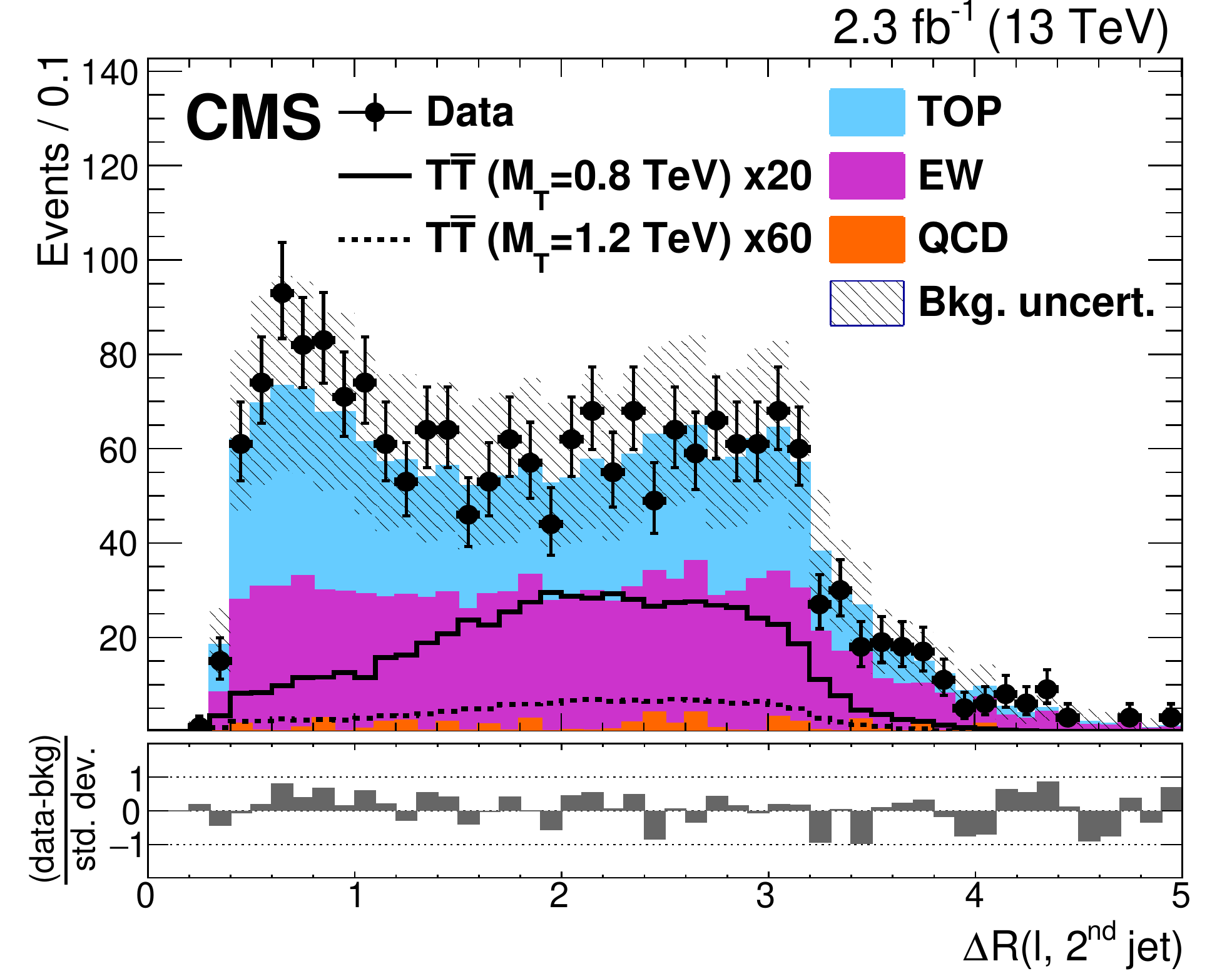}
\caption{Distribution of $\Delta R(\ell,j_2)$ in the \boostedW channel after all selection requirements except for $\Delta R(\ell,j_2)> 1$. Also shown are the distributions of \TTbar signal events with T quark masses of 0.8 and 1.2\TeV, scaled by factors of 20 and 60, respectively. The uncertainty in the background includes the statistical and systematic uncertainties described in \qsec{sec:systs}.}
\label{fig:dr}

\end{figure}

Each event must have three or more AK4 jets, and the three highest \pt jets must satisfy $\pt > 300$, 150, and 100\GeV, respectively. Since a neutrino is expected from a leptonic W boson decay, \MET is required to be greater than 75\GeV, which also significantly reduces the background from multijet events. Control regions are separated from the signal region based on the angular separation between the lepton and the second-highest \pt jet in the event, $\Delta R(\ell,j_2)$.
In both \TTbar and background processes, the lepton is usually observed back-to-back with the highest transverse momentum AK4 jet, and in \TTbar events the second-highest \pt jet also tends to be back-to-back with the lepton, as seen in \qfig{fig:dr}.
The signal region selection requires $\Delta R(\ell,j_2)> 1$.
Figure~\ref{fig:dr} shows the distribution of $\Delta R(\ell,j_2)$ after all selection requirements except for $\Delta R(\ell,j_2) > 1$. All selection efficiency corrections for differences between data and simulation are applied, as well as the \HT-based reweighting described in \qsec{sec:bkg_model}.

\begin{figure}[tb]
\centering
\includegraphics[width=0.49\textwidth]{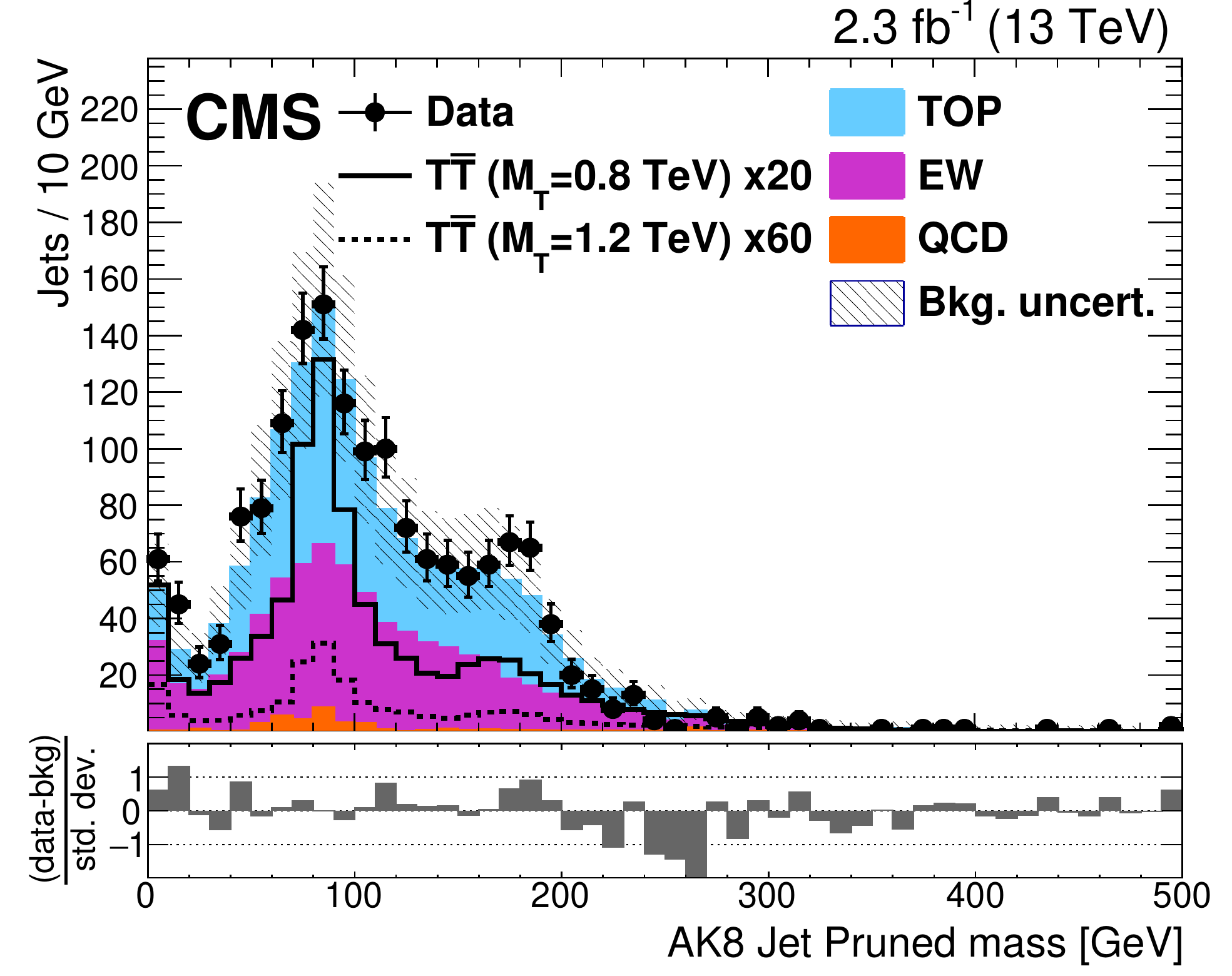}
\includegraphics[width=0.49\textwidth]{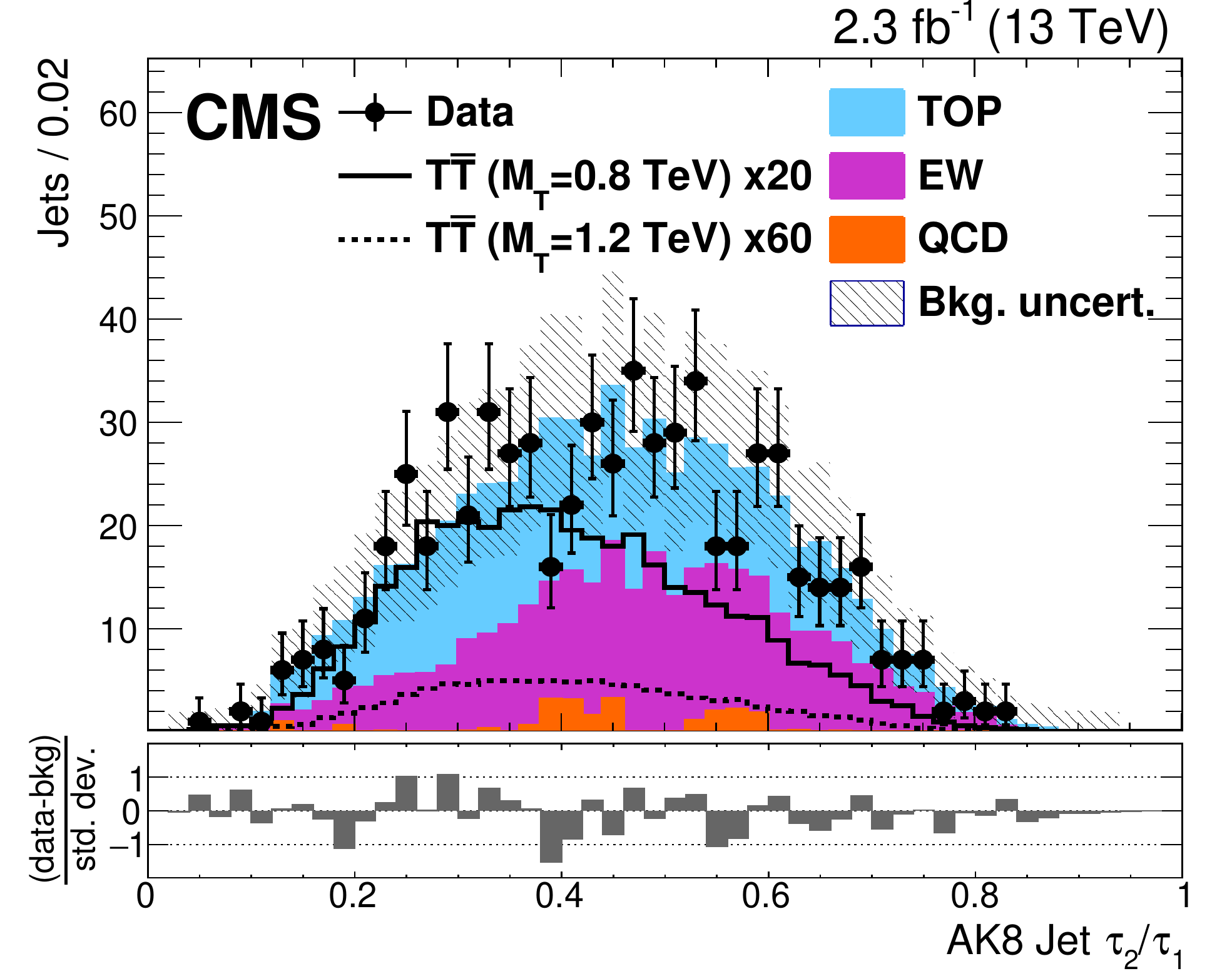}
\includegraphics[width=0.49\textwidth]{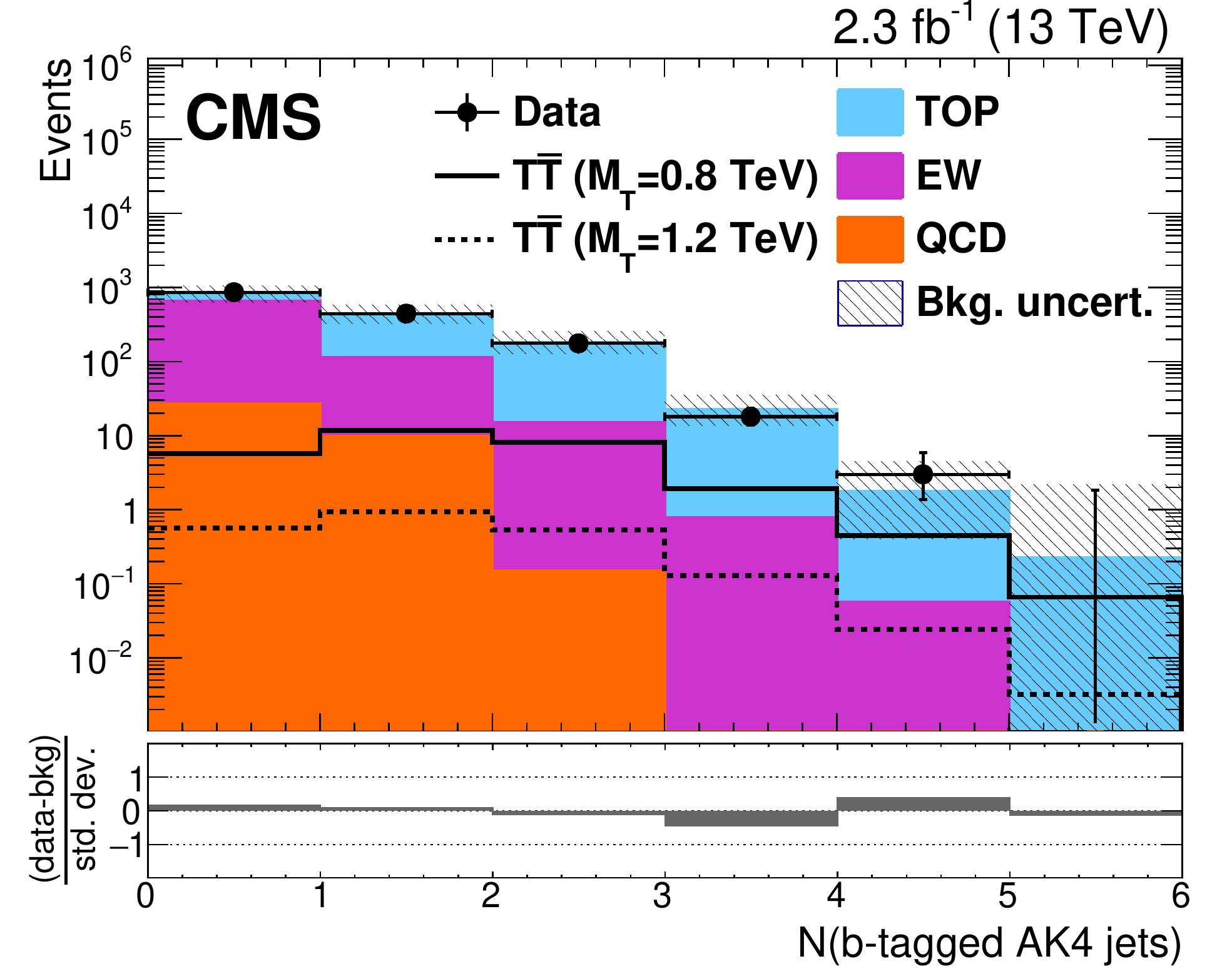}
\includegraphics[width=0.49\textwidth]{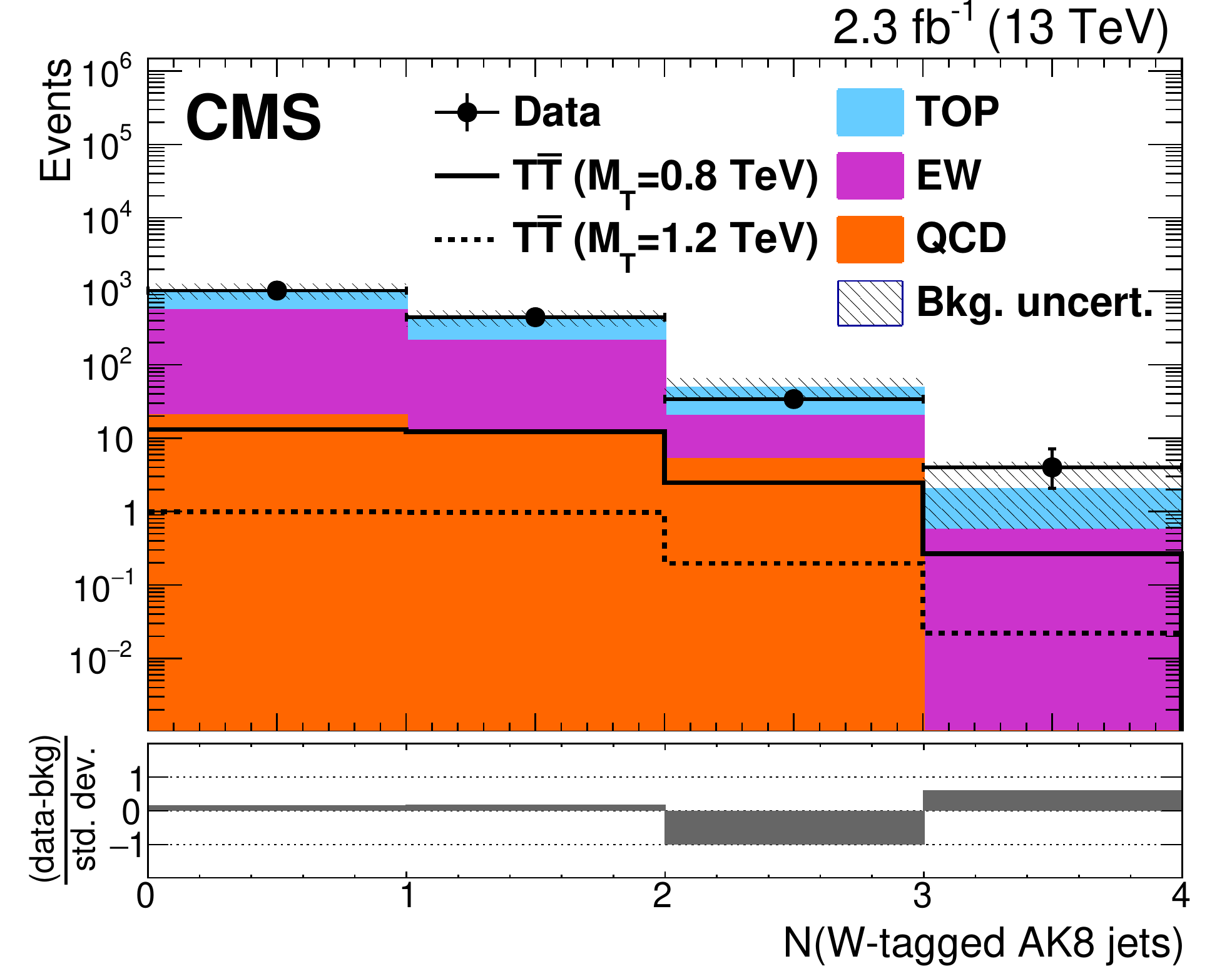}
\caption{Distributions of (left-to-right, upper-to-lower) pruned jet mass for AK8 jets with $\tau_2/\tau_1 < 0.6$, $\tau_2/\tau_1$ for AK8 jets with pruned mass within 65--105\GeV, number of b-tagged AK4 jets, and number of W-tagged AK8 jets in the \boostedW channel with all categories combined. Also shown are the distributions of \TTbar signal events with T quark masses of 0.8 and 1.2\TeV, scaled by factors of 20 and 60, respectively, in the upper figures. The uncertainty in the background includes the statistical and systematic uncertainties described in \qsec{sec:systs}.}
\label{fig:wtag}

\end{figure}

To maximize sensitivity to the presence of \TTbar production, events are divided into 16 categories based on lepton flavor (\Pe, $\mu$), the number of b-tagged jets (0, 1, 2, ${\geq}3$), and the number of boosted W-tagged jets (0, ${\geq}1$). In events with no W-tagged jet, we require a fourth jet with $\pt >30$\GeV. Figure~\ref{fig:wtag} shows the distributions used for tagging boosted W bosons as well as the number of b-tagged and W-tagged jets. The pruned mass distribution for AK8 jets with $\tau_2/\tau_1 < 0.6$ shows a significant contribution of boosted W bosons in signal events weighted to correspond to the singlet branching fraction benchmark. The $\tau_2/\tau_1$ distribution in AK8 jets with pruned mass between 65--105\GeV shows that W\,+\,jets and multijet backgrounds are concentrated at higher values, as expected for jets without substructure.

\begin{figure}[hbp]
\centering
\includegraphics[width=0.49\textwidth]{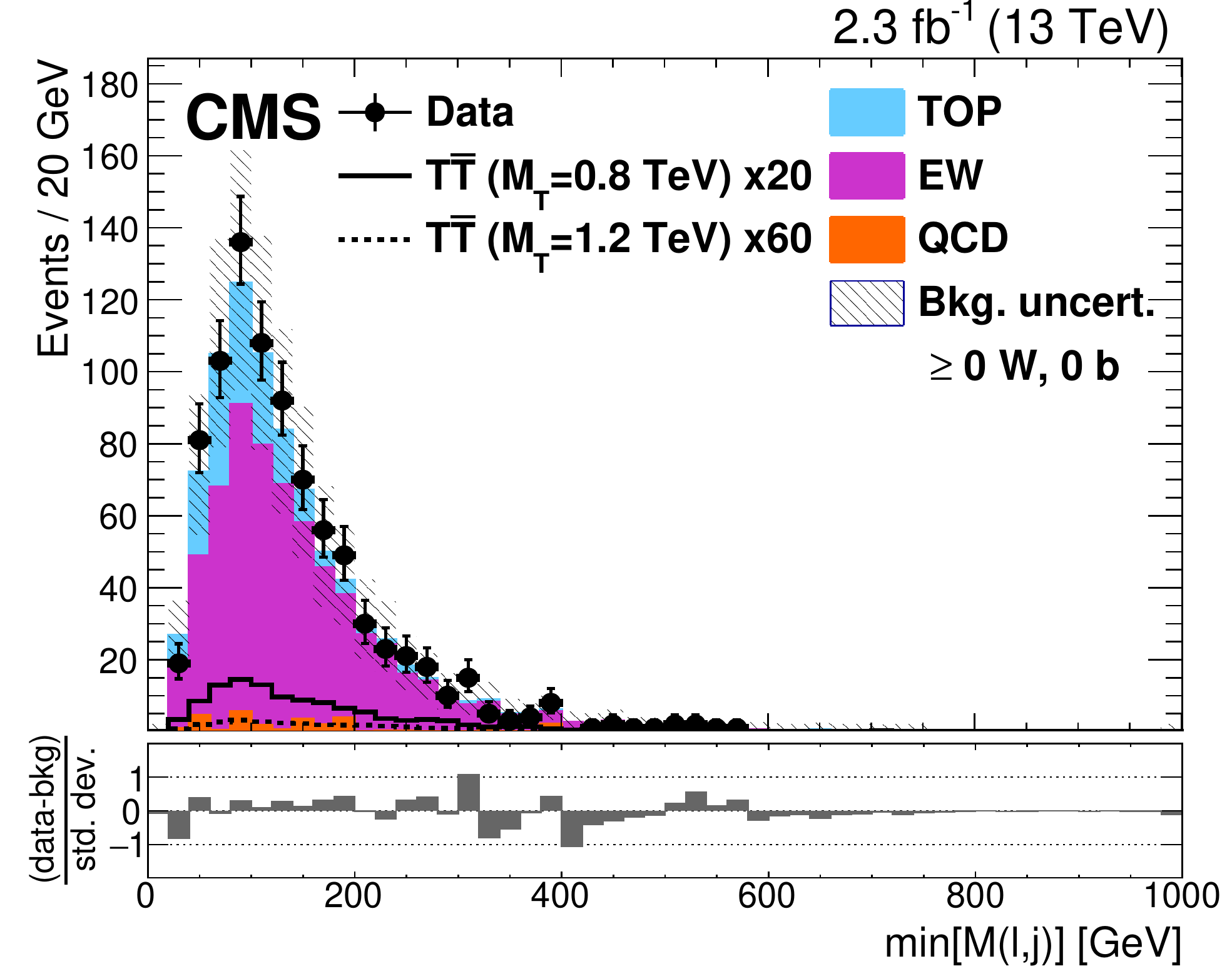}
\includegraphics[width=0.49\textwidth]{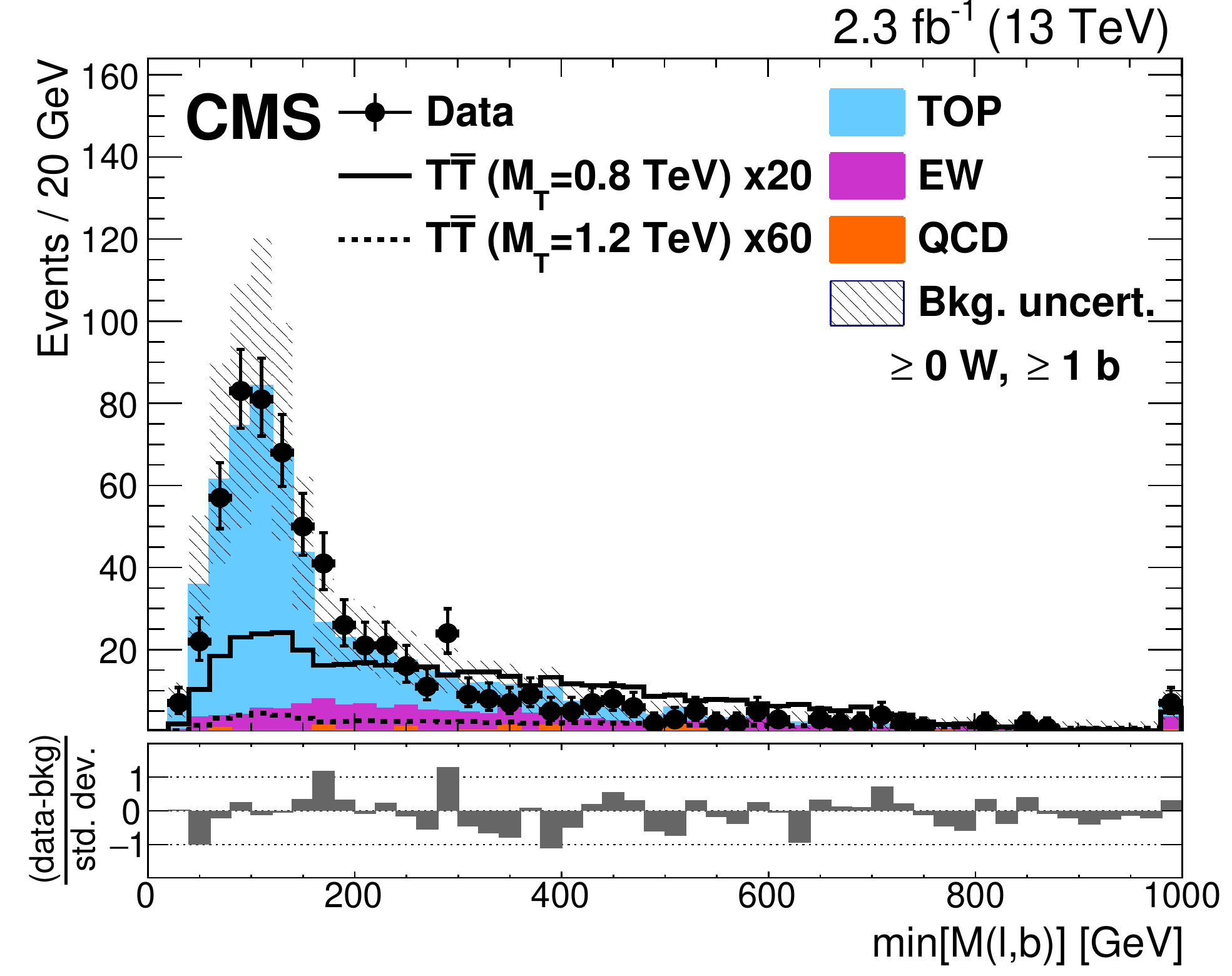}
\caption{Distributions of $\min[M(\ell,\,j)]$ in events without b-tagged AK4 jets (left) and $\min[M(\ell,\,\PQb)]$ in events with ${\geq}$1 b-tagged AK4 jets (right) in the \boostedW channel with all categories combined. Also shown are the distributions of \TTbar signal events with T quark masses of 0.8 and 1.2\TeV, scaled by factors of 20 and 60, respectively. The uncertainty in the background includes the statistical and systematic uncertainties described in \qsec{sec:systs}. }
\label{fig:newdiscrims}

\end{figure}

We finally analyze the minimum mass constructed from the lepton ($\ell$) and a b-tagged AK4 jet, labeled $\min[M(\ell,\,\PQb)]$. In leptonic top quark decays, forming a mass from two of the three decay products, the lepton and b quark jet, produces a sharp edge near the top quark mass. Therefore this distribution is particularly suited to identifying $\PQT\to\PQb\PW$ decays, where the corresponding edge forms at much higher masses, near M(T). In the categories with zero b-tagged AK4 jets, we consider the minimum mass of the lepton and any AK4 jet, denoted $\min[M(\ell,\,j)]$. This combination of discriminating variables provides the best sensitivity to low mass T quark production (${\lesssim}1$\TeV) in the singlet branching fraction scenario.
Figure~\ref{fig:newdiscrims} shows distributions of $\min[M(\ell,\,j)]$ and $\min[M(\ell,\,\PQb)]$ after the final selection but before the likelihood fits described in \qsec{sec:results}.

\subsection{Background modeling}
\label{sec:background}

To cross check the modeling of background processes, we consider two control regions enriched by two dominant background processes, W\,+\,jets and \ttbar. To define these regions we invert the signal region requirement of $\Delta R(\ell,j_2)> 1$ and modify the requirement on the number of b-tagged jets to maximize either W\,+\,jets or \ttbar yield. For an 800\GeV T quark we expect only 3 events in both control regions compared to a total background of 444, for a signal to background ratio that is a factor of ${\approx}3$ smaller than in the signal region.

\begin{figure}[tb]
\centering
\includegraphics[width=0.49\textwidth]{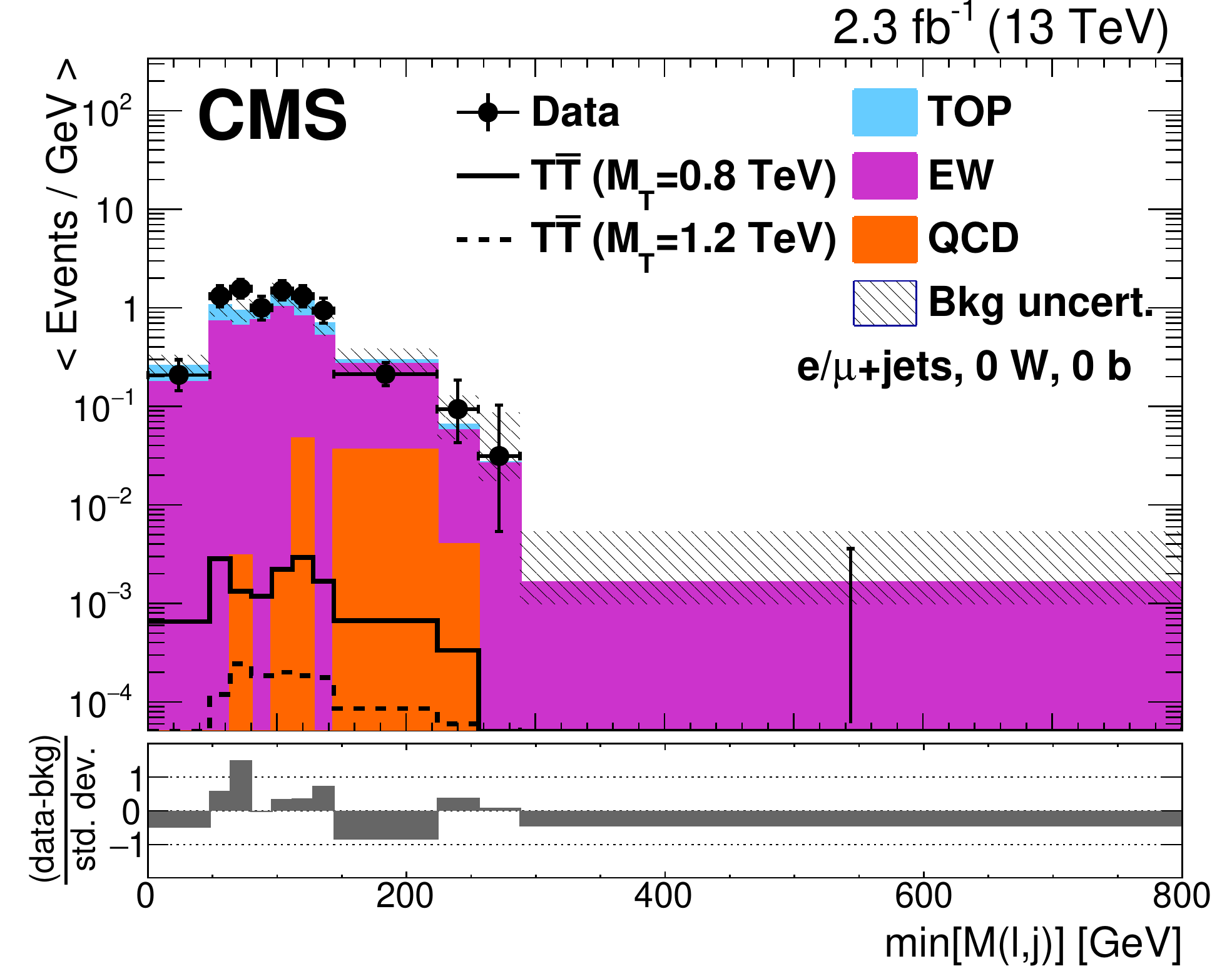}
\includegraphics[width=0.49\textwidth]{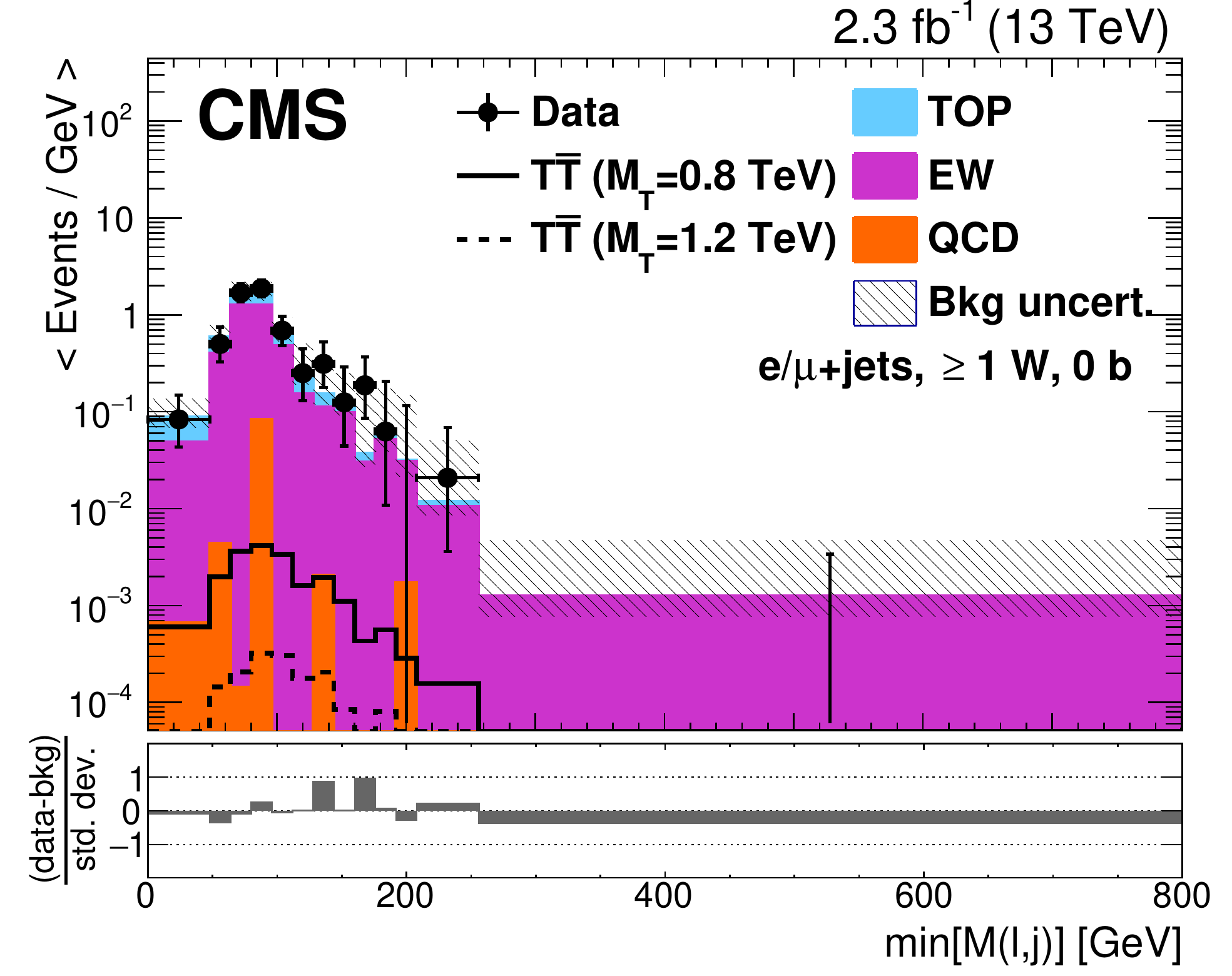}
\includegraphics[width=0.49\textwidth]{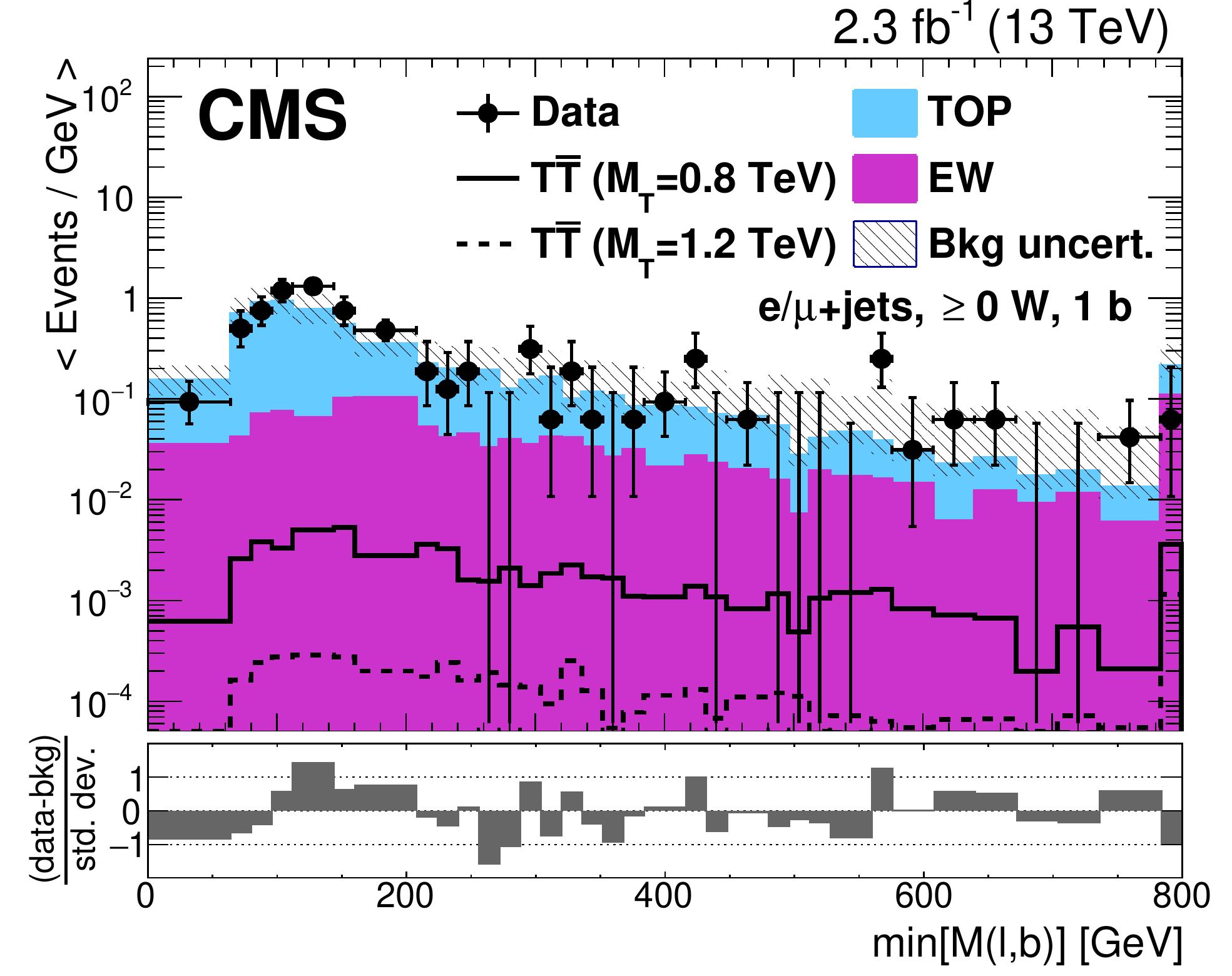}
\includegraphics[width=0.49\textwidth]{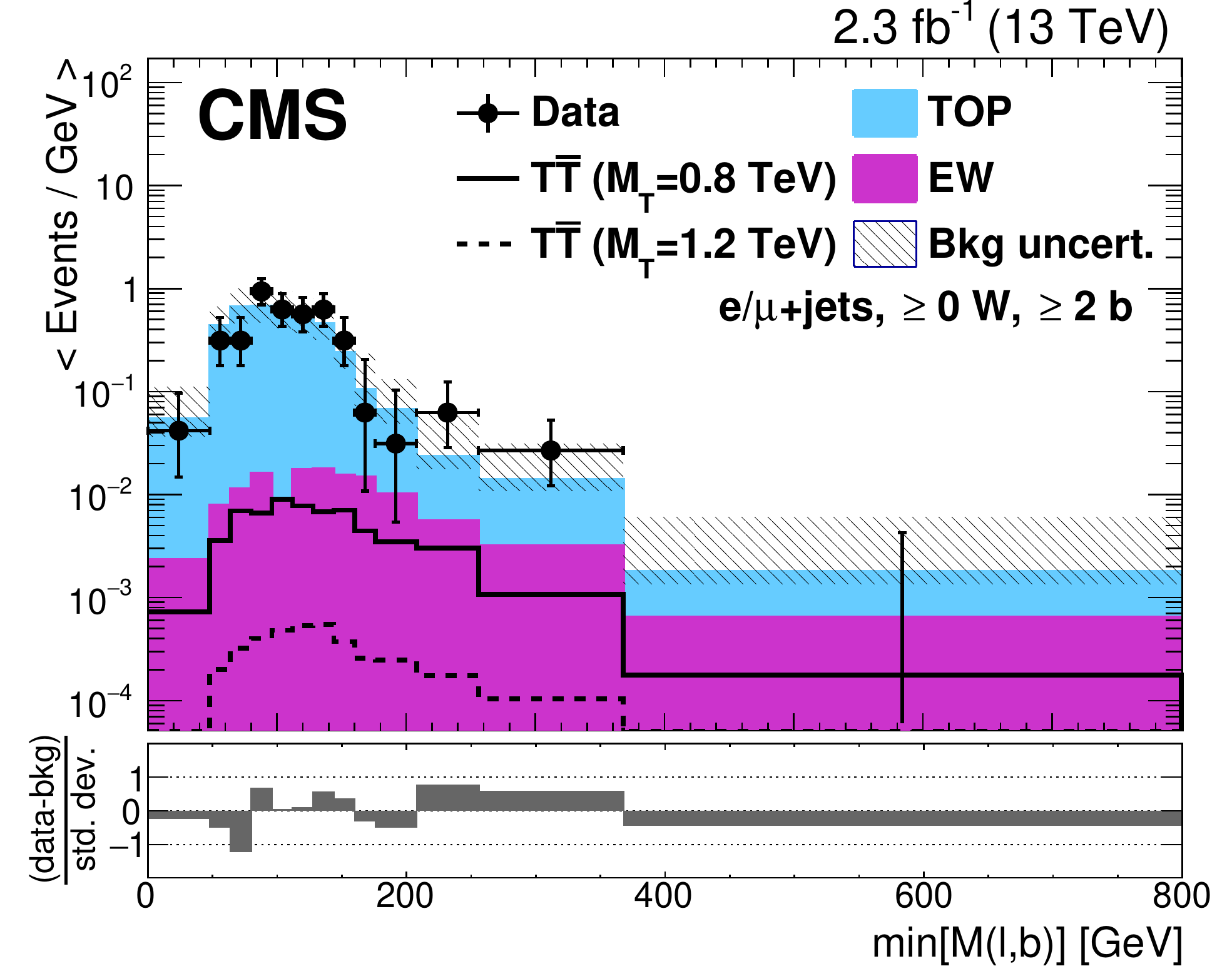}
\caption{Distributions of $\min[M(\ell,\,j)]$ in the W\,+\,jets control region of the \boostedW channel (upper) for 0/${\geq}1$ W tag categories (left/right), and $\min[M(\ell,\,\PQb)]$ in the t$\bar{\rm t}$ control region of the \boostedW channel (lower) for 1/${\geq}2$ b tag categories (left/right). Also shown are the distributions of \TTbar signal events with T quark masses of 0.8 and 1.2\TeV. The uncertainty in the background includes the statistical and systematic uncertainties described in \qsec{sec:systs}.}
\label{fig:CRs}

\end{figure}

The W\,+\,jets control region has zero b-tagged jets and events are categorized according to the number of W-tagged jets (0, ${\geq}1$). The \ttbar region has one or more b-tagged jets and events are categorized according to the number of b-tagged jets (1, ${\geq}2$). Figure~\ref{fig:CRs} shows distributions of $\min[M(\ell,\,j)]$ in the W\,+\,jets control region and $\min[M(\ell,\,\PQb)]$ in the \ttbar control region.
Both regions show that simulation-based background predictions agree with data within the systematic uncertainties described in \qsec{sec:systs}. Observed and predicted event yields in the control regions for all categories are compared as a closure test, and differences in yields are assigned as an additional systematic uncertainty. This uncertainty accounts for any background mismodeling after selection and scale factor application.

\section{Systematic uncertainties}
\label{sec:systs}

We consider sources of systematic uncertainty that can affect the normalization and/or the shape of both background and signal distributions. A summary of these systematic uncertainties along with their numerical values and whether they are applied to signal or background samples can be found in \qtab{tab:systs}.

The uncertainty in the integrated luminosity is 2.3\%~\cite{CMS-PAS-LUM-15-001} and is applied to all simulated samples. Normalization uncertainties in the rates of SM processes include 20\% for single top quark production and 15\% for diboson production, based on CMS measurements~\cite{ZZcms,WZcms}. For multijet production a rate uncertainty of 100\% is assigned in the \boostedHiggs channel since the simulation used in this channel does not contain either the PDF or matrix element scale uncertainties, unlike those used in the \boostedW channel. No rate uncertainty is applied to Z\,+\,jets production since for this process experimental and theoretical uncertainties are small compared to the energy scale and PDF uncertainties described below. Additionally, both channels derive normalization uncertainties for \ttbar and W\,+\,jets samples from control regions, with values of 5--12\% and 4--20\% in the \boostedW channel, and 8.7\% and 6.0\% in the \boostedHiggs channel. Trigger, lepton identification, and lepton isolation efficiency scale factor uncertainties are also applied as normalization uncertainties.

Uncertainties in both channels affecting the shape and normalization of the distributions include uncertainties related to jet energy scale, jet energy resolution, pruned or soft drop jet mass scale and resolution, and b tagging and light-flavor mistagging efficiencies. These are evaluated by raising and lowering their values with respect to the central values by one standard deviation of the respective uncertainties and recreating a distribution using shifted values at each step of the analysis. An additional uncertainty of 5\% is applied in the \boostedHiggs channel to account for potential differences when propagating the jet mass scale and resolution scale factors, measured using hadronic W boson decays, to Higgs boson candidate jets. This uncertainty has been determined by comparing samples simulated with the \PYTHIA~8 and \HERWIGpp~\cite{Bahr:2008pv} (with the CUETP8M1 tune~\cite{Skands:2014pea,Khachatryan:2015pea}) hadronization programs and evaluating the difference between the two programs in the jet mass distributions for hadronically decaying W and Higgs bosons. In the \boostedW channel we also apply shape uncertainties to the W boson tagging corrections for the $\tau_2$/$\tau_1$ selection efficiency and its \pt dependence. To account for small differences in the H-tagging efficiency between the boosted W and \boostedHiggs channel, a 3\% normalization uncertainty is assigned that is correlated with the b tagging uncertainty in the \boostedHiggs channel and anticorrelated in the \boostedW channel.

The uncertainty due to pileup modeling is evaluated by varying by ${\pm}5\%$ the total inelastic cross section used to calculate the pileup distribution. The systematic uncertainty in the \HT-based background reweighting procedure is taken to be the difference between the unweighted distribution and a distribution where the correction factor is applied twice.

The uncertainties in the PDFs used in MC simulation are evaluated from the set of NNPDF3.0 fitted replicas, following the standard procedure~\cite{NNPDF30}.
Renormalization and factorization scale uncertainties are calculated by varying the corresponding scales up or down (either independently or simultaneously) by a factor of two and taking as uncertainty the envelope, or largest spread, of all possible variations. These theoretical uncertainties are applied to the signal simulation as shape uncertainties, together with small normalization uncertainty contributions due to changes in acceptance.

The PDF and scale variation uncertainties affect both the normalization and shape of background distributions for multijet (in the \boostedW channel), Z\,+\,jets, and single top quark MC samples. For the \ttbar and W\,+\,jets backgrounds the theoretical and \HT reweighting uncertainties dominate the total uncertainty in this search, and theoretical uncertainties are treated differently across the two channels.
Changes of energy scale or parton momentum strongly influence \HT and therefore these uncertainties are correlated with the uncertainty in the \HT reweighting method. In the \boostedHiggs channel, only the uncertainty in the \HT reweighting procedure is considered as this uncertainty dominates over energy scale variations and PDF uncertainties, especially in the tails of the \ST distribution. In the \boostedW channel the uncertainty in the \HT reweighting dominates over the PDF uncertainty, but is comparable in shape and magnitude to the scale variation uncertainty, with scale variations providing the dominant uncertainty at low values of $\min[M(\ell,\,\PQb)]$. In this channel both \HT reweighting and scale variation uncertainties are considered for \ttbar and W\,+\,jets backgrounds.
All of these shared uncertainties are treated as correlated between the two analysis channels in the statistical interpretation of the results.

\begin{table}[tb]
  \centering
    \topcaption{Summary of the systematic uncertainties, along with numerical values and application to signal and/or background samples. The second column gives the magnitude of normalization uncertainties or the procedure used to evaluate shape uncertainties. The symbol $\sigma$ indicates one standard deviation of the corresponding systematic uncertainty. Renormalization and factorization energy scale uncertainties are treated as shape-only for signal but include normalization uncertainties in background. Values stated for shape uncertainties indicate a representative range over the categories for the dominant backgrounds and/or signal.}\label{tab:systs}
    \resizebox{\textwidth}{!}{
    \begin{tabular}{lccccc}
      \multirow{2}{*}{Source} & \multicolumn{2}{c}{Uncertainty}  & \multirow{2}{*}{Signal} & \multicolumn{2}{c}{Background} \\\cline{2-3}\cline{5-6}
                              & Boosted W & Boosted H             &                         & Boosted W & Boosted H \\
      \hline
      Int. luminosity      &  \multicolumn{2}{c}{2.3\%} & Yes & All     & All     \\
      Diboson rate         &  \multicolumn{2}{c}{15\%}  & No  & diboson & diboson \\
      Single t quark rate  &  \multicolumn{2}{c}{20\%}  & No  & t       & t \\
      QCD rate             &  ---      & 100\% & No  & ---      & QCD \\
      \ttbar rate          &  5--12\% & 8.7\% & No  & \ttbar  & \ttbar \\
      W\,+\,jets rate      &  4--20\% & 6.0\% & No  & W\,+\,jets & W\,+\,jets \\
      \hline
      Trigger (\Pe)              & 5\% & 2\% & Yes & All & All \\
      Trigger ($\mu$)            & 5\% & 1\% & Yes & All & All \\
      Identification (\Pe,$\mu$) & 1\% & 2\% & Yes & All & All \\
      Isolation (\Pe,$\mu$)      & 1\% & ---  & Yes & All & --- \\
      \hline
      Pileup           & \multicolumn{2}{c}{$\sigma_{\mathrm{inel.}}\pm5$\%} & Yes & All (0--3\%)  & All (0--3\%)\\
      Jet energy scale & \multicolumn{2}{c}{${\pm}\sigma(p_T,\eta)$}         & Yes & All (0--12\%) & All (0--4)\%\\
      Jet energy res.  & \multicolumn{2}{c}{${\pm}\sigma(\eta)$}             & Yes & All (0--8\%)  & All (0--1)\%\\
      \hline
      \multirow{2}{*}{\HT reweighting}  & \multicolumn{2}{c}{envelope(no weight,}  & \multirow{2}{*}{No}  & \ttbar, W\,+\,jets, QCD & \ttbar, W\,+\,jets\\
                                        & \multicolumn{2}{c}{weight squared)}      &                      & (17--34\%)              & (13--21\%) \\
      \hline
      b tag: b                   & \multicolumn{2}{c}{${\pm}\sigma(\pt)$}      & Yes & All (0--16\%) & All (3--8\%)\\
      b tag: light flavors       & \multicolumn{2}{c}{${\pm}\sigma$}           & Yes & All (0--6\%)  & All (1--4\%)\\
      W/H tag: mass scale        & \multicolumn{2}{c}{${\pm}\sigma(\pt,\eta)$} & Yes & All (0--3\%)  & All (0--7\%) \\
      W/H tag: mass res.         & \multicolumn{2}{c}{${\pm}\sigma(\eta)$}     & Yes & All (0--5\%)  & All (0--7\%) \\
      H tag: efficiency          & \multicolumn{2}{c}{3\%}                     & Yes & All           & All \\
      H tag: propagation         & ---                 & 5\%   & Yes & ---            & All \\
      W tag: $\tau_2/\tau_1$     & ${\pm}\sigma$      & ---    & Yes & All (0--2\%)  & --- \\
      W tag: $\tau_2/\tau_1$ \pt & ${\pm}\sigma(\pt)$ & ---    & Yes & All (0--2\%)  & --- \\
      \hline
      Renorm./fact. scale & \multicolumn{2}{c}{envelope\,($\times$2,\,$\times$0.5)} & Shape & All (22--44\%)               & Z\,+\,jets, t (2--23\%)\\
      PDF                 & \multicolumn{2}{c}{${\pm}\sigma$}                     & Shape & Z\,+\,jets, t, QCD (1--7\%)  & Z\,+\,jets, t (0--13\%)\\
    \end{tabular}
    }
\end{table}

\section{Results}
\label{sec:results}

Signal efficiencies for all possible final states of \TTbar and \BBbar production in the boosted W and \boostedHiggs channels (after combining all categories in each channel) are listed in \qtab{tab_ev_effs_fs} for two signal hypotheses with a high and a low vector-like quark mass. The values are derived by dividing the number of signal events that have the corresponding decay mode in each category by the number of expected events in the same decay mode before any selection. It can be seen that the selection applied in the \boostedHiggs channel is most efficient if a Higgs boson is present in the final state, whereas the selection in the \boostedW channel favors $\PQT\to\PQb\PW$ decays, thus showing how the combination of the two channels improves sensitivity to most branching fraction combinations of the T quark. For B quark decays the \boostedW channel has high efficiency for the tW decays and reduced efficiency for the bZ/bH decays owing to the lack of semileptonic top quark decays. Similarly, the \boostedHiggs channel is most efficient for the bHtW final state since a leptonic decay is required as well as an H-tag.

\begin{table}[tb]
\centering
\topcaption{Signal efficiencies in the boosted W and \boostedHiggs event categories, split into the six possible final states, of both \TTbar and \BBbar production for two illustrative mass points. Efficiencies are calculated with respect to the expected number of events in the corresponding final state before any selection. The relative uncertainty in the efficiencies after combining systematic and statistical uncertainties in the MC samples is about 8\% in the boosted W categories and about 12\% in the boosted H categories.}\label{tab_ev_effs_fs}
\begin{tabular}{cccc}Production process & Decay mode & Boosted W categories & Boosted H categories\\ \hline
\multirow{6}{*}{\TTbar (0.8\TeV)} & tHtH & $ 2.9\%$ & $      8.7\%$ \\
 & tHtZ & $ 3.2\%$ & $      7.3\%$ \\
 & tHbW & $ 5.8\%$ & $      6.3\%$ \\
 & tZtZ & $ 3.7\%$ & $      5.6\%$ \\
 & tZbW & $ 6.3\%$ & $      4.2\%$ \\
 & bWbW & $10.0\%$ & $      2.5\%$ \\
\hline
\multirow{6}{*}{\TTbar (1.2\TeV)} & tHtH & $ 3.6\%$ & $     10.5\%$ \\
 & tHtZ & $ 4.1\%$ & $      9.0\%$ \\
 & tHbW & $ 7.3\%$ & $      7.1\%$ \\
 & tZtZ & $ 4.7\%$ & $      6.7\%$ \\
 & tZbW & $ 8.3\%$ & $      4.8\%$ \\
 & bWbW & $13.2\%$ & $      2.5\%$ \\
\hline
\multirow{6}{*}{\BBbar (0.8\TeV)} & bHbH & $1.7\%$  & $      1.9\%$ \\
 & bHbZ & $1.3\%$  & $      1.9\%$ \\
 & bHtW & $5.8\%$  & $      6.1\%$ \\
 & bZbZ & $0.8\%$  & $      1.4\%$ \\
 & bZtW & $6.4\%$  & $      4.2\%$ \\
 & tWtW & $7.9\%$  & $      5.7\%$ \\
\hline
\multirow{6}{*}{\BBbar (1.2\TeV)} & bHbH & $1.7 \%$  & $      2.1\%$ \\
 & bHbZ & $1.4 \%$  & $      1.9\%$ \\
 & bHtW & $7.3 \%$  & $      7.1\%$ \\
 & bZbZ & $0.8 \%$  & $      1.5\%$ \\
 & bZtW & $8.2 \%$  & $      4.7\%$ \\
 & tWtW & $11.4\%$  & $      7.0\%$ \\
\end{tabular}
\end{table}

\begin{figure}[tbp]
\centering
\includegraphics[width=0.425\textwidth]{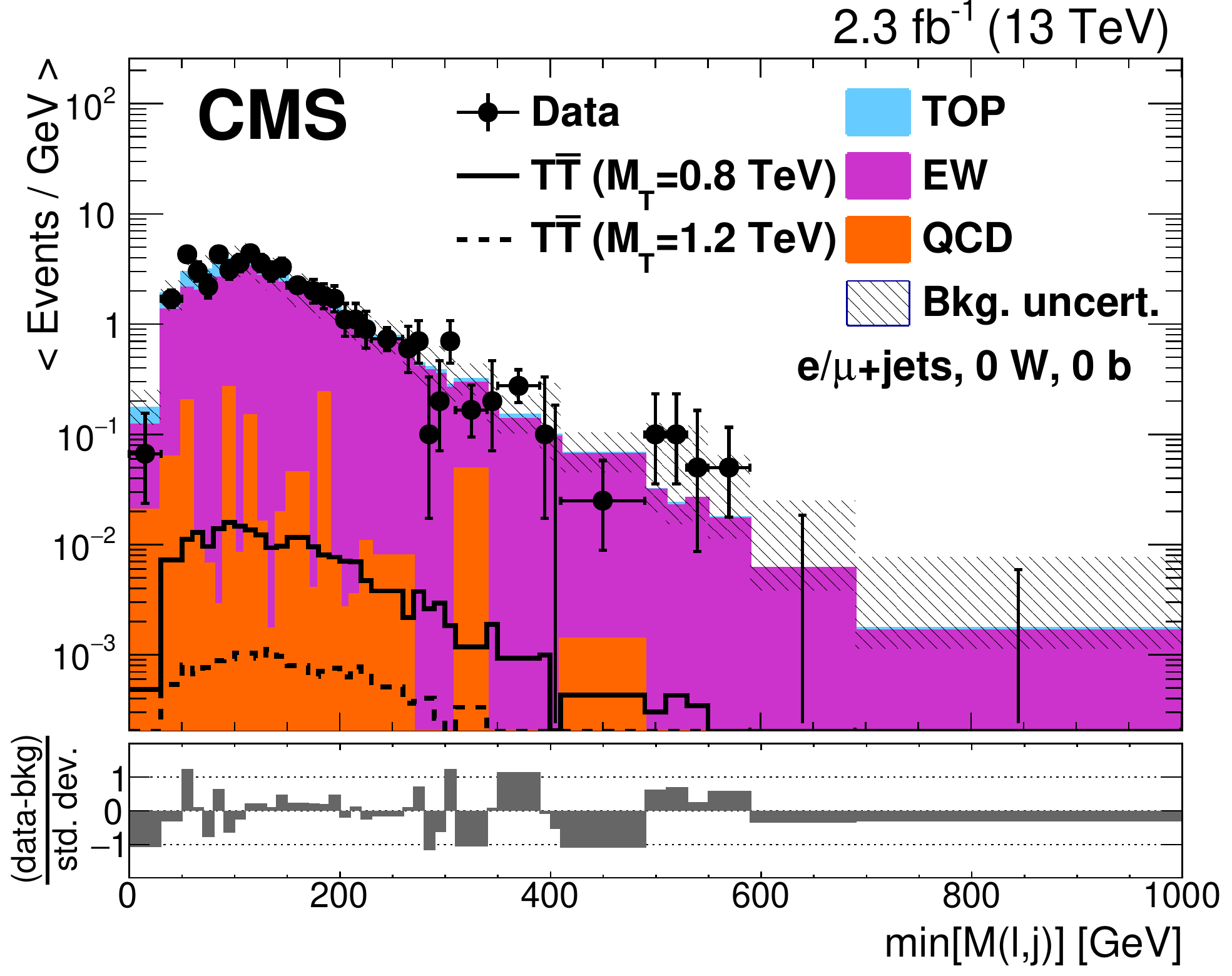}
\includegraphics[width=0.425\textwidth]{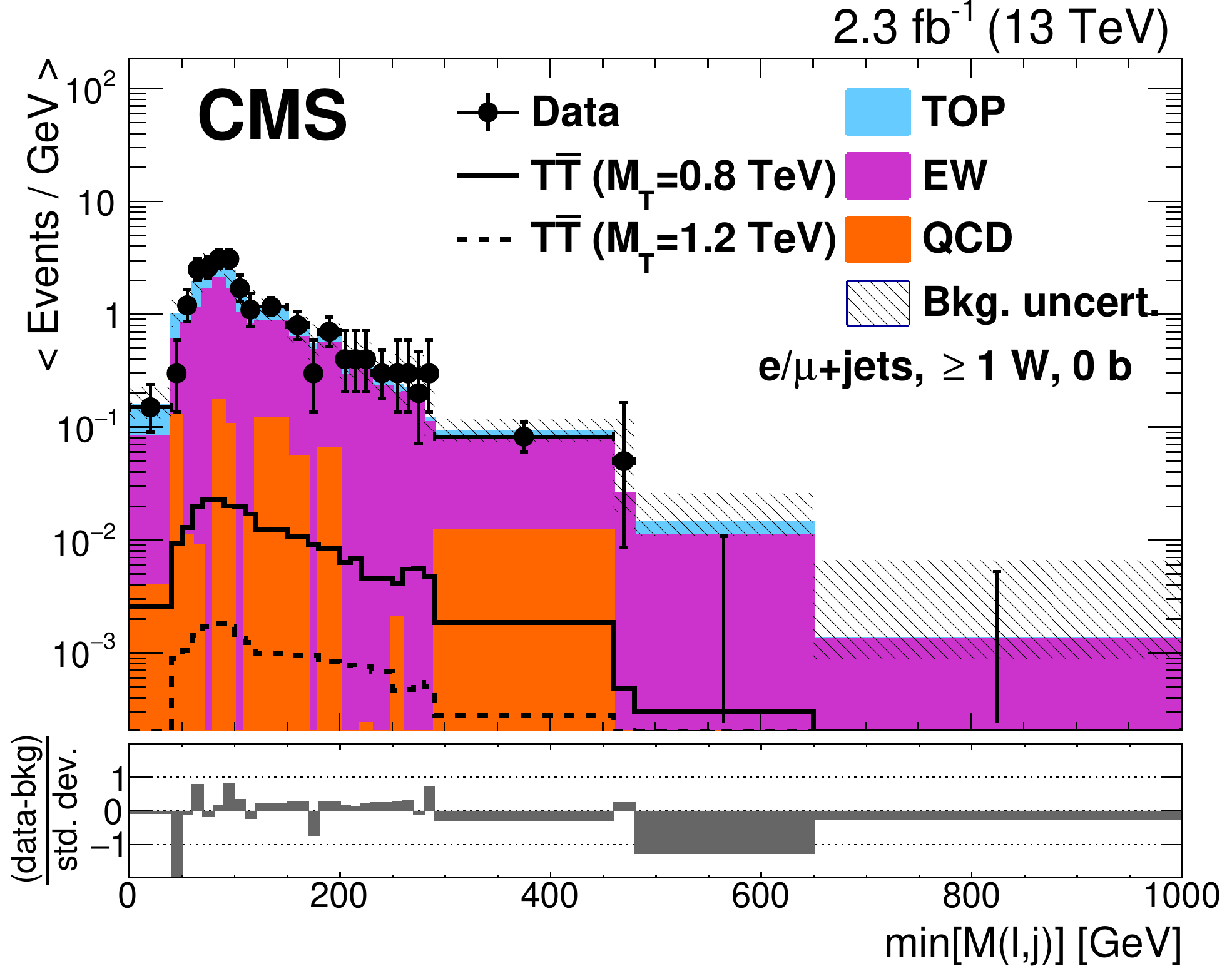}
\includegraphics[width=0.425\textwidth]{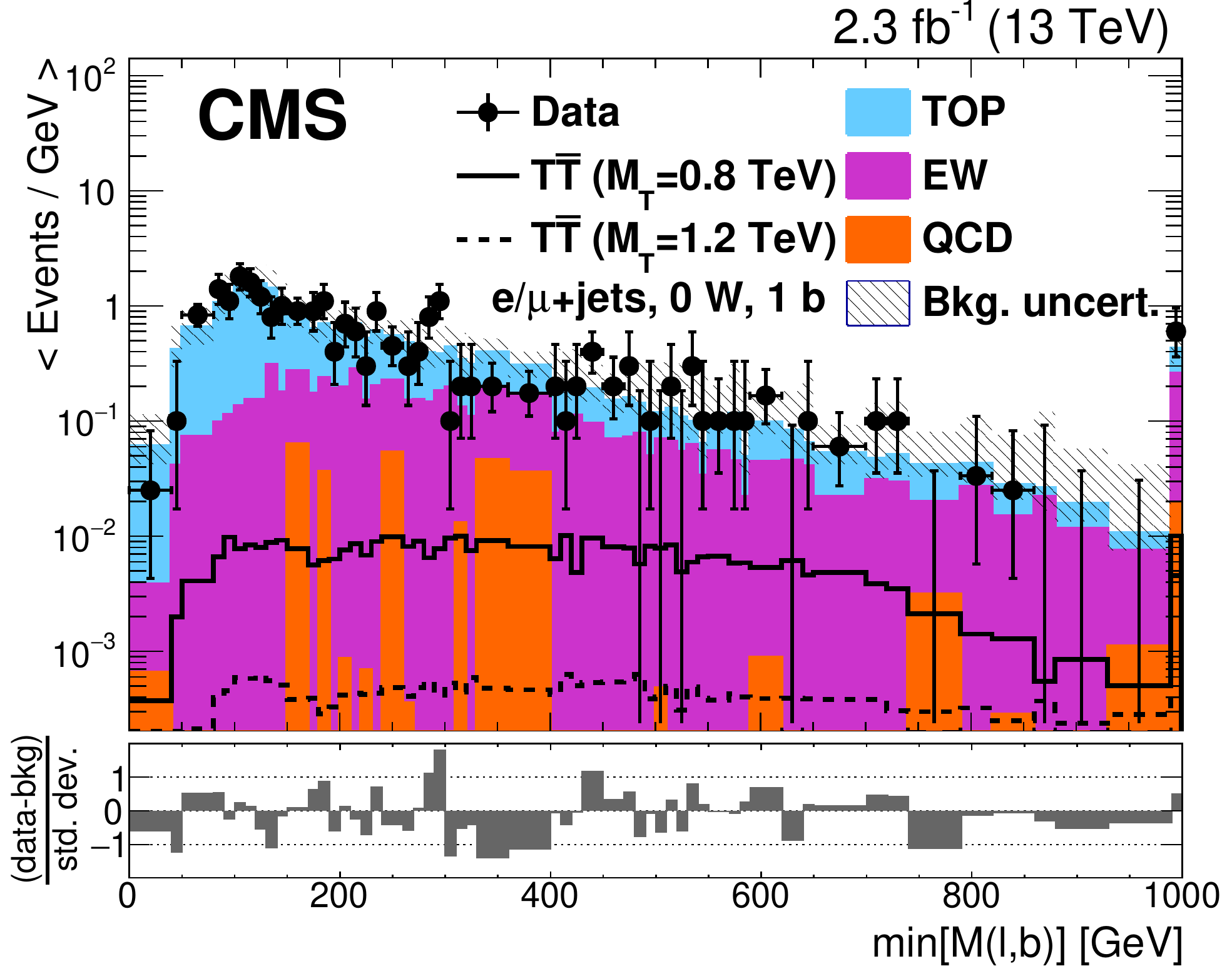}
\includegraphics[width=0.425\textwidth]{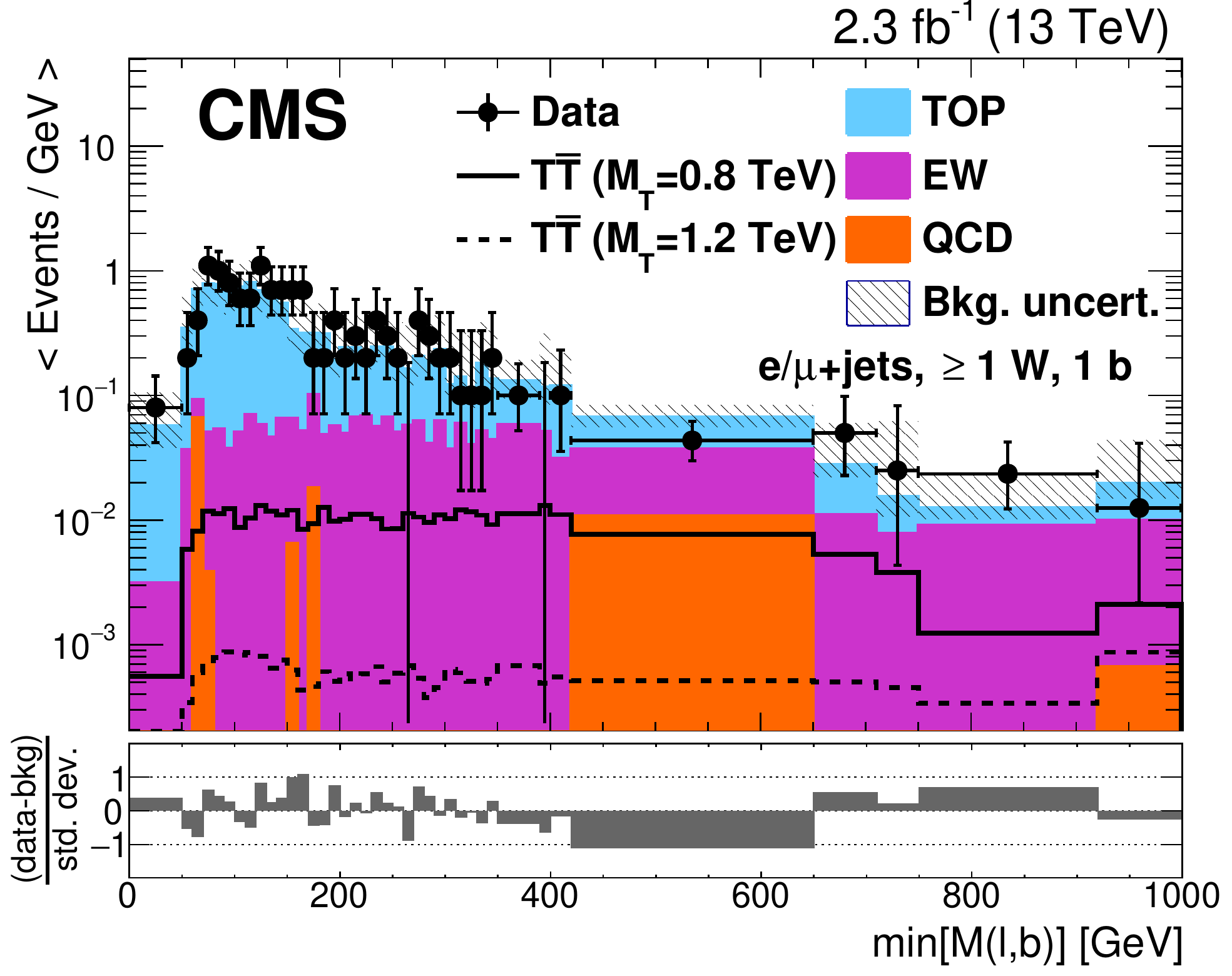}
\includegraphics[width=0.425\textwidth]{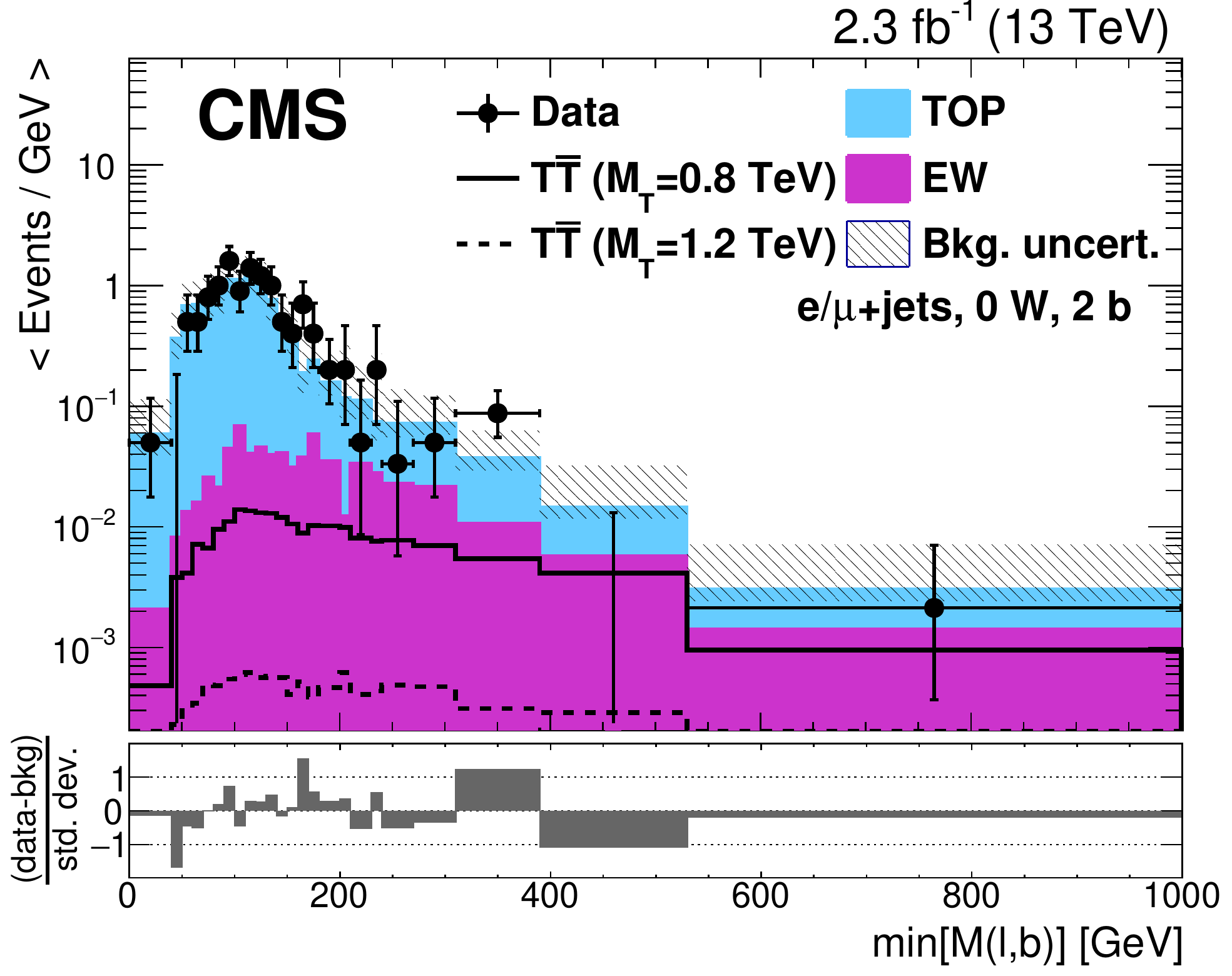}
\includegraphics[width=0.425\textwidth]{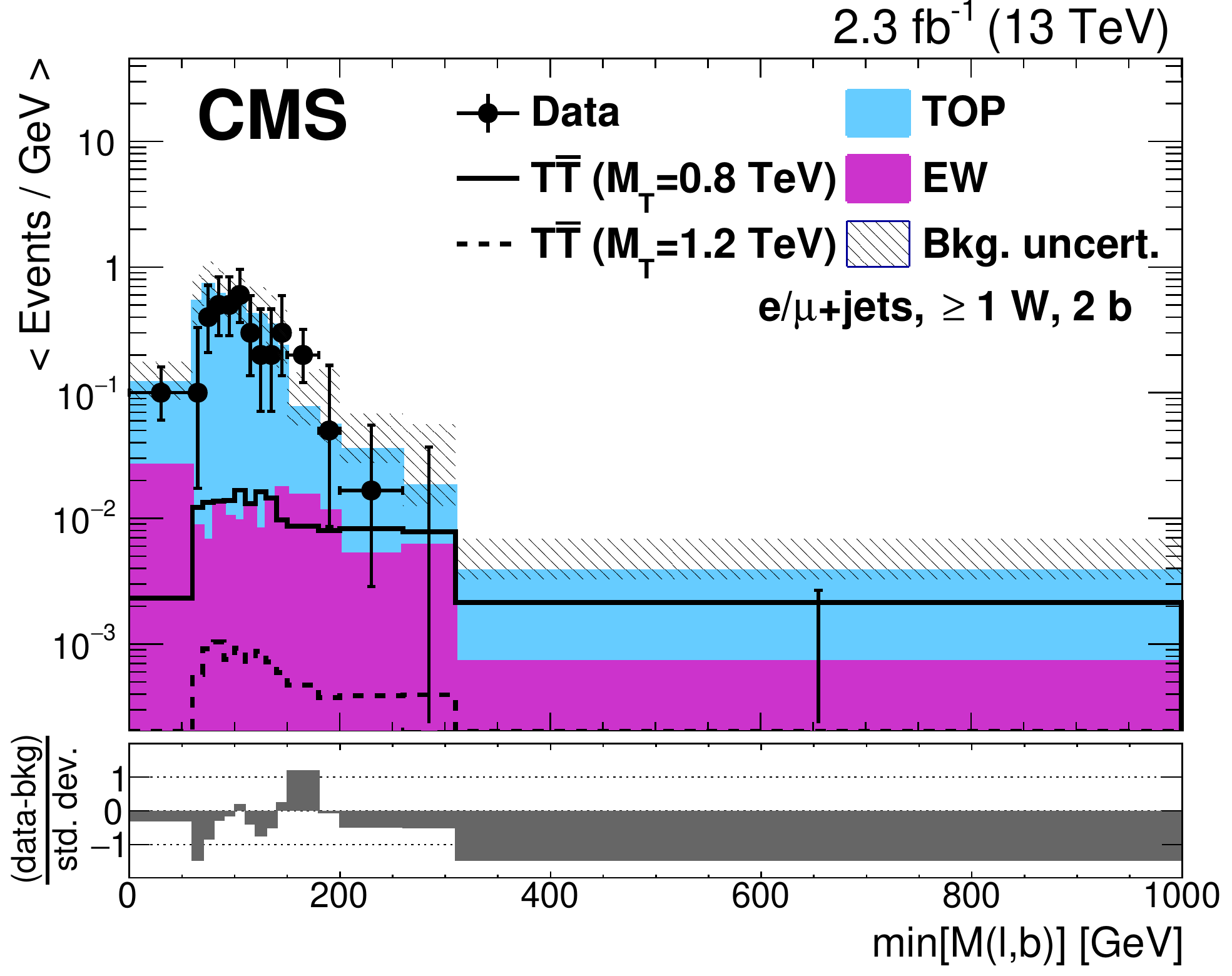}
\includegraphics[width=0.425\textwidth]{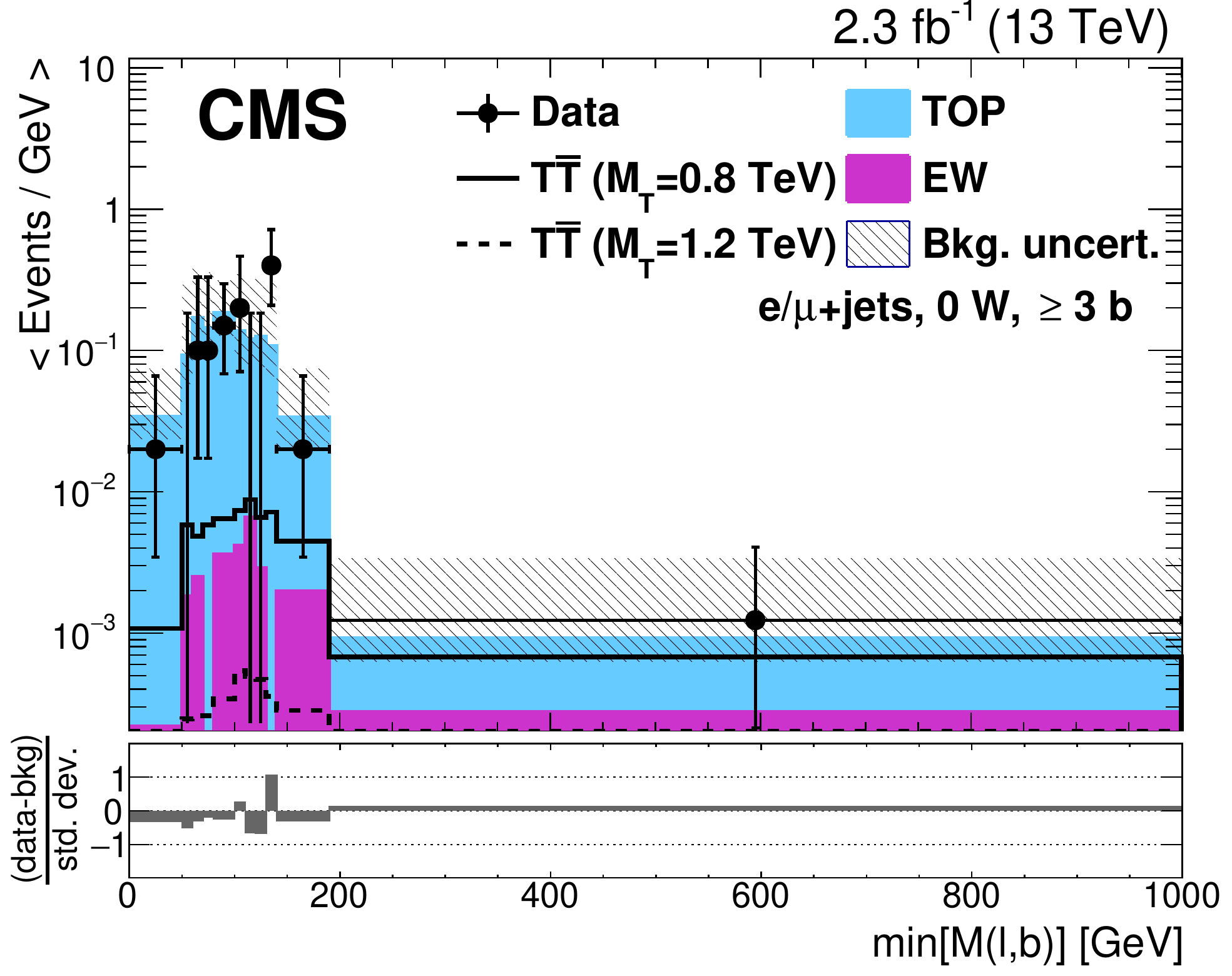}
\includegraphics[width=0.425\textwidth]{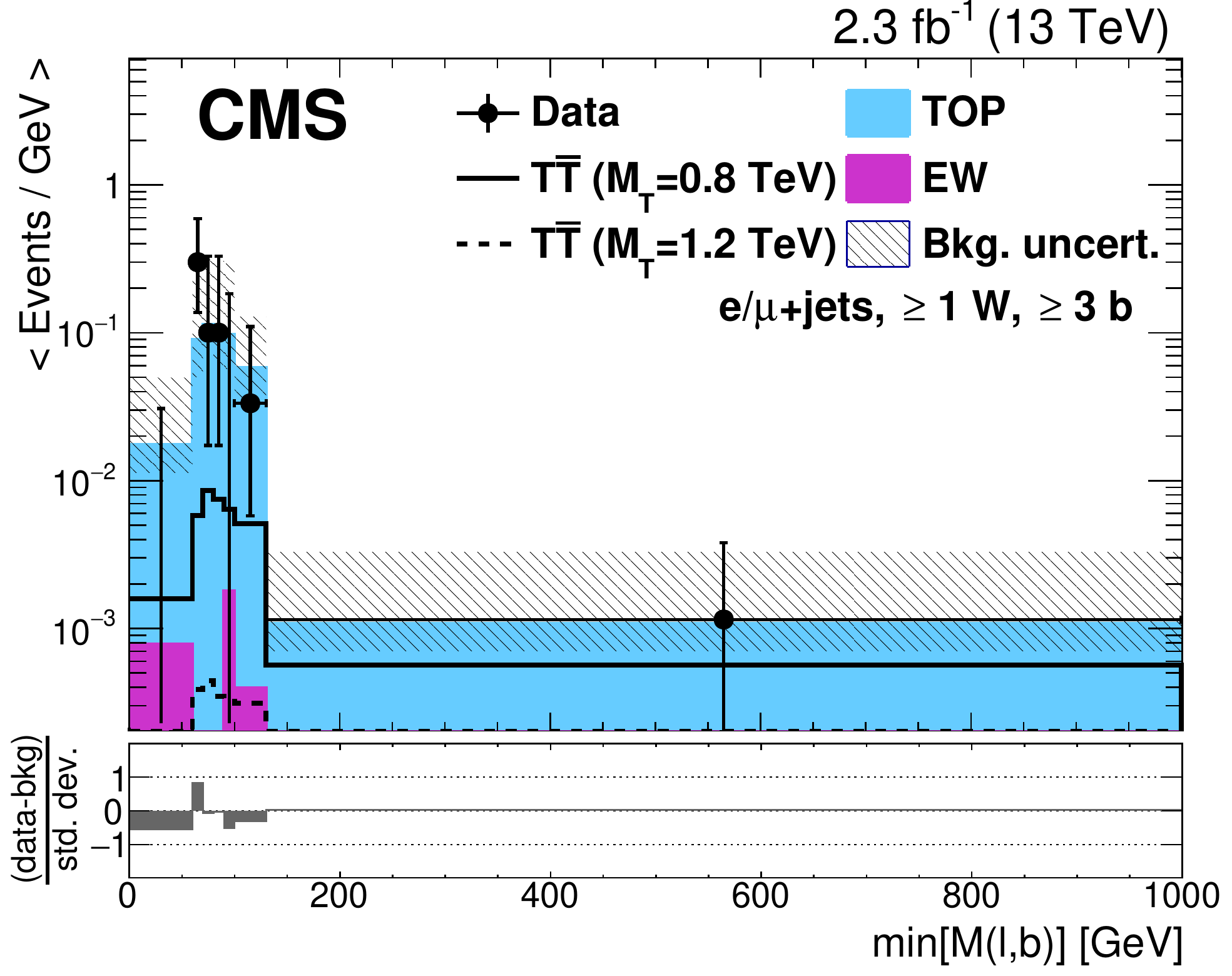}
\caption{Distributions of $\min[M(\ell,\,j)]$ or $\min[M(\ell,\,\PQb)]$ in the combination of electron and muon channels in the \boostedW categories with 0 (left) or ${\geq}1$ (right) W-tagged jets and (upper to lower) 0, 1, 2,  or ${\geq}3$ b-tagged jets. Also shown are the distributions of \TTbar signal events with T quark masses of 0.8 and 1.2\TeV. The uncertainty in the background includes the statistical and systematic uncertainties described in \qsec{sec:systs}.}
\label{fig:templates}

\end{figure}

\begin{figure}[tbp]
\centering
\includegraphics[width=0.48\textwidth]{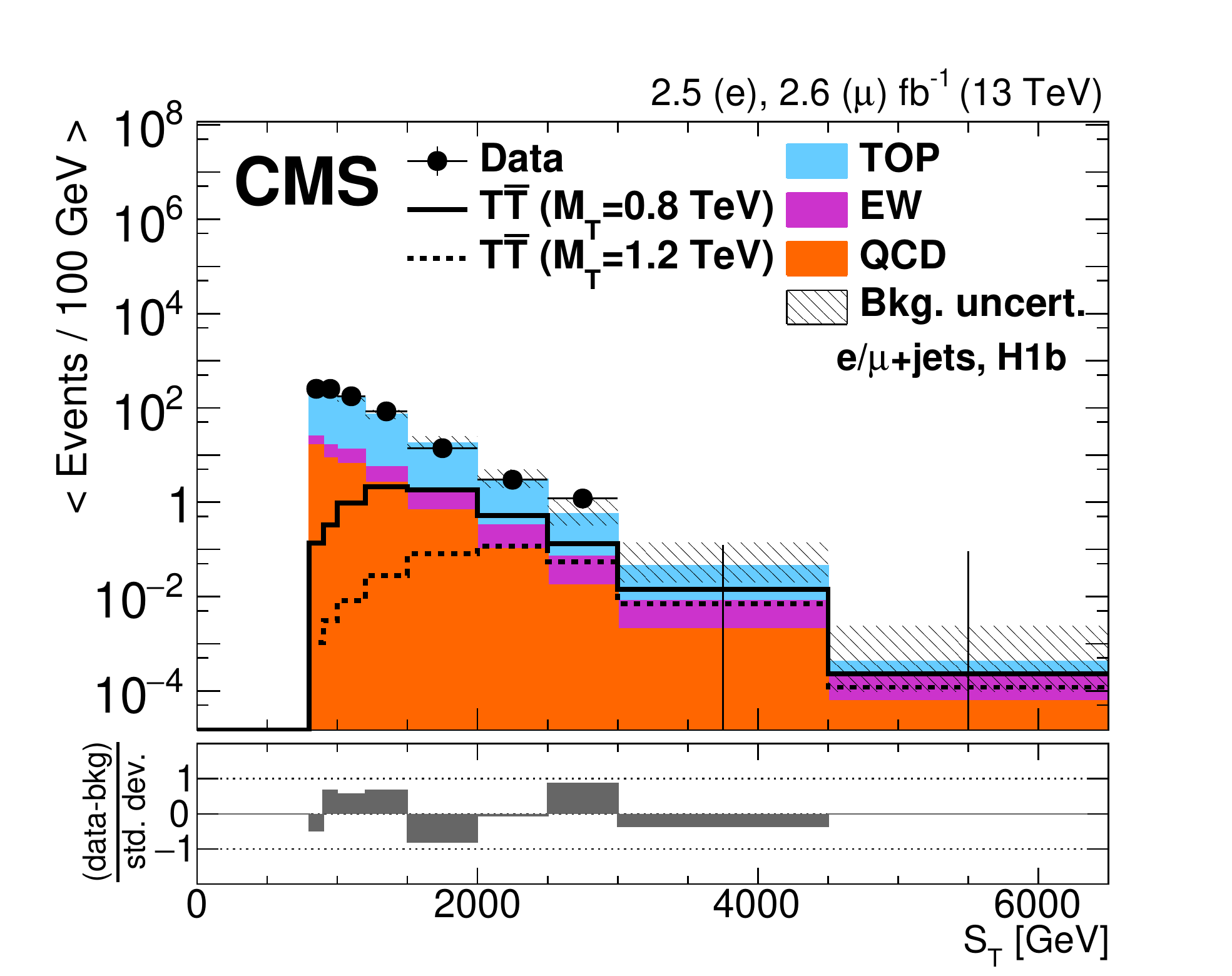}
\includegraphics[width=0.48\textwidth]{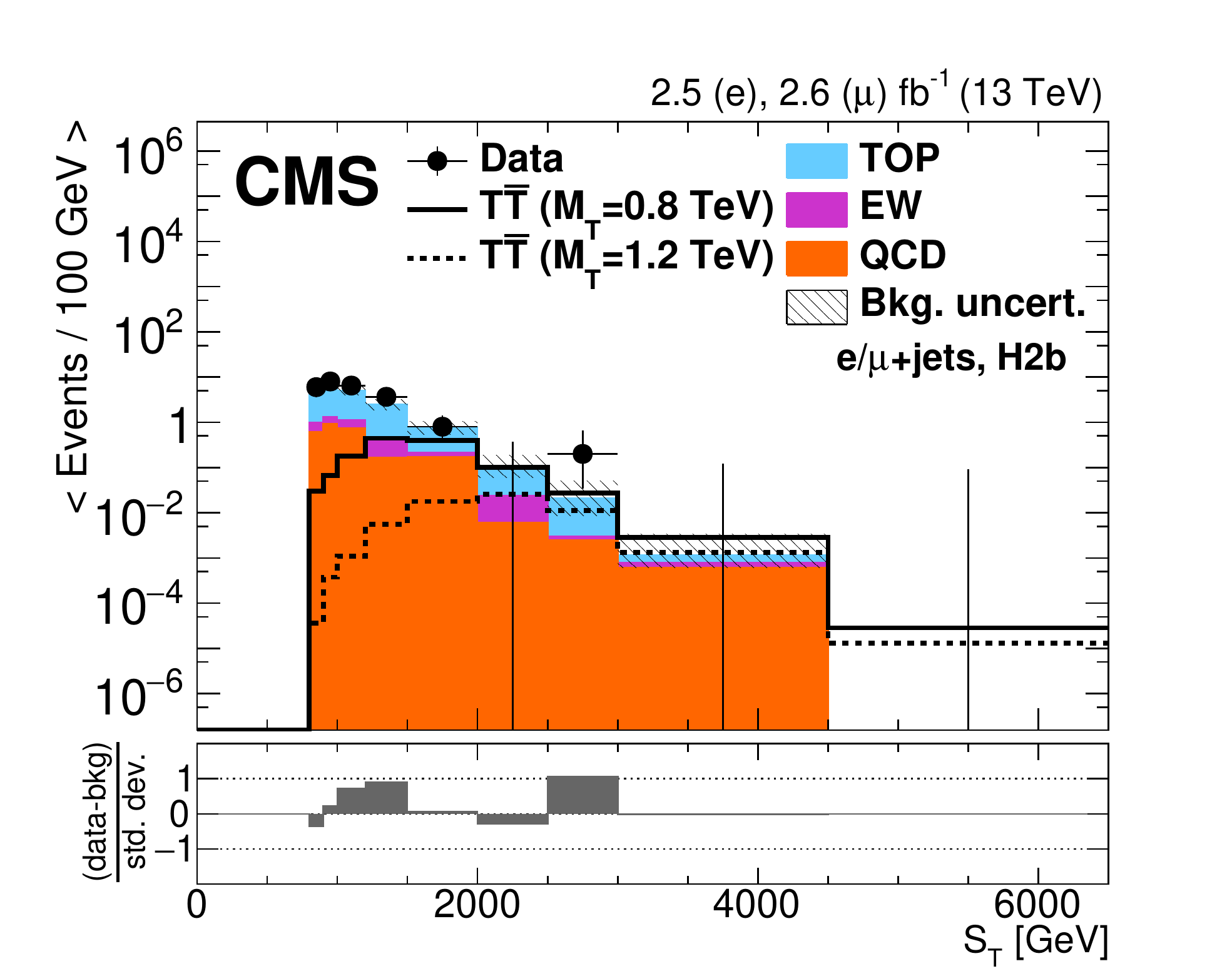}
\caption{
Distributions of \ST in the H1b (left) and H2b (right) categories in the combination of electron and muon channels. The \TTbar signal, shown for T quark masses of 0.8 and 1.2\TeV, is normalized to the theoretical cross section and the singlet benchmark branching fractions are assumed. The uncertainty in the background includes the statistical and systematic uncertainties described in \qsec{sec:systs}.
}
\label{fig_st_final}
\end{figure}

In \qfig{fig:templates}, $\min[M(\ell,\,j)]$ or $\min[M(\ell,\,\PQb)]$ distributions are shown for each of the 8 tagging categories in the \boostedW channel after the final event selection, with the electron and muon channels combined.
Figure~\ref{fig_st_final} shows distributions of \ST in the H1b and H2b categories after combining the electron and muon channels.
As these two variables provide good discrimination between signal and background in their respective categories, they are used for the final statistical interpretation of the data.
In all plots, the \TTbar signal distributions assume the singlet benchmark branching fractions. The event yields are given in \qtab{tab:nevents}.

\begin{table}[tb]
\centering
\topcaption{Number of events in each category after combining the electron and muon channels. Uncertainties include statistical and systematic components from \qtab{tab:systs}, with uncertainty in the total background yield accounting for correlations across background processes. Yields of \TTbar signal assume the theoretically predicted production cross section within the singlet branching fraction scenario.}\label{tab:nevents}\begin{tabular}{l|c|c|c|c}
{ Sample} & { 0 W, 0 b} &  { 0 W, 1 b} & { 0 W, 2 b} &  { 0 W, ${\geq}3$ b} \\
\hline
\TTbar (0.8\TeV)  & 2.5 $\pm$ 0.7          & 5.3 $\pm$ 1.3          & 3.9 $\pm$ 1.0          & 1.4 $\pm$ 0.4        \\
\TTbar (1.2\TeV)  & 0.23 $\pm$ 0.06        & 0.42 $\pm$ 0.11        & 0.26 $\pm$ 0.07        & 0.09 $\pm$ 0.02       \\
\hline
TOP               & 103 $\pm$ 41           & 205 $\pm$ 78           & 111 $\pm$ 41           & 16.3 $\pm$ 6.8         \\
EW                & 460 $\pm$ 160          & 80 $\pm$ 30            & 10.7 $\pm$ 4.0         & 0.6 $\pm$ 0.2         \\
QCD               & 14.1 $\pm$ 6.3         & 6.2 $\pm$ 3.7          & $<$1                  & $<$1                  \\
\hline
Total bkg.        & 570 $\pm$ 170          & 292 $\pm$ 84           & 122 $\pm$ 41           & 16.9 $\pm$ 6.8         \\
Data              & 588                    & 288                    & 131                    & 14                    \\
\hline
\noalign{\bigskip}
\multicolumn{1}{c|}{ Sample} & { ${\geq}1$ W, 0 b} &  { ${\geq}1$ W, 1 b} & { ${\geq}1$ W, 2 b} &  { ${\geq}1$ W, ${\geq}3$ b} \\
\hline
\TTbar (0.8\TeV)  & 3.3 $\pm$ 0.9          & 6.6 $\pm$ 1.7          & 4.2 $\pm$ 1.1          & 1.0 $\pm$ 0.3        \\
\TTbar (1.2\TeV)  & 0.34 $\pm$ 0.09        & 0.52 $\pm$ 0.13        & 0.27 $\pm$ 0.07        & 0.06 $\pm$ 0.02      \\
\hline
TOP               & 71 $\pm$ 26            & 111 $\pm$ 42           & 56 $\pm$ 20            & 7.6 $\pm$ 3.3          \\
EW                & 180 $\pm$ 50           & 29.0 $\pm$ 8.4         & 4.4 $\pm$ 2.0          & 0.2 $\pm$ 0.1          \\
QCD               & 12.6 $\pm$ 7.0         & 3.5 $\pm$ 2.6          & 0.2 $\pm$ 0.2          & $<$ 1                  \\
\hline
Total bkg.        & 263 $\pm$ 57           & 143 $\pm$ 43           & 60 $\pm$ 20            & 7.8 $\pm$ 3.3          \\
Data              & 274                    & 155                    & 45                     & 7                      \\
\hline
\noalign{\bigskip}
\multicolumn{1}{c|}{ Sample} & \multicolumn{2}{c|}{ H1b category} & \multicolumn{2}{c}{ H2b category}\\
\hline
\TTbar (0.8\TeV) & \multicolumn{2}{c|}{$     21.5 \pm      2.1 $} & \multicolumn{2}{c}{$      4.4 \pm      0.7 $} \\
\TTbar (1.2\TeV) & \multicolumn{2}{c|}{$      1.5 \pm      0.2 $} & \multicolumn{2}{c}{$     0.31 \pm     0.05 $} \\
\hline
TOP & \multicolumn{2}{c|}{$   1050 \pm    220 $} & \multicolumn{2}{c}{$     29.6 \pm      8.6 $} \\
EW  & \multicolumn{2}{c|}{$     45 \pm     11 $} & \multicolumn{2}{c}{$      2.5 \pm      0.9 $} \\
QCD & \multicolumn{2}{c|}{$     50 \pm     55 $} & \multicolumn{2}{c}{$      4.4 \pm      5.1 $} \\
\hline
Total bkg. & \multicolumn{2}{c|}{$   1150 \pm    260 $} & \multicolumn{2}{c}{$     37 \pm     12 $} \\
Data & \multicolumn{2}{c|}{1204} & \multicolumn{2}{c}{43} \\
\end{tabular}
\end{table}

After the final event selection, no significant excess above the SM expectations is observed in data. We set 95\% CL upper limits on the cross section of \TTbar production in various branching fraction scenarios. These limits are defined as Bayesian credible intervals~\cite{PDGSTATS} and are derived using the \textsc{Theta}~\cite{THETA} program.
Statistical uncertainties due to the finite size of the MC samples are accounted for using the Barlow--Beeston lite method~\cite{BBLITE1}. Systematic uncertainties are treated as nuisance parameters with log-normal priors for normalization uncertainties, Gaussian priors for shape uncertainties with shifted templates, and a flat prior on the signal cross section. The limits are then calculated by simultaneously fitting the binned marginal likelihoods obtained from the $\min[M(\ell,\,\PQb)]$ distributions in all \boostedW categories and the \ST distributions in all \boostedHiggs categories. This creates a combined search with 20 categories after dividing into electron and muon channels: 16 categories from the \boostedW channel and 4 categories with a boosted Higgs boson. The systematic uncertainties for these categories are correlated, as described in \qsec{sec:systs}.

\begin{figure}[tbp]
\centering
\includegraphics[width=0.49\textwidth]{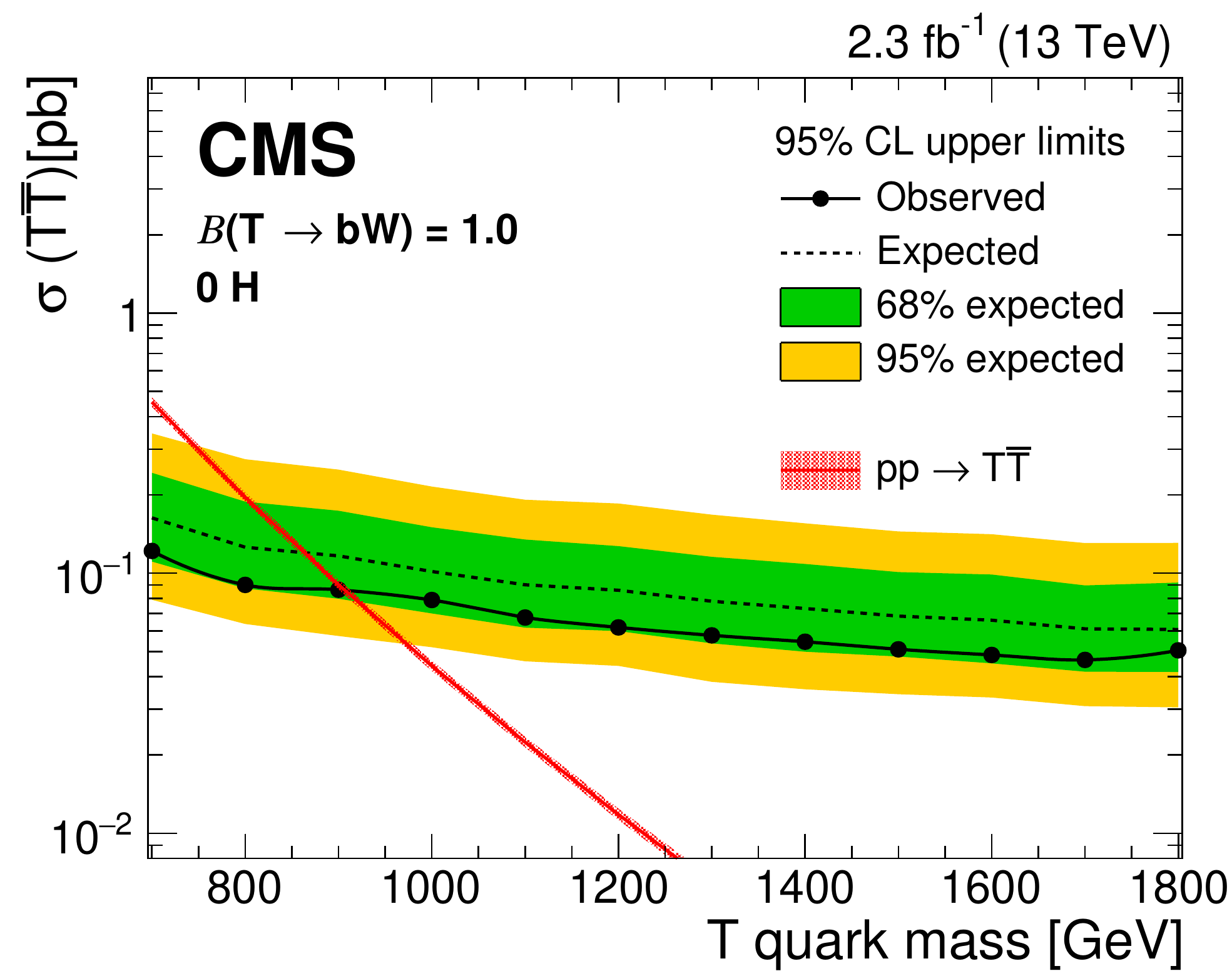}
\includegraphics[width=0.49\textwidth]{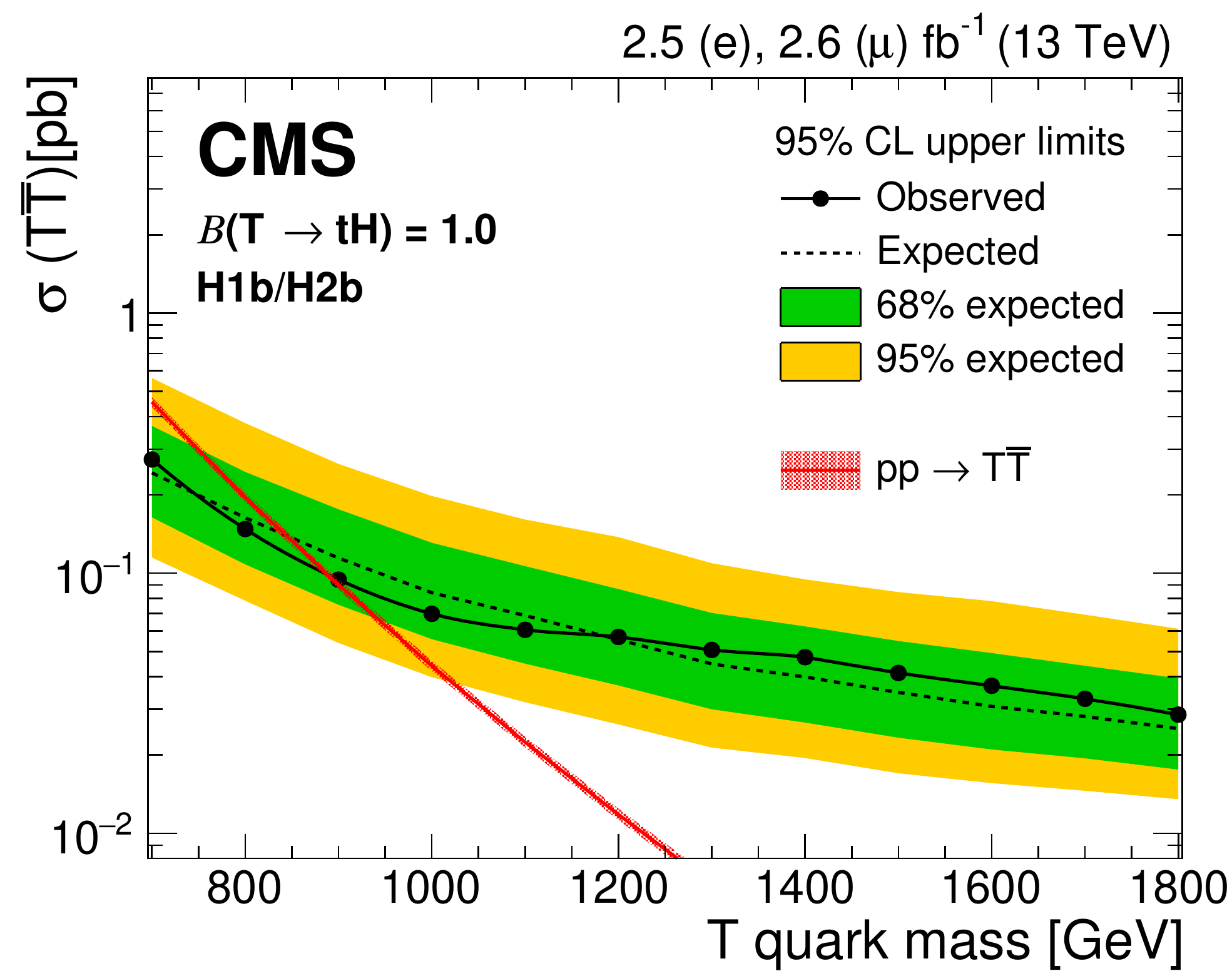}
\caption{The expected and observed upper limits (Bayesian) at 95\% CL on the cross section of \TTbar production for 100\% $\PQT\to\PQb\PW$ in the \boostedW channel (left), and 100\% $\PQT\to\PQt\PH$ in the \boostedHiggs channel (right). The theoretically predicted cross section for \TTbar production calculated at NNLO is shown as red line, with the uncertainties in the PDFs and renormalization and factorization scales indicated by the shaded area. Masses below 700\GeV were excluded previously.}
\label{fig:individual}
\end{figure}

\begin{figure}[tbp]
\centering
\includegraphics[width=0.49\textwidth]{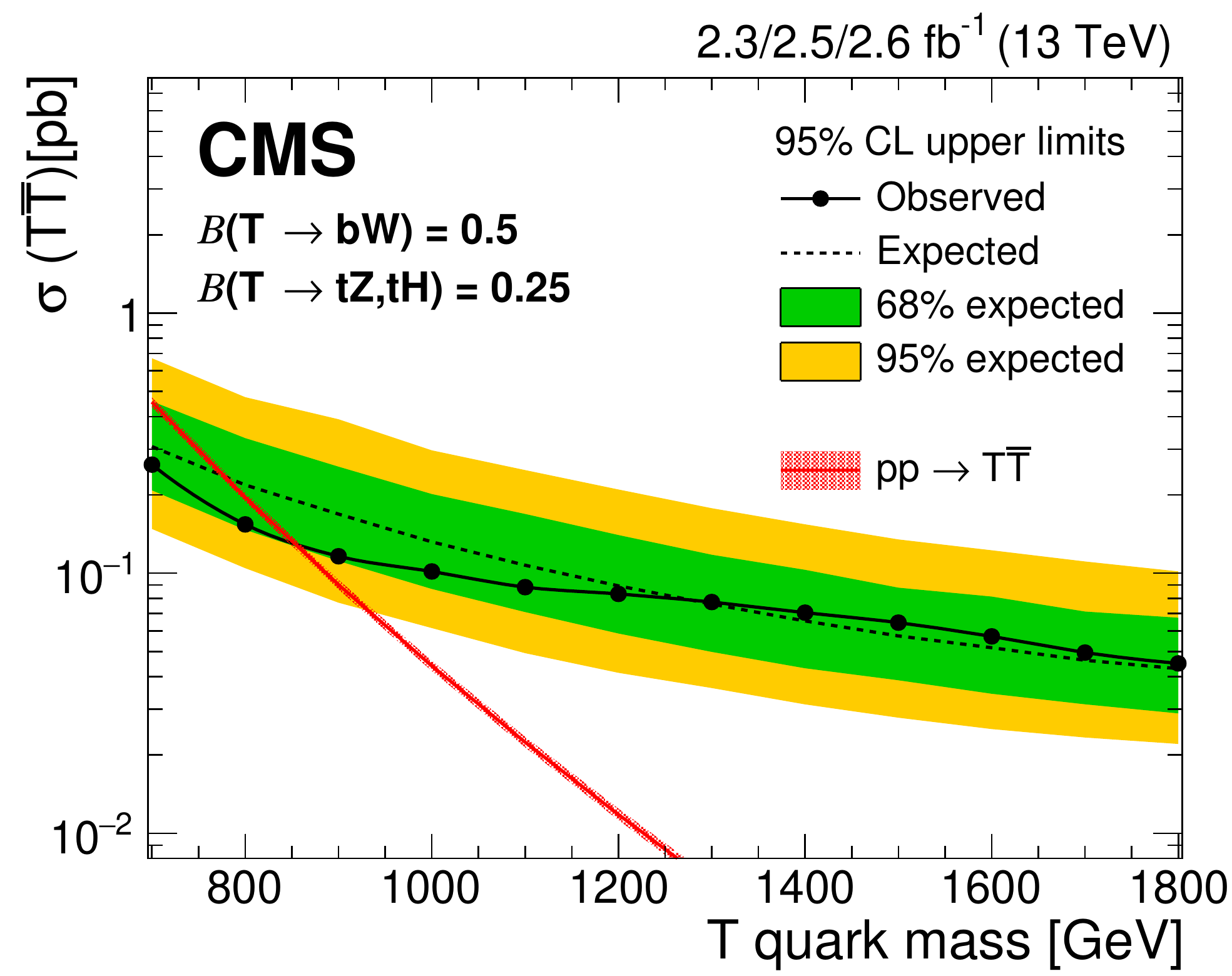}
\includegraphics[width=0.49\textwidth]{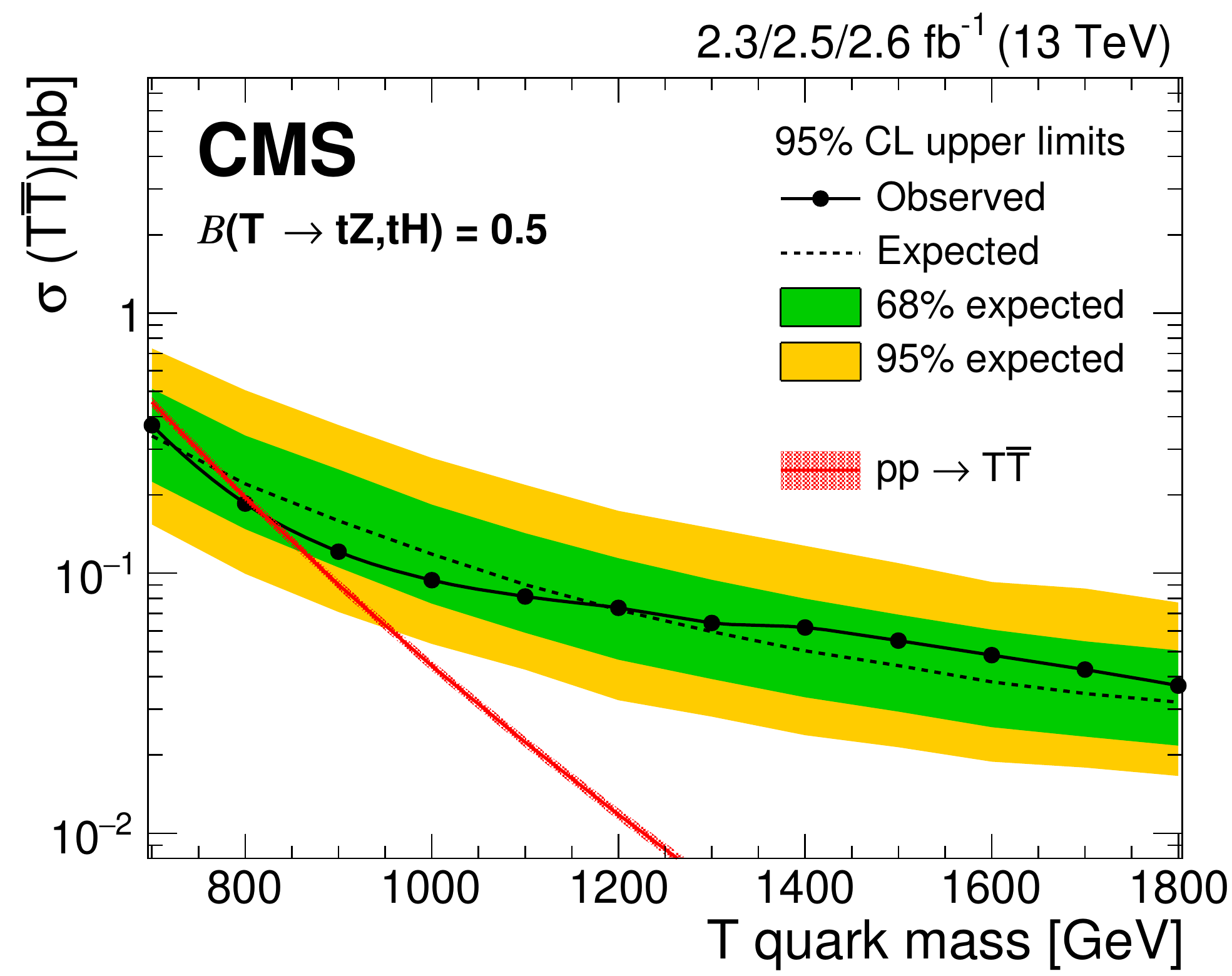}
\caption{The expected and observed upper limits (Bayesian) at 95\% CL on the cross section of \TTbar production for the singlet benchmark (left) and the doublet benchmark (right) after combining the \boostedW and \boostedHiggs channels. The theoretically predicted cross section for \TTbar production calculated at NNLO is shown as red line, with the uncertainties in the PDFs and renormalization and factorization scales indicated by the shaded area. Masses below 700\GeV were excluded previously.}
\label{fig:combinedSD}

\end{figure}

\begin{figure}[tbp]
\centering
\includegraphics[width=0.49\textwidth]{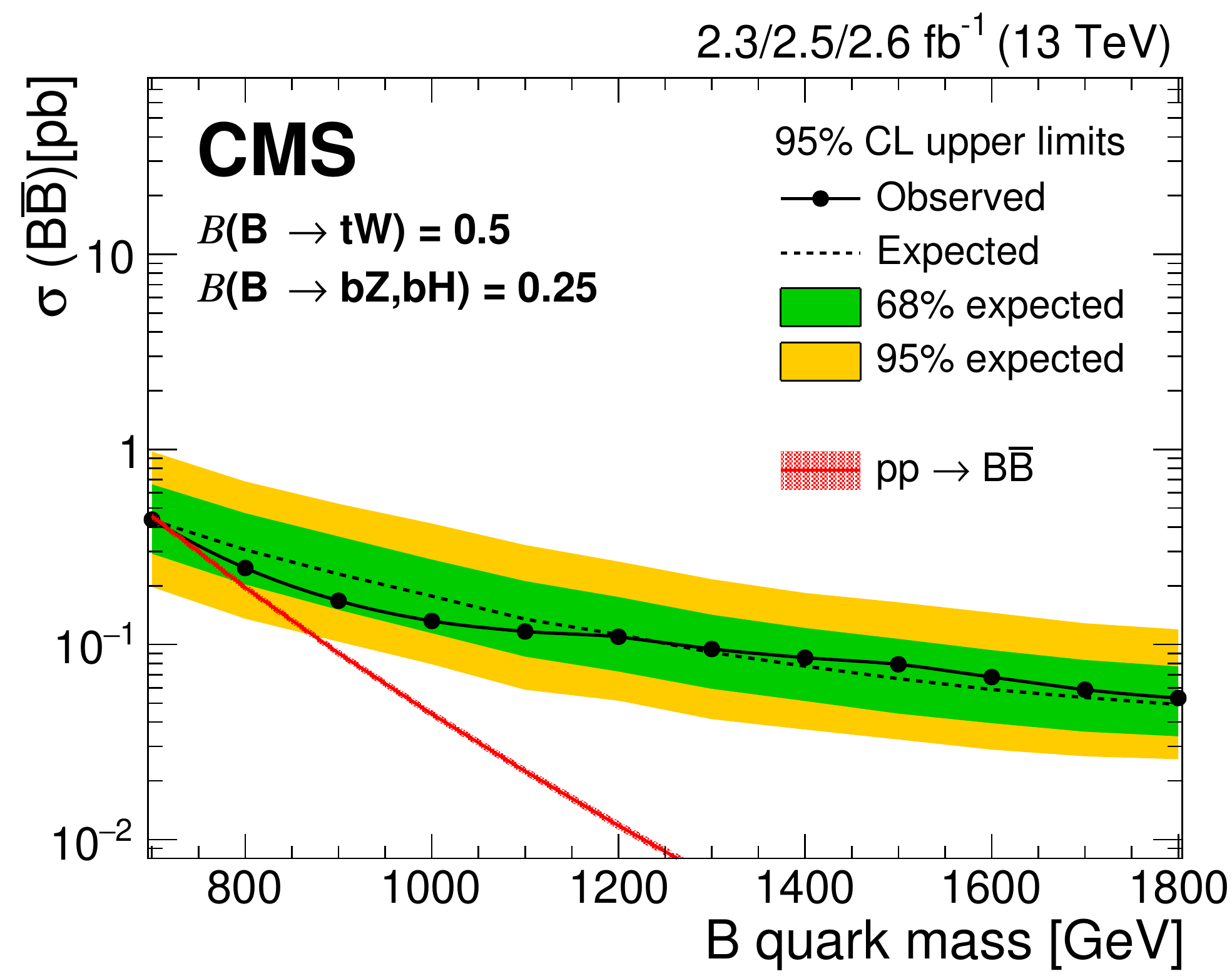}
\includegraphics[width=0.49\textwidth]{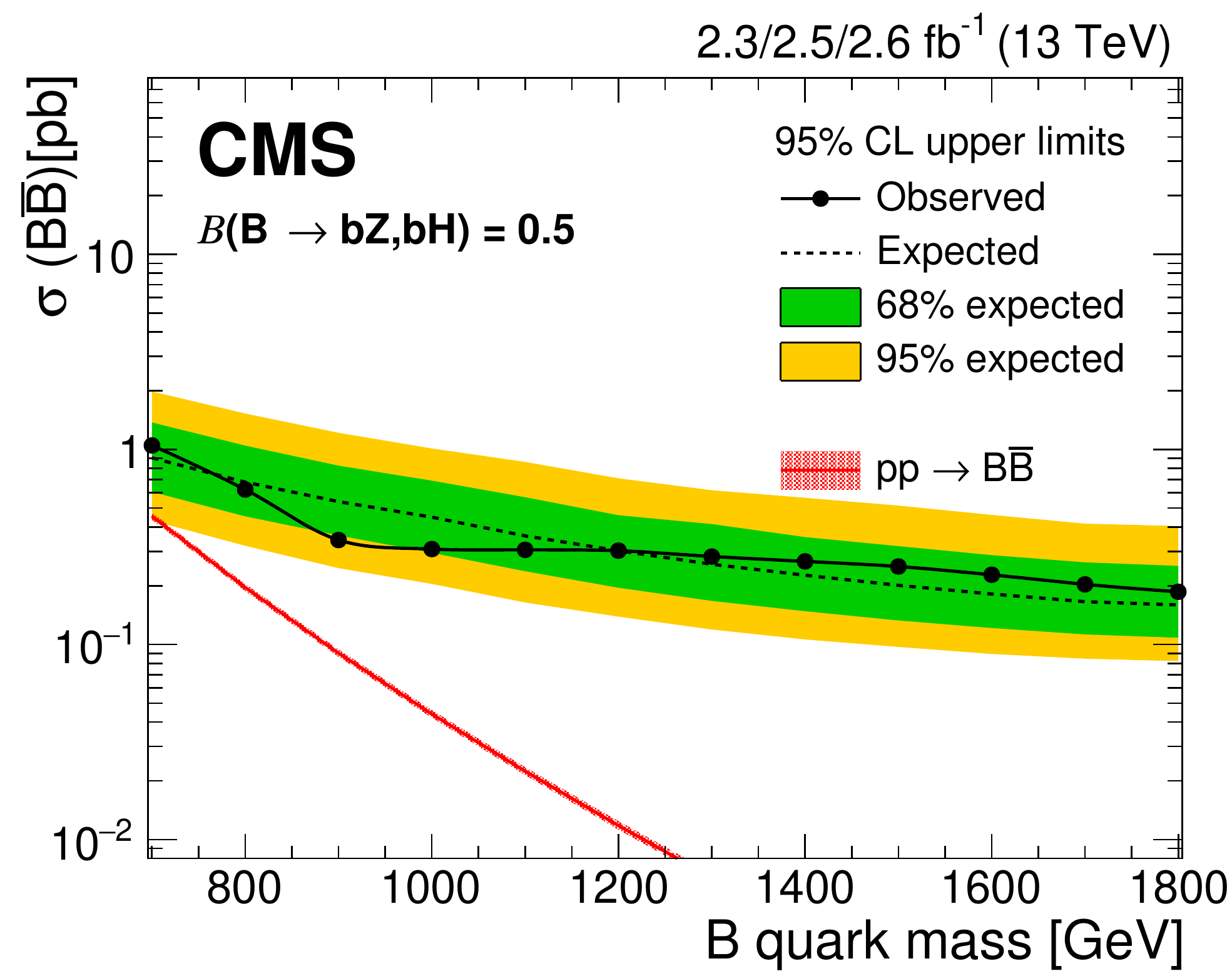}
\caption{The expected and observed upper limits (Bayesian) at 95\% CL on the cross section of $\BBbar$ production for the singlet benchmark (left) and the doublet benchmark (right) after combining the \boostedW and \boostedHiggs channels. The theoretically predicted cross section for $\BBbar$ production calculated at NNLO is shown as red line, with the uncertainties in the PDFs and renormalization and factorization scales indicated by the shaded area. Masses below 700\GeV were excluded previously.}
\label{fig:combinedSDBB}

\end{figure}

Results for the individual channels are shown in \qfig{fig:individual}. The \boostedW channel excludes T quarks decaying only to bW with masses below 910\GeV (870\GeV expected), and the \boostedHiggs channel excludes T quarks decaying only to tH for masses below 890\GeV (860\GeV expected). In \qfig{fig:combinedSD} we present combined 95\% CL upper limits on the \TTbar production cross section for two VLQ benchmark branching fraction combinations: singlet (50\% bW, 25\% tZ/tH) and doublet (50\% tZ/tH). For an electroweak singlet T quark, the observed (expected) upper limits on the production cross section range from 0.26 to 0.04\unit{pb} (0.31 to 0.04\unit{pb})  and we exclude masses below 860\GeV (790\GeV). For a doublet T quark, the observed (expected) upper limits on the production cross section range from 0.37 to 0.04\unit{pb} (0.34 to 0.03\unit{pb}) and we exclude masses below 830\GeV (780\GeV). The corresponding benchmarks for B quark production are shown in \qfig{fig:combinedSDBB}, and we can exclude masses below 730\GeV (720\GeV expected) for the singlet branching fraction combination while for the doublet scenario, no lower mass limit above 700\GeV was observed. Sensitivity to \BBbar production in this search is limited by the single lepton selection efficiency for $\PQb\PZ$ and $\PQb\PH$ decays, as noted above. The combinations benefit from the difference in discriminating variables between the channels: the min[M($\ell$,b)] distributions used in the \boostedW channel provide good sensitivity to low-mass T quarks, while the peaking signal shape in the \ST distribution drives the combination at high masses. The observed exclusion limits are stronger than expected due to an over-prediction of the background that remains after the \HT-based reweighting, particularly in categories with a W-tagged jet and several b-tagged jets. This effect is not significant given the systematic uncertainty in the reweighting procedure.

\begin{figure}[tbp]
\centering
\includegraphics[width=0.49\textwidth]{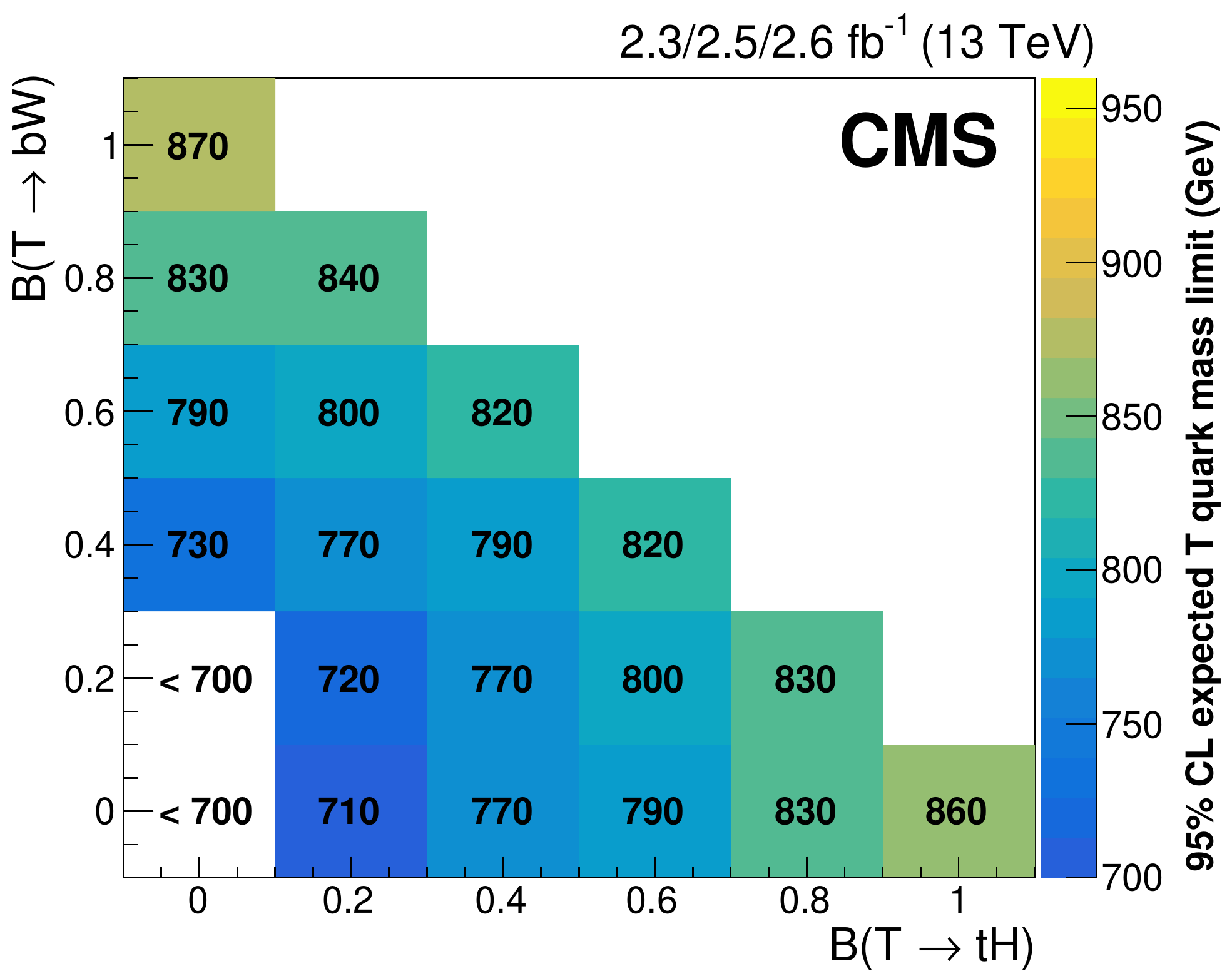}
\includegraphics[width=0.49\textwidth]{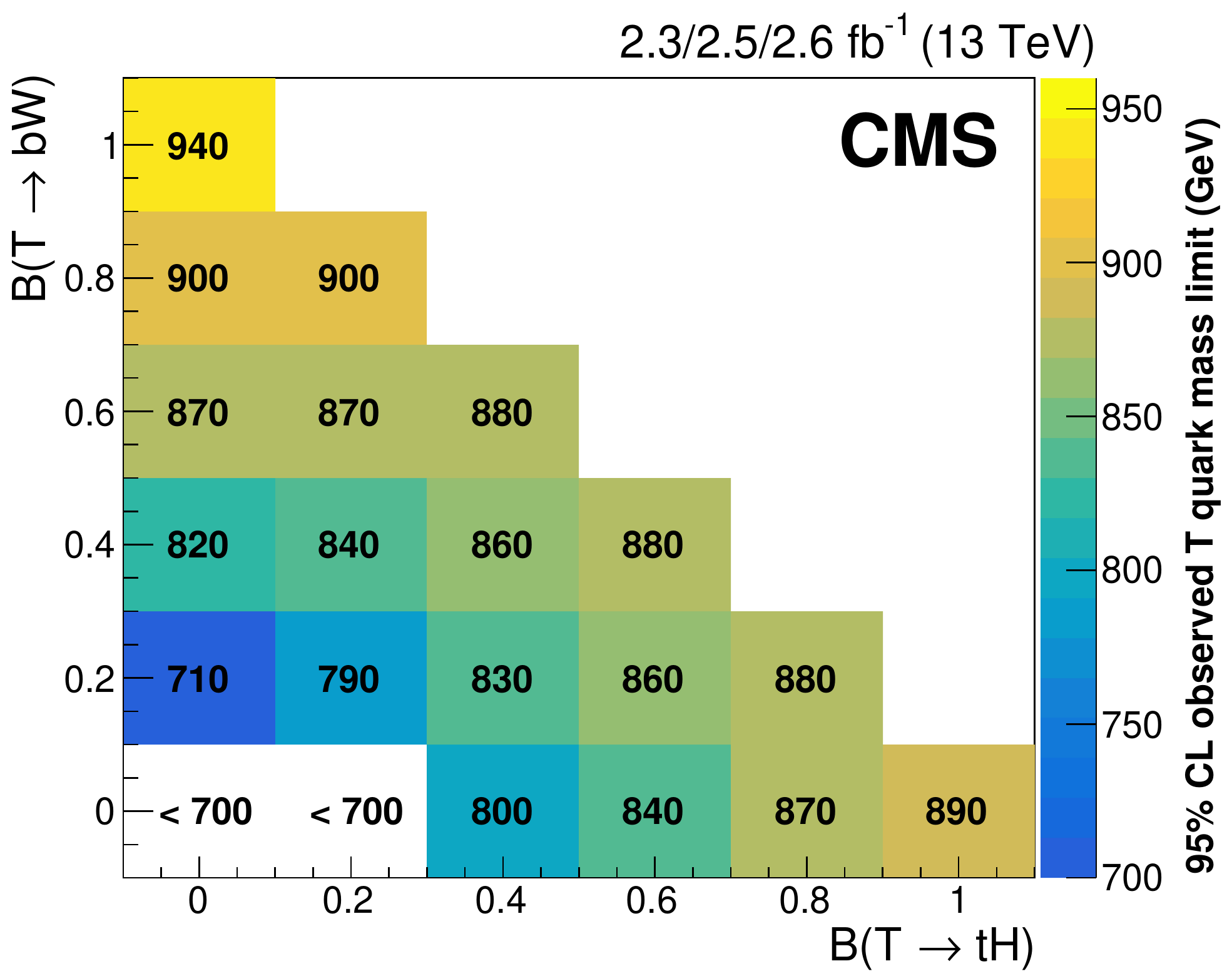}
\caption{The expected (left) and observed (right) at 95\% CL lower limits (Bayesian) on the T quark mass for a variety of $\PQT\to\PQt\PH$ and $\PQT\to\PQb\PW$ branching fraction combinations, indicated by the coordinates at the center of each box, after combining the \boostedW and \boostedHiggs channels. A limit of $<$700\GeV indicates that this search is not sensitive to T quark decays with that branching fraction combination.}
\label{fig:Tscan}

\end{figure}

Figure~\ref{fig:Tscan} shows expected and observed exclusion limits at 95\% CL on the T quark mass, for a scan of possible branching fractions: we set lower mass limits with values ranging from 790 to 940\GeV for combinations with $\mathcal{B}(\PQT\to\PQt\PH) + \mathcal{B}(\PQT\to\PQb\PW) \geq 0.4$. Compared to the combination of many leptonic and hadronic search channels in $\sqrt{s} = 8$\TeV collision data corresponding to an integrated luminosity of 19.7\fbinv, the current combination of two single lepton channels produces similar expected exclusion limits. This represents an improved sensitivity to \TTbar pair production at $\sqrt{s} = 13$\TeV due to the increase in the \TTbar production cross section from 8 to 13\TeV as well as to significant improvements in techniques for identifying boosted hadronic massive-particle decays. For branching fraction scenarios with $\mathcal{B}(\PQT\to\PQt\PH) + \mathcal{B}(\PQT\to\PQb\PW) \geq 0.4$ these results extend the excluded mass range of the 8\TeV search by up to 160\GeV.

\section{Summary}
The first search by CMS for pair-produced vector-like T and B quarks at $\sqrt{s}= 13$\TeV is presented, using data from proton-proton collisions recorded in 2015 corresponding to integrated luminosities of 2.3--2.6\fbinv. The search requires at least one lepton in the final state and is optimized for cases where a T quark decays to a boosted W or Higgs boson. No excess above the standard model background is observed and 95\% confidence level upper limits are placed on the cross section of \TTbar and \BBbar production. For an electroweak singlet T quark, masses below 860\GeV are excluded, and for a doublet T quark, masses below 830\GeV are excluded. Considering other possible branching fraction combinations for T quarks, and assuming that the sum of the branching fractions to bW, tH and tZ is equal to unity, we set lower mass limits that range from 790 to 940\GeV for combinations with $\mathcal{B}(\PQT\to\PQt\PH) + \mathcal{B}(\PQT\to\PQb\PW) \geq 0.4$. These results extend the sensitivity of previous CMS searches for many possible T quark decay scenarios, and
showcase the importance of new techniques for understanding highly-boosted final states in extending searches for new particles to higher masses.

\begin{acknowledgments}
We congratulate our colleagues in the CERN accelerator departments for the excellent performance of the LHC and thank the technical and administrative staffs at CERN and at other CMS institutes for their contributions to the success of the CMS effort. In addition, we gratefully acknowledge the computing centers and personnel of the Worldwide LHC Computing Grid for delivering so effectively the computing infrastructure essential to our analyses. Finally, we acknowledge the enduring support for the construction and operation of the LHC and the CMS detector provided by the following funding agencies: BMWFW and FWF (Austria); FNRS and FWO (Belgium); CNPq, CAPES, FAPERJ, and FAPESP (Brazil); MES (Bulgaria); CERN; CAS, MoST, and NSFC (China); COLCIENCIAS (Colombia); MSES and CSF (Croatia); RPF (Cyprus); SENESCYT (Ecuador); MoER, ERC IUT, and ERDF (Estonia); Academy of Finland, MEC, and HIP (Finland); CEA and CNRS/IN2P3 (France); BMBF, DFG, and HGF (Germany); GSRT (Greece); OTKA and NIH (Hungary); DAE and DST (India); IPM (Iran); SFI (Ireland); INFN (Italy); MSIP and NRF (Republic of Korea); LAS (Lithuania); MOE and UM (Malaysia); BUAP, CINVESTAV, CONACYT, LNS, SEP, and UASLP-FAI (Mexico); MBIE (New Zealand); PAEC (Pakistan); MSHE and NSC (Poland); FCT (Portugal); JINR (Dubna); MON, RosAtom, RAS, RFBR and RAEP (Russia); MESTD (Serbia); SEIDI, CPAN, PCTI and FEDER (Spain); Swiss Funding Agencies (Switzerland); MST (Taipei); ThEPCenter, IPST, STAR, and NSTDA (Thailand); TUBITAK and TAEK (Turkey); NASU and SFFR (Ukraine); STFC (United Kingdom); DOE and NSF (USA).

\hyphenation{Rachada-pisek} Individuals have received support from the Marie-Curie program and the European Research Council and Horizon 2020 Grant, contract No. 675440 (European Union); the Leventis Foundation; the A. P. Sloan Foundation; the Alexander von Humboldt Foundation; the Belgian Federal Science Policy Office; the Fonds pour la Formation \`a la Recherche dans l'Industrie et dans l'Agriculture (FRIA-Belgium); the Agentschap voor Innovatie door Wetenschap en Technologie (IWT-Belgium); the Ministry of Education, Youth and Sports (MEYS) of the Czech Republic; the Council of Science and Industrial Research, India; the HOMING PLUS program of the Foundation for Polish Science, cofinanced from European Union, Regional Development Fund, the Mobility Plus program of the Ministry of Science and Higher Education, the National Science Center (Poland), contracts Harmonia 2014/14/M/ST2/00428, Opus 2014/13/B/ST2/02543, 2014/15/B/ST2/03998, and 2015/19/B/ST2/02861, Sonata-bis 2012/07/E/ST2/01406; the National Priorities Research Program by Qatar National Research Fund; the Programa Clar\'in-COFUND del Principado de Asturias; the Thalis and Aristeia programs cofinanced by EU-ESF and the Greek NSRF; the Rachadapisek Sompot Fund for Postdoctoral Fellowship, Chulalongkorn University and the Chulalongkorn Academic into Its 2nd Century Project Advancement Project (Thailand); and the Welch Foundation, contract C-1845.

\end{acknowledgments}
\bibliography{auto_generated}

\cleardoublepage \appendix\section{The CMS Collaboration \label{app:collab}}\begin{sloppypar}\hyphenpenalty=5000\widowpenalty=500\clubpenalty=5000\input{B2G-16-024-authorlist.tex}\end{sloppypar}
\end{document}

%% file: B2G-16-024-authorlist.tex
\textbf{Yerevan Physics Institute,  Yerevan,  Armenia}\\*[0pt]
A.M.~Sirunyan, A.~Tumasyan
\vskip\cmsinstskip
\textbf{Institut f\"{u}r Hochenergiephysik,  Wien,  Austria}\\*[0pt]
W.~Adam, F.~Ambrogi, E.~Asilar, T.~Bergauer, J.~Brandstetter, E.~Brondolin, M.~Dragicevic, J.~Er\"{o}, M.~Flechl, M.~Friedl, R.~Fr\"{u}hwirth\cmsAuthorMark{1}, V.M.~Ghete, J.~Grossmann, J.~Hrubec, M.~Jeitler\cmsAuthorMark{1}, A.~K\"{o}nig, N.~Krammer, I.~Kr\"{a}tschmer, D.~Liko, T.~Madlener, I.~Mikulec, E.~Pree, D.~Rabady, N.~Rad, H.~Rohringer, J.~Schieck\cmsAuthorMark{1}, R.~Sch\"{o}fbeck, M.~Spanring, D.~Spitzbart, J.~Strauss, W.~Waltenberger, J.~Wittmann, C.-E.~Wulz\cmsAuthorMark{1}, M.~Zarucki
% ...
\vskip\cmsinstskip
% ...
1:~~Also at Vienna University of Technology, Vienna, Austria\\
% ...


%% file: B2G-16-024_temp.bbl
\providecommand{\href}[2]{#2}\begingroup\raggedright\begin{thebibliography}{10}%
\makeatletter
\providecommand{\hrefCMSnoop }[0]{\@secondoftwo}%
\makeatother
\providecommand{\doi}{\texttt{doi:}\begingroup \urlstyle{tt}\Url}

\bibitem{Aad20121}
\hrefCMSnoop {}{{ATLAS Collaboration}, ``{Observation of a new particle in the
  search for the Standard Model Higgs boson with the ATLAS detector at the
  LHC}'',} \textit{ Phys. Lett. B} \textbf{ 716} (2012) 1,
  \href{http://dx.doi.org/10.1016/j.physletb.2012.08.020}{\doi{10.1016/j.physletb.2012.08.020}},
\href{http://www.arXiv.org/abs/1207.7214}{\texttt{arXiv:1207.7214}}.
%%CITATION = ARXIV:1207.7214;%%.

\bibitem{Chatrchyan201230}
\hrefCMSnoop {}{{CMS Collaboration}, ``{Observation of a new boson at a mass of
  125 GeV with the CMS experiment at the LHC}'',} \textit{ Phys. Lett. B}
  \textbf{ 716} (2012) 30,
  \href{http://dx.doi.org/10.1016/j.physletb.2012.08.021}{\doi{10.1016/j.physletb.2012.08.021}},
\href{http://www.arXiv.org/abs/1207.7235}{\texttt{arXiv:1207.7235}}.
%%CITATION = ARXIV:1207.7235;%%.

\bibitem{Chatrchyan:2013lba}
\hrefCMSnoop {}{{CMS Collaboration}, ``{Observation of a new boson with mass
  near 125 GeV in pp collisions at $\sqrt{s}$ = 7 and 8 TeV}'',} \textit{ JHEP}
  \textbf{ 06} (2013) 081,
  \href{http://dx.doi.org/10.1007/JHEP06(2013)081}{\doi{10.1007/JHEP06(2013)081}},
\href{http://www.arXiv.org/abs/1303.4571}{\texttt{arXiv:1303.4571}}.
%%CITATION = ARXIV:1303.4571;%%.

\bibitem{LHC}
\hrefCMSnoop {}{L.~Evans and P.~Bryant, ``{LHC Machine}'',} \textit{ JINST}
  \textbf{ 3} (2008) S08001,
\href{http://dx.doi.org/10.1088/1748-0221/3/08/S08001}{\doi{10.1088/1748-0221/3/08/S08001}}.
%%CITATION = JINST,3,S08001;%%.

\bibitem{PhysRevD.69.075002}
\hrefCMSnoop {}{M.~Perelstein, M.~E. Peskin, and A.~Pierce, ``Top quarks and
  electroweak symmetry breaking in little {H}iggs models'',} \textit{ Phys.
  Rev. D} \textbf{ 69} (2004) 075002,
  \href{http://dx.doi.org/10.1103/PhysRevD.69.075002}{\doi{10.1103/PhysRevD.69.075002}},
  \href{http://www.arXiv.org/abs/hep-ph/0310039}{\texttt{arXiv:hep-ph/0310039}}.

\bibitem{Matsedonskyi2013}
\hrefCMSnoop {}{O.~Matsedonskyi, G.~Panico, and A.~Wulzer, ``Light top partners
  for a light composite {H}iggs'',} \textit{ JHEP} \textbf{ 01} (2013) 164,
  \href{http://dx.doi.org/10.1007/JHEP01(2013)164}{\doi{10.1007/JHEP01(2013)164}},
  \href{http://www.arXiv.org/abs/1204.6333}{\texttt{arXiv:1204.6333}}.

\bibitem{PhysRevD.75.055014}
\hrefCMSnoop {}{R.~Contino, L.~Da~Rold, and A.~Pomarol, ``Light custodians in
  natural composite {H}iggs models'',} \textit{ Phys. Rev. D} \textbf{ 75}
  (2007) 055014,
  \href{http://dx.doi.org/10.1103/PhysRevD.75.055014}{\doi{10.1103/PhysRevD.75.055014}},
  \href{http://www.arXiv.org/abs/hep-ph/0612048}{\texttt{arXiv:hep-ph/0612048}}.

\bibitem{compHiggs}
\hrefCMSnoop {}{R.~Contino, T.~Kramer, M.~Son, and R.~Sundrum,
  ``Warped/composite phenomenology simplified'',} \textit{ JHEP} \textbf{ 05}
  (2007) 074,
  \href{http://dx.doi.org/10.1088/1126-6708/2007/05/074}{\doi{10.1088/1126-6708/2007/05/074}},
  \href{http://www.arXiv.org/abs/hep-ph/0612180}{\texttt{arXiv:hep-ph/0612180}}.

\bibitem{KAPLAN1991259}
\hrefCMSnoop {}{D.~B. Kaplan, ``Flavor at {SSC} energies: A new mechanism for
  dynamically generated fermion masses'',} \textit{ Nucl. Phys. B} \textbf{
  365} (1991) 259,
  \href{http://dx.doi.org/10.1016/S0550-3213(05)80021-5}{\doi{10.1016/S0550-3213(05)80021-5}}.

\bibitem{Dugan:1984hq}
\hrefCMSnoop {}{M.~J. Dugan, H.~Georgi, and D.~B. Kaplan, ``Anatomy of a
  composite {H}iggs model'',} \textit{ Nucl. Phys. B} \textbf{ 254} (1985) 299,
\href{http://dx.doi.org/10.1016/0550-3213(85)90221-4}{\doi{10.1016/0550-3213(85)90221-4}}.
%%CITATION = NUPHA,B254,299;%%.

\bibitem{vecQuarkMix}
\hrefCMSnoop {}{J.~A. Aguilar-Saavedra, ``Mixing with vector-like quarks:
  constraints and expectations'',} in \textit{ {LHCP} 2013 -- {Large Hadron
  Collider Physics} 2013}, p.~16012.
\newblock 2013.
\newblock \href{http://www.arXiv.org/abs/1306.4432}{\texttt{arXiv:1306.4432}}.
\newblock
  \href{http://dx.doi.org/10.1051/epjconf/20136016012}{\doi{10.1051/epjconf/20136016012}}.

\bibitem{PhysRevLett.82.1628}
\hrefCMSnoop {}{F.~del Aguila, J.~A. Aguilar-Saavedra, and R.~Miquel,
  ``Constraints on top couplings in models with exotic quarks'',} \textit{
  Phys. Rev. Lett.} \textbf{ 82} (1999) 1628,
  \href{http://dx.doi.org/10.1103/PhysRevLett.82.1628}{\doi{10.1103/PhysRevLett.82.1628}},
  \href{http://www.arXiv.org/abs/hep-ph/9808400}{\texttt{arXiv:hep-ph/9808400}}.

\bibitem{LEP-2}
\hrefCMSnoop {}{{{ALEPH, LEP Electroweak, L3, DELPHI, and OPAL
  Collaborations}}, ``Electroweak measurements in electron-positron collisions
  at {W}-boson-pair energies at {LEP}'',} \textit{ Phys. Rept.} \textbf{ 532}
  (2013) 119,
  \href{http://dx.doi.org/10.1016/j.physrep.2013.07.004}{\doi{10.1016/j.physrep.2013.07.004}},
  \href{http://www.arXiv.org/abs/1302.3415}{\texttt{arXiv:1302.3415}}.

\bibitem{PhysRevLett.109.241802}
O.~Eberhardt\hrefCMSnoop {}{ {et~al.}, ``Impact of a {H}iggs boson at a mass of
  126 {GeV} on the standard model with three and four fermion generations'',}
  \textit{ Phys. Rev. Lett.} \textbf{ 109} (2012) 241802,
  \href{http://dx.doi.org/10.1103/PhysRevLett.109.241802}{\doi{10.1103/PhysRevLett.109.241802}},
  \href{http://www.arXiv.org/abs/1209.1101}{\texttt{arXiv:1209.1101}}.

\bibitem{Djouadi2012310}
\hrefCMSnoop {}{A.~Djouadi and A.~Lenz, ``Sealing the fate of a fourth
  generation of fermions'',} \textit{ Phys. Lett. B} \textbf{ 715} (2012) 310,
  \href{http://dx.doi.org/10.1016/j.physletb.2012.07.060}{\doi{10.1016/j.physletb.2012.07.060}},
  \href{http://www.arXiv.org/abs/1204.1252}{\texttt{arXiv:1204.1252}}.

\bibitem{PhysRevD.88.094010}
\hrefCMSnoop {}{J.~A. Aguilar-Saavedra, R.~Benbrik, S.~Heinemeyer, and
  M.~P\'erez-Victoria, ``Handbook of vectorlike quarks: Mixing and single
  production'',} \textit{ Phys. Rev. D} \textbf{ 88} (2013) 094010,
  \href{http://dx.doi.org/10.1103/PhysRevD.88.094010}{\doi{10.1103/PhysRevD.88.094010}}.

\bibitem{Ellis:2016jkw}
\hrefCMSnoop {}{J.~Ellis, ``{TikZ-Feynman}: Feynman diagrams with {TikZ}'',}
  \textit{ Comput. Phys. Commun.} \textbf{ 210} (2017) 103,
  \href{http://dx.doi.org/10.1016/j.cpc.2016.08.019}{\doi{10.1016/j.cpc.2016.08.019}},
\href{http://www.arXiv.org/abs/1601.05437}{\texttt{arXiv:1601.05437}}.
%%CITATION = ARXIV:1601.05437;%%.

\bibitem{DeSimone2013}
\hrefCMSnoop {}{A.~De~Simone, O.~Matsedonskyi, R.~Rattazzi, and A.~Wulzer, ``A
  first top partner hunter's guide'',} \textit{ JHEP} \textbf{ 04} (2013) 004,
  \href{http://dx.doi.org/10.1007/JHEP04(2013)004}{\doi{10.1007/JHEP04(2013)004}},
  \href{http://www.arXiv.org/abs/1211.5663}{\texttt{arXiv:1211.5663}}.

\bibitem{delAguila:1989rq}
\hrefCMSnoop {}{F.~del Aguila, L.~Ametller, G.~L. Kane, and J.~Vidal, ``{Vector
  like fermion and standard Higgs production at hadron colliders}'',} \textit{
  Nucl. Phys. B} \textbf{ 334} (1990) 1,
\href{http://dx.doi.org/10.1016/0550-3213(90)90655-W}{\doi{10.1016/0550-3213(90)90655-W}}.
%%CITATION = NUPHA,B334,1;%%.

\bibitem{signalxsec}
\hrefCMSnoop {}{O.~Matsedonskyi, G.~Panico, and A.~Wulzer, ``On the
  interpretation of top partners searches'',} \textit{ JHEP} \textbf{ 12}
  (2014) 097,
  \href{http://dx.doi.org/10.1007/JHEP12(2014)097}{\doi{10.1007/JHEP12(2014)097}},
\href{http://www.arXiv.org/abs/1409.0100}{\texttt{arXiv:1409.0100}}.
%%CITATION = ARXIV:1409.0100;%%.

\bibitem{CMScombo2014}
\hrefCMSnoop {}{{CMS Collaboration}, ``{Search for vector-like charge 2/3 T
  quarks in proton-proton collisions at $\sqrt{s}$ = 8 TeV}'',} \textit{ Phys.
  Rev. D} \textbf{ 93} (2016) 012003,
  \href{http://dx.doi.org/10.1103/PhysRevD.92.012003}{\doi{10.1103/PhysRevD.92.012003}},
  \href{http://www.arXiv.org/abs/1509.04177}{\texttt{arXiv:1509.04177}}.

\bibitem{Run1anal}
\hrefCMSnoop {}{{CMS Collaboration}, ``{Inclusive search for a vector-like T
  quark with charge $\frac{2}{3}$ in pp collisions at $\sqrt{s}$ = 8 TeV}'',}
  \textit{ Phys. Lett. B} \textbf{ 729} (2014) 149,
  \href{http://dx.doi.org/10.1016/j.physletb.2014.01.006}{\doi{10.1016/j.physletb.2014.01.006}},
\href{http://www.arXiv.org/abs/1311.7667}{\texttt{arXiv:1311.7667}}.
%%CITATION = ARXIV:1311.7667;%%.

\bibitem{PhysRevD.92.112007}
\hrefCMSnoop {}{{ATLAS Collaboration}, ``Search for pair production of a new
  heavy quark that decays into a $w$ boson and a light quark in pp collisions
  at $\sqrt{s}=8\text{ }\text{ }\mathrm{TeV}$ with the atlas detector'',}
  \textit{ Phys. Rev. D} \textbf{ 92} (2015) 112007,
  \href{http://dx.doi.org/10.1103/PhysRevD.92.112007}{\doi{10.1103/PhysRevD.92.112007}},
  \href{http://www.arXiv.org/abs/1509.04261}{\texttt{arXiv:1509.04261}}.

\bibitem{Aad2015}
\hrefCMSnoop {}{{ATLAS Collaboration}, ``Search for production of vector-like
  quark pairs and of four top quarks in the lepton-plus-jets final state in pp
  collisions at $\sqrt{s}=8\text{ }\text{ }\mathrm{TeV}$ with the atlas
  detector'',} \textit{ JHEP} \textbf{ 08} (2015) 105,
  \href{http://dx.doi.org/10.1007/JHEP08(2015)105}{\doi{10.1007/JHEP08(2015)105}},
  \href{http://www.arXiv.org/abs/1505.04306}{\texttt{arXiv:1505.04306}}.

\bibitem{Khachatryan:2015gza}
\hrefCMSnoop {}{{CMS Collaboration}, ``{Search for pair-produced vectorlike B
  quarks in proton-proton collisions at $\sqrt{s}$ = 8 TeV}'',} \textit{ Phys.
  Rev. D} \textbf{ 93} (2016) 112009,
  \href{http://dx.doi.org/10.1103/PhysRevD.93.112009}{\doi{10.1103/PhysRevD.93.112009}},
\href{http://www.arXiv.org/abs/1507.07129}{\texttt{arXiv:1507.07129}}.
%%CITATION = ARXIV:1507.07129;%%.

\bibitem{Aad:2015mba}
\hrefCMSnoop {}{{ATLAS Collaboration}, ``{Search for vector-like $B$ quarks in
  events with one isolated lepton, missing transverse momentum and jets at
  $\sqrt{s}=$ 8 TeV with the ATLAS detector}'',} \textit{ Phys. Rev. D}
  \textbf{ 91} (2015) 112011,
  \href{http://dx.doi.org/10.1103/PhysRevD.91.112011}{\doi{10.1103/PhysRevD.91.112011}},
\href{http://www.arXiv.org/abs/1503.05425}{\texttt{arXiv:1503.05425}}.
%%CITATION = ARXIV:1503.05425;%%.

\bibitem{Aaboud:2017qpr}
\hrefCMSnoop {}{{ATLAS Collaboration}, ``Search for pair production of
  vector-like top quarks in events with one lepton, jets, and missing
  transverse momentum in {$\sqrt{s} = 13$ TeV} pp collisions with the {ATLAS}
  detector'',} (2017).
  \href{http://www.arXiv.org/abs/1705.10751}{\texttt{arXiv:1705.10751}}.
Submitted to {JHEP}.
%%CITATION = ARXIV:1705.10751;%%.

\bibitem{Aaboud:2017zfn}
\hrefCMSnoop {}{{ATLAS Collaboration}, ``{Search for pair production of heavy
  vector-like quarks decaying to high-$p_{\mathrm{T}}$ $W$ bosons and $b$
  quarks in the lepton-plus-jets final state in $pp$ collisions at
  $\sqrt{s}$=13 TeV with the ATLAS detector}'',} (2017).
  \href{http://www.arXiv.org/abs/1707.03347}{\texttt{arXiv:1707.03347}}.
Submitted to {JHEP}.
%%CITATION = ARXIV:1707.03347;%%.

\bibitem{Chatrchyan:2008zzk}
\hrefCMSnoop {}{{CMS Collaboration}, ``The {CMS} experiment at the {CERN}
  {LHC}'',} \textit{ JINST} \textbf{ 3} (2008) S08004,
\href{http://dx.doi.org/10.1088/1748-0221/3/08/S08004}{\doi{10.1088/1748-0221/3/08/S08004}}.
%%CITATION = JINST,3,S08004;%%.

\bibitem{CMS-PRF-14-001}
\hrefCMSnoop {}{{CMS Collaboration}, ``Particle-flow reconstruction and global
  event description with the {CMS} detector'',} (2017).
  \href{http://www.arXiv.org/abs/1706.04965}{\texttt{arXiv:1706.04965}}.
Submitted to {JINST}.
%%CITATION = ARXIV:1706.04965;%%.

\bibitem{Khachatryan:2015hwa}
\hrefCMSnoop {}{{CMS Collaboration}, ``{Performance of electron reconstruction
  and selection with the CMS detector in proton-proton collisions at $\sqrt{s}
  = 8$\TeV}'',} \textit{ JINST} \textbf{ 10} (2015) P06005,
  \href{http://dx.doi.org/10.1088/1748-0221/10/06/P06005}{\doi{10.1088/1748-0221/10/06/P06005}},
\href{http://www.arXiv.org/abs/1502.02701}{\texttt{arXiv:1502.02701}}.
%%CITATION = ARXIV:1502.02701;%%.

\bibitem{Chatrchyan:2012xi}
\hrefCMSnoop {}{{CMS Collaboration}, ``{Performance of CMS muon reconstruction
  in pp collision events at $\sqrt{s} = 7$\TeV}'',} \textit{ JINST} \textbf{ 7}
  (2012) P10002,
  \href{http://dx.doi.org/10.1088/1748-0221/7/10/P10002}{\doi{10.1088/1748-0221/7/10/P10002}},
\href{http://www.arXiv.org/abs/1206.4071}{\texttt{arXiv:1206.4071}}.
%%CITATION = ARXIV:1206.4071;%%.

\bibitem{Cacciari:2008gp}
\hrefCMSnoop {}{M.~Cacciari, G.~P. Salam, and G.~Soyez, ``The
  anti-$k_\mathrm{t}$ jet clustering algorithm'',} \textit{ JHEP} \textbf{ 04}
  (2008) 063,
  \href{http://dx.doi.org/10.1088/1126-6708/2008/04/063}{\doi{10.1088/1126-6708/2008/04/063}},
  \href{http://www.arXiv.org/abs/0802.1189}{\texttt{arXiv:0802.1189}}.

\bibitem{Cacciari:2011ma}
\hrefCMSnoop {}{M.~Cacciari, G.~P. Salam, and G.~Soyez, ``{FastJet user
  manual}'',} \textit{ Eur. Phys. J. C} \textbf{ 72} (2012) 1896,
  \href{http://dx.doi.org/10.1140/epjc/s10052-012-1896-2}{\doi{10.1140/epjc/s10052-012-1896-2}},
\href{http://www.arXiv.org/abs/1111.6097}{\texttt{arXiv:1111.6097}}.
%%CITATION = ARXIV:1111.6097;%%.

\bibitem{Cacciari:2008gn}
\hrefCMSnoop {}{M.~Cacciari, G.~P. Salam, and G.~Soyez, ``{The catchment area
  of jets}'',} \textit{ JHEP} \textbf{ 04} (2008) 005,
  \href{http://dx.doi.org/10.1088/1126-6708/2008/04/005}{\doi{10.1088/1126-6708/2008/04/005}},
\href{http://www.arXiv.org/abs/0802.1188}{\texttt{arXiv:0802.1188}}.
%%CITATION = ARXIV:0802.1188;%%.

\bibitem{Khachatryan:2016kdb}
\hrefCMSnoop {}{{CMS Collaboration}, ``{Jet energy scale and resolution in the
  CMS experiment in pp collisions at 8 TeV}'',} \textit{ JINST} \textbf{ 12}
  (2017) P02014,
  \href{http://dx.doi.org/10.1088/1748-0221/12/02/P02014}{\doi{10.1088/1748-0221/12/02/P02014}},
\href{http://www.arXiv.org/abs/1607.03663}{\texttt{arXiv:1607.03663}}.
%%CITATION = ARXIV:1607.03663;%%.

\bibitem{Chatrchyan:2011ds}
\hrefCMSnoop {}{{CMS Collaboration}, ``{Determination of jet energy calibration
  and transverse momentum resolution in CMS}'',} \textit{ JINST} \textbf{ 6}
  (2011) P11002,
  \href{http://dx.doi.org/10.1088/1748-0221/6/11/P11002}{\doi{10.1088/1748-0221/6/11/P11002}},
\href{http://www.arXiv.org/abs/1107.4277}{\texttt{arXiv:1107.4277}}.
%%CITATION = ARXIV:1107.4277;%%.

\bibitem{Nason:2004rx}
\hrefCMSnoop {}{P.~Nason, ``A new method for combining {NLO QCD} with shower
  {Monte Carlo} algorithms'',} \textit{ JHEP} \textbf{ 11} (2004) 040,
  \href{http://dx.doi.org/10.1088/1126-6708/2004/11/040}{\doi{10.1088/1126-6708/2004/11/040}},
\href{http://www.arXiv.org/abs/hep-ph/0409146}{\texttt{arXiv:hep-ph/0409146}}.
%%CITATION = HEP-PH/0409146;%%.

\bibitem{Frixione:2007vw}
\hrefCMSnoop {}{S.~Frixione, P.~Nason, and C.~Oleari, ``Matching {NLO QCD}
  computations with parton shower simulations: the {POWHEG} method'',} \textit{
  JHEP} \textbf{ 11} (2007) 070,
  \href{http://dx.doi.org/10.1088/1126-6708/2007/11/070}{\doi{10.1088/1126-6708/2007/11/070}},
\href{http://www.arXiv.org/abs/0709.2092}{\texttt{arXiv:0709.2092}}.
%%CITATION = ARXIV:0709.2092;%%.

\bibitem{Alioli:2010xd}
\hrefCMSnoop {}{S.~Alioli, P.~Nason, C.~Oleari, and E.~Re, ``A general
  framework for implementing {NLO} calculations in shower {Monte Carlo}
  programs: the {POWHEG BOX}'',} \textit{ JHEP} \textbf{ 06} (2010) 043,
  \href{http://dx.doi.org/10.1007/JHEP06(2010)043}{\doi{10.1007/JHEP06(2010)043}},
\href{http://www.arXiv.org/abs/1002.2581}{\texttt{arXiv:1002.2581}}.
%%CITATION = ARXIV:1002.2581;%%.

\bibitem{Frixione:2007nw}
\hrefCMSnoop {}{S.~Frixione, P.~Nason, and G.~Ridolfi, ``A positive-weight
  next-to-leading-order {Monte Carlo} for heavy flavour hadroproduction'',}
  \textit{ JHEP} \textbf{ 09} (2007) 126,
  \href{http://dx.doi.org/10.1088/1126-6708/2007/09/126}{\doi{10.1088/1126-6708/2007/09/126}},
\href{http://www.arXiv.org/abs/0707.3088}{\texttt{arXiv:0707.3088}}.
%%CITATION = ARXIV:0707.3088;%%.

\bibitem{MADGRAPH}
J.~Alwall\hrefCMSnoop {}{ {et~al.}, ``{The automated computation of tree-level
  and next-to-leading order differential cross sections, and their matching to
  parton shower simulations}'',} \textit{ JHEP} \textbf{ 07} (2014) 079,
  \href{http://dx.doi.org/10.1007/JHEP07(2014)079}{\doi{10.1007/JHEP07(2014)079}},
  \href{http://www.arXiv.org/abs/hep-ph/1405.0301}{\texttt{arXiv:hep-ph/1405.0301}}.

\bibitem{FXFX}
\hrefCMSnoop {}{R.~Frederix and S.~Frixione, ``{Merging meets matching in
  MC@NLO}'',} \textit{ JHEP} \textbf{ 12} (2012) 061,
  \href{http://dx.doi.org/10.1007/JHEP12(2012)061}{\doi{10.1007/JHEP12(2012)061}},
  \href{http://www.arXiv.org/abs/1209.6215}{\texttt{arXiv:1209.6215}}.

\bibitem{MLMmatching}
J.~Alwall\hrefCMSnoop {}{ {et~al.}, ``Comparative study of various algorithms
  for the merging of parton showers and matrix elements in hadronic
  collisions'',} \textit{ Eur. Phys. J. C} \textbf{ 53} (2008) 473,
  \href{http://dx.doi.org/10.1140/epjc/s10052-007-0490-5}{\doi{10.1140/epjc/s10052-007-0490-5}},
\href{http://www.arXiv.org/abs/0706.2569}{\texttt{arXiv:0706.2569}}.
%%CITATION = ARXIV:0706.2569;%%.

\bibitem{Sjostrand:2006za}
\hrefCMSnoop {}{{T. Sj\"ostrand, S. Mrenna and P. Skands}, ``{PYTHIA} 6.4
  physics and manual'',} \textit{ JHEP} \textbf{ 05} (2006) 026,
  \href{http://dx.doi.org/10.1088/1126-6708/2006/05/026}{\doi{10.1088/1126-6708/2006/05/026}},
\href{http://www.arXiv.org/abs/hep-ph/0603175}{\texttt{arXiv:hep-ph/0603175}}.
%%CITATION = HEP-PH/0603175;%%.

\bibitem{Sjostrand:2014zea}
T.~Sj{\"o}strand\hrefCMSnoop {}{ {et~al.}, ``An introduction to pythia 8.2'',}
  \textit{ Comput. Phys. Commun.} \textbf{ 191} (2015) 159,
  \href{http://dx.doi.org/10.1016/j.cpc.2015.01.024}{\doi{10.1016/j.cpc.2015.01.024}},
\href{http://www.arXiv.org/abs/1410.3012}{\texttt{arXiv:1410.3012}}.
%%CITATION = ARXIV:1410.3012;%%.

\bibitem{TPRIMEXSEC}
\hrefCMSnoop {}{M.~Czakon and A.~Mitov, ``{Top++}: A program for the
  calculation of the top-pair cross-section at hadron colliders'',} \textit{
  Comput. Phys. Commun.} \textbf{ 185} (2014) 2930,
  \href{http://dx.doi.org/10.1016/j.cpc.2014.06.021}{\doi{10.1016/j.cpc.2014.06.021}},
  \href{http://www.arXiv.org/abs/1112.5675}{\texttt{arXiv:1112.5675}}.

\bibitem{MITOV1}
\hrefCMSnoop {}{M.~Czakon, P.~Fiedler, and A.~Mitov, ``Total top-quark
  pair-production cross section at hadron colliders through
  $\mathcal{O}({\ensuremath{\alpha}}_{S}^{4})$'',} \textit{ Phys. Rev. Lett.}
  \textbf{ 110} (2013) 252004,
  \href{http://dx.doi.org/10.1103/PhysRevLett.110.252004}{\doi{10.1103/PhysRevLett.110.252004}},
  \href{http://www.arXiv.org/abs/1303.6254}{\texttt{arXiv:1303.6254}}.

\bibitem{MITOV2}
\hrefCMSnoop {}{M.~Czakon and A.~Mitov, ``{NNLO corrections to top pair
  production at hadron colliders: the quark-gluon reaction}'',} \textit{ JHEP}
  \textbf{ 01} (2013) 080,
  \href{http://dx.doi.org/10.1007/JHEP01(2013)080}{\doi{10.1007/JHEP01(2013)080}},
  \href{http://www.arXiv.org/abs/1210.6832}{\texttt{arXiv:1210.6832}}.

\bibitem{MITOV3}
\hrefCMSnoop {}{M.~Czakon and A.~Mitov, ``{NNLO corrections to top-pair
  production at hadron colliders: the all-fermionic scattering channels}'',}
  \textit{ JHEP} \textbf{ 12} (2012) 054,
  \href{http://dx.doi.org/10.1007/JHEP12(2012)054}{\doi{10.1007/JHEP12(2012)054}},
  \href{http://www.arXiv.org/abs/1207.0236}{\texttt{arXiv:1207.0236}}.

\bibitem{BARNREUTHER}
\hrefCMSnoop {}{P.~B{\"a}rnreuther, M.~Czakon, and A.~Mitov, ``Percent level
  precision physics at the {Tevatron}: First genuine {NNLO QCD} corrections to
  $\mathrm{q \bar{q} \to t \bar{t}} + x$'',} \textit{ Phys. Rev. Lett.}
  \textbf{ 109} (2012) 132001,
  \href{http://dx.doi.org/10.1103/PhysRevLett.109.132001}{\doi{10.1103/PhysRevLett.109.132001}},
  \href{http://www.arXiv.org/abs/1204.5201}{\texttt{arXiv:1204.5201}}.

\bibitem{NNLL}
M.~Cacciari\hrefCMSnoop {}{ {et~al.}, ``{Top-pair production at hadron
  colliders with next-to-next-to-leading logarithmic soft-gluon
  resummation}'',} \textit{ Phys. Lett. B} \textbf{ 710} (2012) 612,
  \href{http://dx.doi.org/10.1016/j.physletb.2012.03.013}{\doi{10.1016/j.physletb.2012.03.013}},
  \href{http://www.arXiv.org/abs/1111.5869}{\texttt{arXiv:1111.5869}}.

\bibitem{Skands:2014pea}
\hrefCMSnoop {}{P.~Skands, S.~Carrazza, and J.~Rojo, ``Tuning {PYTHIA} 8.1: the
  {Monash} 2013 tune'',} \textit{ Eur. Phys. J. C} \textbf{ 74} (2014) 3024,
  \href{http://dx.doi.org/10.1140/epjc/s10052-014-3024-y}{\doi{10.1140/epjc/s10052-014-3024-y}},
\href{http://www.arXiv.org/abs/1404.5630}{\texttt{arXiv:1404.5630}}.
%%CITATION = ARXIV:1404.5630;%%.

\bibitem{Khachatryan:2015pea}
\hrefCMSnoop {}{{CMS Collaboration}, ``{Event generator tunes obtained from
  underlying event and multiparton scattering measurements}'',} \textit{ Eur.
  Phys. J. C} \textbf{ 76} (2016) 155,
  \href{http://dx.doi.org/10.1140/epjc/s10052-016-3988-x}{\doi{10.1140/epjc/s10052-016-3988-x}},
\href{http://www.arXiv.org/abs/1512.00815}{\texttt{arXiv:1512.00815}}.
%%CITATION = ARXIV:1512.00815;%%.

\bibitem{NNPDF30}
\hrefCMSnoop {}{{NNPDF} Collaboration, ``{Parton distributions for the LHC Run
  II}'',} \textit{ JHEP} \textbf{ 04} (2015) 040,
  \href{http://dx.doi.org/10.1007/JHEP04(2015)040}{\doi{10.1007/JHEP04(2015)040}},
  \href{http://www.arXiv.org/abs/1410.8849}{\texttt{arXiv:1410.8849}}.

\bibitem{GEANT4}
\hrefCMSnoop {}{{GEANT4} Collaboration, ``{GEANT4: A simulation toolkit}'',}
  \textit{ Nucl. Instrum. Meth. A} \textbf{ 506} (2003) 250,
  \href{http://dx.doi.org/10.1016/S0168-9002(03)01368-8}{\doi{10.1016/S0168-9002(03)01368-8}}.

\bibitem{Chatrchyan:2014fea}
\hrefCMSnoop {}{{CMS Collaboration}, ``{Description and performance of track
  and primary-vertex reconstruction with the CMS tracker}'',} \textit{ JINST}
  \textbf{ 9} (2014) P10009,
  \href{http://dx.doi.org/10.1088/1748-0221/9/10/P10009}{\doi{10.1088/1748-0221/9/10/P10009}},
\href{http://www.arXiv.org/abs/1405.6569}{\texttt{arXiv:1405.6569}}.
%%CITATION = ARXIV:1405.6569;%%.

\bibitem{ATLASpileup}
\hrefCMSnoop {}{{ATLAS Collaboration}, ``Measurement of the inelastic
  proton-proton cross section at $\sqrt{s}$ = 13~{TeV} with the {ATLAS}
  detector at the {LHC}'',} \textit{ Phys. Rev. Lett.} \textbf{ 117} (2016)
  182002,
  \href{http://dx.doi.org/10.1103/PhysRevLett.117.182002}{\doi{10.1103/PhysRevLett.117.182002}},
\href{http://www.arXiv.org/abs/1606.02625}{\texttt{arXiv:1606.02625}}.
%%CITATION = ARXIV:1606.02625;%%.

\bibitem{Khachatryan:2010xn}
\hrefCMSnoop {}{{CMS Collaboration}, ``Measurements of inclusive {W} and {Z}
  cross sections in pp collisions at $\sqrt{s}=7$ {TeV}'',} \textit{ JHEP}
  \textbf{ 01} (2011) 080,
  \href{http://dx.doi.org/10.1007/JHEP01(2011)080}{\doi{10.1007/JHEP01(2011)080}},
\href{http://www.arXiv.org/abs/1012.2466}{\texttt{arXiv:1012.2466}}.
%%CITATION = ARXIV:1012.2466;%%.

\bibitem{NSUBJETS}
\hrefCMSnoop {}{J.~Thaler and K.~Van~Tilburg, ``Maximizing boosted top
  identification by minimizing {$N$-subjettiness}'',} \textit{ JHEP} \textbf{
  02} (2012) 093,
  \href{http://dx.doi.org/10.1007/JHEP02(2012)093}{\doi{10.1007/JHEP02(2012)093}},
\href{http://www.arXiv.org/abs/1108.2701}{\texttt{arXiv:1108.2701}}.
%%CITATION = ARXIV:1108.2701;%%.

\bibitem{PRUNING}
\hrefCMSnoop {}{S.~D. Ellis, C.~K. Vermilion, and J.~R. Walsh, ``Techniques for
  improved heavy particle searches with jet substructure'',} \textit{ Phys.
  Rev. D} \textbf{ 80} (2009) 051501,
  \href{http://dx.doi.org/10.1103/PhysRevD.80.051501}{\doi{10.1103/PhysRevD.80.051501}},
  \href{http://www.arXiv.org/abs/0903.5081}{\texttt{arXiv:0903.5081}}.

\bibitem{SOFTDROP}
\hrefCMSnoop {}{A.~J. Larkoski, S.~Marzani, G.~Soyez, and J.~Thaler, ``Soft
  drop'',} \textit{ JHEP} \textbf{ 05} (2014) 146,
  \href{http://dx.doi.org/10.1007/JHEP05(2014)146}{\doi{10.1007/JHEP05(2014)146}},
  \href{http://www.arXiv.org/abs/hep-ph/1402.2657}{\texttt{arXiv:hep-ph/1402.2657}}.

\bibitem{BTAG}
\hrefCMSnoop {}{{CMS Collaboration}, ``Identification of b-quark jets with the
  {CMS} experiment'',} \textit{ JINST} \textbf{ 8} (2013) P04013,
  \href{http://dx.doi.org/10.1088/1748-0221/8/04/P04013}{\doi{10.1088/1748-0221/8/04/P04013}},
\href{http://www.arXiv.org/abs/1211.4462}{\texttt{arXiv:1211.4462}}.
%%CITATION = ARXIV:1211.4462;%%.

\bibitem{WTAGSFS}
\href {http://cdsweb.cern.ch/record/2256875}{{CMS Collaboration}, ``{Jet
  algorithms performance in 13 TeV data}'',} CMS Physics Analysis Summary
  CMS-PAS-JME-16-003, 2016.

\bibitem{B2G16005}
\hrefCMSnoop {}{{CMS Collaboration}, ``{Search for electroweak production of a
  vector-like quark decaying to a top quark and a Higgs boson using boosted
  topologies in fully hadronic final states}'',} \textit{ JHEP} \textbf{ 04}
  (2017) 136,
  \href{http://dx.doi.org/10.1007/JHEP04(2017)136}{\doi{10.1007/JHEP04(2017)136}},
\href{http://www.arXiv.org/abs/1612.05336}{\texttt{arXiv:1612.05336}}.
%%CITATION = ARXIV:1612.05336;%%.

\bibitem{B2G16006}
\hrefCMSnoop {}{{CMS Collaboration}, ``{Search for single production of
  vector-like quarks decaying into a b quark and a W boson in proton-proton
  collisions at $\sqrt{s} = $ 13 TeV}'',} (2017).
  \href{http://www.arXiv.org/abs/1701.08328}{\texttt{arXiv:1701.08328}}.
Submitted to {Phys. Lett. B}.
%%CITATION = ARXIV:1701.08328;%%.

\bibitem{PDGSTATS}
\hrefCMSnoop {}{G.~Cowan, ``{PDG} review on statistics (chap. 39)'',} \textit{
  Chin. Phys. C} \textbf{ 40} (2016) 100001,
\href{http://dx.doi.org/10.1088/1674-1137/40/10/100001}{\doi{10.1088/1674-1137/40/10/100001}}.
%%CITATION = CHPHD,C40,100001;%%.

\bibitem{THETA}
\href
  {http://www-ekp.physik.uni-karlsruhe.de/~ott/theta/theta-auto/index.html}{T.~M{\"u}ller,
  J.~Ott, and J.~Wagner-Kuhr, ``theta -- a framework for template-based
  modeling and inference'',} 2012.
\newblock
  \url{http://www-ekp.physik.uni-karlsruhe.de/~ott/theta/theta-auto/index.html}.

\bibitem{CMS-PAS-LUM-15-001}
\href {http://cds.cern.ch/record/2138682}{{CMS Collaboration}, ``{CMS
  luminosity measurement for the 2015 data-taking period}'',} Technical Report
  CMS-PAS-LUM-15-001, 2017.

\bibitem{ZZcms}
\hrefCMSnoop {}{{CMS Collaboration}, ``{Measurement of the ZZ production cross
  section and Z $\to \ell^+\ell^-\ell'^+\ell'^-$ branching fraction in pp
  collisions at $\sqrt{s}$ = 13 TeV}'',} \textit{ Phys. Lett. B} \textbf{ 763}
  (2016) 280,
  \href{http://dx.doi.org/10.1016/j.physletb.2016.10.054}{\doi{10.1016/j.physletb.2016.10.054}},
  \href{http://www.arXiv.org/abs/1607.08834}{\texttt{arXiv:1607.08834}}.

\bibitem{WZcms}
\hrefCMSnoop {}{{CMS Collaboration}, ``{Measurement of the WZ production cross
  section in pp collisions at $\sqrt{s} = 13$\TeV}'',} \textit{ Phys. Lett. B}
  \textbf{ 766} (2017) 268,
  \href{http://dx.doi.org/10.1016/j.physletb.2017.01.011}{\doi{10.1016/j.physletb.2017.01.011}},
\href{http://www.arXiv.org/abs/1607.06943}{\texttt{arXiv:1607.06943}}.
%%CITATION = ARXIV:1607.06943;%%.

\bibitem{Bahr:2008pv}
M.~B{\"a}hr\hrefCMSnoop {}{ {et~al.}, ``Herwig++ physics and manual'',}
  \textit{ Eur. Phys. J. C} \textbf{ 58} (2008) 639,
  \href{http://dx.doi.org/10.1140/epjc/s10052-008-0798-9}{\doi{10.1140/epjc/s10052-008-0798-9}},
\href{http://www.arXiv.org/abs/0803.0883}{\texttt{arXiv:0803.0883}}.
%%CITATION = ARXIV:0803.0883;%%.

\bibitem{BBLITE1}
\hrefCMSnoop {}{R.~J. Barlow and C.~Beeston, ``Fitting using finite monte carlo
  samples'',} \textit{ Comput. Phys. Commun.} (1993) 219,
  \href{http://dx.doi.org/10.1016/0010-4655(93)90005-W}{\doi{10.1016/0010-4655(93)90005-W}}.

\end{thebibliography}\endgroup
